\documentclass[11pt]{article}

\usepackage[a4paper,margin=25mm]{geometry}
\usepackage[T1]{fontenc}
\usepackage[utf8]{inputenc}
\usepackage{lmodern}
\usepackage{microtype}
\usepackage{amsmath,amssymb,mathtools,bm}
\usepackage{graphicx}
\usepackage{booktabs}
\usepackage{enumitem}
\usepackage[dvipsnames]{xcolor}
\usepackage{caption}
\usepackage{subcaption}
\usepackage{float}
\usepackage{tcolorbox}
\usepackage[numbers,sort&compress]{natbib}
\usepackage[colorlinks=true,linkcolor=MidnightBlue,citecolor=OliveGreen,urlcolor=BrickRed]{hyperref}
\usepackage[nameinlink,noabbrev]{cleveref}
\usepackage{fancyhdr}

\graphicspath{{figures/}}
\setlist[itemize]{leftmargin=5.5mm,itemsep=2pt,topsep=4pt}
\numberwithin{equation}{section}

\definecolor{draftgreen}{RGB}{25,92,68}
\definecolor{draftback}{RGB}{243,248,245}
\newtcolorbox{draftscope}{
  colback=draftback,
  colframe=draftgreen,
  boxrule=0.6pt,
  arc=1mm,
  left=2mm,
  right=2mm,
  top=1.5mm,
  bottom=1.5mm
}

\newcommand{\paperfigure}[2][]{%
  \IfFileExists{figures/#2}{%
    \includegraphics[#1]{#2}%
  }{%
    \IfFileExists{#2}{%
      \includegraphics[#1]{#2}%
    }{%
      \fbox{\parbox[c][42mm][c]{0.92\textwidth}{\centering
        \textbf{Figure file to be supplied}\\[1mm]
        \texttt{\detokenize{#2}}}}%
    }%
  }%
}

\newcommand{\ads}{\ell_{\mathrm{AdS}}}
\newcommand{\angL}{\mathfrak{L}}
\newcommand{\wallr}{\mathfrak{r}}
\newcommand{\rph}{r_{\mathrm{ph}}}
\newcommand{\OmR}{\Omega_{\wallr}}
\newcommand{\Omph}{\Omega_{\mathrm{ph}}}
\newcommand{\dd}{\mathrm{d}}
\newcommand{\ii}{\mathrm{i}}
\newcommand{\e}{\mathrm{e}}
\newcommand{\Order}{\mathcal{O}}

\title{\textbf{Conformal boundaries and the photon sphere of AdS black holes}}
\author{%
  Raúl Arias$^{1}$, Valentino Di Pierro$^{1}$ and Silvia Georgescu$^{2}$\\[0.8em]
  \parbox{0.9\textwidth}{\centering\small
    $^{1}$ Instituto de F\'isica La Plata -- CONICET and Departamento de F\'isica,\\
Universidad Nacional de La Plata, C.C. 67, 1900 La Plata, Argentina\\
   \vspace{0.1cm}
    $^{2}$ Department of Mathematics, King's College London, Strand, London WC2R 2LS, UK
  }
}

\providecommand{\bomega}{\bm{\omega}}

\date{}

\begin{document}

\maketitle

\begin{abstract}
We study linearized gravitational perturbations of four-dimensional Schwarzschild-AdS black holes with finite timelike boundaries subject to conformal boundary conditions. These fix the conformal class of the induced metric and the mean curvature, while allowing the boundary Weyl mode to fluctuate. The photon sphere plays a distinguished role in this problem: a static boundary at this radius is totally umbilic, and the scalar problem separates exactly into independent bulk and boundary sectors. We prove the absence of exponentially growing scalar and vector modes for spherical harmonics of degree two and higher. We determine how large angular momentum instabilities are suppressed as the boundary approaches umbilicity. Near the horizon, we recover the universal Rindler spectrum, including sound and shear excitations. We numerically follow these fluid modes as the boundary is moved outwards and find that they approach standard gravitational quasinormal modes at the asymptotic AdS$_4$ boundary. For large black holes, these asymptotic endpoints correspond to the sound and shear poles of the dual three-dimensional conformal field theory. We also study perturbations of the AdS$_4$ black brane, where we find an analogous continuation of the fluid modes in the absence of an intermediate umbilic boundary.
\end{abstract}

\newpage
\tableofcontents

\newpage

\section{Introduction}
\label{sec:introduction}

Gravity in a finite region of spacetime requires a choice of boundary data. This choice determines the admissible gravitational perturbations and hence the dynamical problem to be solved. One possibility, going back to York's formulation of gravitational boundary data \cite{York1972ConformalVfive}, is to fix the conformal class of the induced metric together with the trace of the extrinsic curvature, $K$. We refer to these as conformal boundary conditions (CBCs). These boundary conditions leave the local scale of the boundary metric unfixed. In Lorentzian signature, this degree of freedom becomes a boundary Weyl mode, whose evolution is determined together with the bulk gravitational field \cite{AnAnderson2021,AnninosAriasGalanteManeerat2025}. In Euclidean signature, CBCs have been shown, in contrast with the  Dirichlet choice, to define an elliptic boundary-value problem for the Einstein equations \cite{Anderson2008,Witten2021}. In Lorentzian signature, the question of well-posedness remains open. Although CBCs were conjectured to yield a well-posed initial-boundary-value problem \cite{AnAnderson2021}, subsequent analyses have revealed features of the linearized dynamics that make this question more subtle \cite{LiuReallSantosWiseman2025}, as we discuss below.

Over the past few years, gravity in finite regions with conformal boundaries has been explored in a variety of settings, including Minkowski and de Sitter space, vacuum AdS, near-extremal black holes, and higher-curvature gravity, together with its thermodynamic and holographic aspects \cite{AnninosGalanteManeerat2023,AnninosGalanteManeerat2024,AnninosAriasGalanteManeerat2025,GalanteManeeratSvesko2025,GalanteGeorgescuSvesko,GalanteMyersZikopoulos2026,Banihashemi:2025qqi,BanihashemiShaghoulianShashi2025,Allameh:2025gsa,Hamdan:2026olq}. Across these analyses, a recurring feature of the linearized gravitational dynamics with CBCs is the presence of modes whose growth rates become unbounded at large angular momentum \cite{AnninosGalanteManeerat2023,LiuReallSantosWiseman2025}. The fate of these instabilities depends on the location of the conformal boundary. In global AdS$_4$, the finite-boundary growing modes are suppressed as the boundary is moved towards the asymptotic AdS boundary and disappear in the strict asymptotic limit \cite{AnninosAriasGalanteManeerat2025}, where the standard fixed-boundary-metric problem has a stable gravitational spectrum \cite{IshibashiWald2004}. A similar suppression of the growing modes occurs as the boundary approaches a non-degenerate horizon, the so-called stretched-horizon limit \cite{AnninosGalanteGeorgescuManeeratSvesko2026}. In this case, a universal spectrum of allowed modes emerges, tied to the near-horizon Rindler geometry. This spectrum includes sound and shear modes described by an effective relativistic conformal fluid \cite{AnninosGalanteManeerat2024,AnninosGalanteGeorgescuManeeratSvesko2026}\footnote{Earlier related fluid descriptions of horizon dynamics include \cite{Bredberg:2010ky,Anninos:2011zn}.}.

These observations raise two questions that we address in this article. First, which geometric properties of a finite conformal boundary control the rapidly growing modes? Second, in an AdS setting, how are the near-horizon fluid excitations related to the sound and shear modes of the dual CFT$_3$ at the asymptotic boundary? We address these questions by analyzing linearized gravitational perturbations of the four-dimensional Schwarzschild-AdS$_4$ black hole. This geometry contains a near-horizon region, an asymptotic AdS region and a photon sphere in between, defined by the presence of circular null geodesics. Familiar from black-hole shadows and the geometric description of high-frequency quasinormal modes \cite{PerlickTsupko2022,CardosoEtAl2009}, the photon sphere also singles out a distinguished boundary geometry: a static spherical timelike surface located at this radius is totally umbilic, meaning that its extrinsic curvature is proportional to its induced metric \cite{ClaudelVirbhadraEllis2001}.

Our first result concerns gravitational perturbations with a timelike conformal boundary located at the photon sphere. At this special location, the scalar conformal boundary conditions factorize into a boundary branch, with vanishing gravitational master field, and a bulk branch, with vanishing boundary Weyl factor. The boundary branch is particularly simple and allows for an exact determination of the allowed frequencies. All frequencies are real. We present an argument showing that the bulk branch contains no unstable modes for any angular momentum $l\geq 2$. In the large angular momentum limit, we find analytic expressions for these stable modes. This statement concerns perturbations about the umbilic background surface, whereas the perturbed boundary is not required to remain umbilic.\footnote{A similar setting is provided by four-dimensional Minkowski spacetime foliated by dS$_3$ hypersurfaces, which are totally umbilic. Gravitational dynamics with CBCs on these hypersurfaces is studied in \cite{AnninosAriasToAppear}.} A similar argument shows that the vector sector contains no growing modes for any boundary position. Therefore, umbilicity provides a geometric mechanism for eliminating the large angular momentum instability\footnote{Local well-posedness for the vacuum Einstein equations
with an umbilic timelike boundary was established in
\cite{Fournodavlos:2021eye}. That boundary condition requires the
full evolving boundary to remain umbilic, whereas here umbilicity
is a property of the background and is not imposed on general
perturbations.}.

Moving the conformal boundary away from the photon sphere restores the rapidly growing scalar sector. Close to this surface, its growth rate scales as $|K-K_{\rm ph}|^{2/3}l^{1/3}$ in the large angular momentum limit, where $K_{\rm ph}$ is the mean curvature of the static photon-sphere boundary. The angular momentum required to reach this regime therefore diverges as the inverse square of the departure from umbilicity.

Our second result is a continuous connection between the near-horizon fluid modes and the hydrodynamic sound and shear modes of the dual CFT$_3$ at the asymptotic AdS$_4$ boundary. We numerically follow individual stretched-horizon sound and shear modes as the conformal boundary is moved outwards. We find that they approach standard gravitational quasinormal modes at the asymptotic AdS boundary and, in the large-black-hole limit, the hydrodynamic sound and shear poles of the dual CFT$_3$ \cite{MichalogiorgakisPufu2007,Siopsis2007}. Along this evolution, the sound branch passes through the bulk sector of the factorized photon-sphere problem, while the shear branch remains smooth across the same surface. We perform the same analysis for the AdS$_4$ black brane, following its near-horizon fluid modes to the asymptotic AdS boundary. In this case, the continuation connects the two fluid regimes directly, without an intermediate photon sphere, and the asymptotic hydrodynamic limit is reached by taking small planar momentum. These results extend earlier relations between near-horizon and asymptotic boundary fluid dynamics \cite{Kovtun:2003wp,Iqbal:2008by,Bredberg:2010ky,Donos:2015gia} by following the frequencies of individual modes as a finite conformal boundary is moved through the bulk.

\medskip
\noindent\textbf{Plan of the article.}
In Section \ref{sec:setup}, we introduce the Schwarzschild-AdS$_4$ black hole and the AdS$_4$ black brane and formulate the scalar and vector perturbation problems with conformal boundary conditions. We then focus on the umbilic case, deriving the exact factorization of the scalar problem at the photon sphere and establishing mode stability there for $l\geq2$. We also derive the asymptotic boundary conditions that recover the standard AdS$_4$ quasinormal-mode problem. In Section \ref{sec:large L analysis}, we analyze the large-angular-momentum spectrum for conformal boundaries on either side of the photon sphere and through the near-photon-sphere regime, including the exactly umbilic case. In Section \ref{sec:str-horizon}, we derive the stretched-horizon spectra of the black hole and black brane, including the sound and shear modes of the near-horizon fluid sector.
In Section \ref{sec:global-sound}, we numerically follow the sound and shear branches to the asymptotic AdS$_4$ boundary and compare their endpoints with the hydrodynamic sound and shear modes of the dual CFT\(_3\). We conclude in Section \ref{sec:conclusions}. Appendices \ref{app:reconstruction}-\ref{app:wkb} contain the metric reconstruction, the nonlinear homogeneous dynamics, and details of the planar and asymptotic calculations.

\section{Gravitational dynamics with conformal boundaries}
\label{sec:setup}

In this section, we formulate the problem of linearized gravitational perturbations around Schwarzschild-AdS$_4$ and the planar AdS$_4$ black brane in the presence of a timelike boundary on which we impose conformal boundary conditions. We describe the background solutions and the corresponding boundary data, before turning to the linearized scalar and vector perturbations. We then consider the special case of umbilic conformal boundaries. In Schwarzschild-AdS$_4$, the boundary is exactly umbilic when it is placed at the photon sphere. We show that, in this case, the scalar spectral problem factorizes into independent boundary and bulk sectors. An analogous factorization emerges in the limit in which the conformal boundary approaches the asymptotic boundary of AdS.

\subsection{Conformal boundary conditions and boundary dynamics}

\label{sec:background-cbc}
We consider the theory of general relativity with negative cosmological constant $\Lambda$ in four spacetime
dimensions, on a manifold $\mathcal{M}$ with codimension-1 timelike boundary $\Gamma$,
\begin{equation}
  I=
  \frac{1}{16\pi G_N}
  \int_{\mathcal M}\dd^4x\sqrt{-\det g_{\mu\nu}}\,(\mathcal R-2\Lambda)
  +\frac{1}{24\pi G_N}
  \int_\Gamma\dd^3x\sqrt{-\det g_{mn}}\,K,
  \quad
  \Lambda=-\frac{3}{\ads^2}.
  \label{eq:action}
\end{equation}
The boundary term is chosen so that the variational problem is well-posed when conformal boundary conditions are imposed at $\Gamma$ \cite{Witten2021,OdakSpeziale2021,AnninosAriasGalanteManeerat2025}. These boundary conditions fix both the conformal class of the induced metric and the trace of the extrinsic curvature, $K=g^{mn}K_{mn}$, at $\Gamma$. Writing
\begin{equation}
  \left.\dd s^2\right|_\Gamma
  =g_{mn}\dd x^m\dd x^n
  =\e^{2\bm{\omega}}[g]_{mn}\dd x^m\dd x^n,
  \label{eq:boundary-conformal-representative}
\end{equation}
we use $[g]_{mn}$ to denote a chosen representative of the fixed conformal class, corresponding to $\bm{\omega}=0$ in this parametrization. The Weyl factor $\bm{\omega}$ is allowed to fluctuate. Around a background whose induced metric satisfies $\bar{g}_{mn}=[g]_{mn}$, the linearized boundary conditions therefore read
\begin{equation}
  \delta g_{mn}
  =2\,\delta\bm{\omega}\,\bar g_{mn},
  \qquad
  \delta K=0.
  \label{eq:CBC-linear}
\end{equation}
The Weyl perturbation $\delta\bm{\omega}$ is not prescribed as boundary data, but is instead determined dynamically as part of the solution.

Varying the action \eqref{eq:action} on-shell with respect to the boundary conformal metric $[g]_{mn}$  defines the conformal version of the Brown-York stress tensor density \cite{OdakSpeziale2021,AnninosAriasGalanteManeerat2025}
\begin{align}
  T_{mn}&=-\frac{e^{\bm{\omega}}}{8\pi G_N}
  \left(K_{mn}-\frac{K}{3}g_{mn}\right) \bigg|_\Gamma .
  \label{eq:CBY-definition}
\end{align}
The two nontrivial projections of the Einstein equations onto $\Gamma$ determine important properties of the boundary dynamics. Assuming $K$ is constant, the normal-tangential projection implies that $T_{mn}$ is covariantly conserved with respect to $[g]_{mn}$. On the other hand, the normal-normal projection of the Einstein equations at $\Gamma$,
\begin{equation}
  \mathcal R_{g}-K^2+K_{mn}K^{mn}-2\Lambda=0,
  \label{eq:Hamiltonian-constraint-background}
\end{equation}
where $\mathcal{R}_g$ is the Ricci scalar of the induced metric,
characterizes the dynamics of the Weyl mode \cite{AnninosGalanteGeorgescuManeeratSvesko2026}
\begin{align}\label{eq: weyl mode eom}
  \mathcal{D}_m\mathcal{D}^m \bm{\omega} + \frac{1}{2}\mathcal{D}_m \bm{\omega} \mathcal{D}^m \bm{\omega} - \frac{\mathcal R_\Gamma}{4} - 16\pi^2G_N^2 T^{mn}T_{mn} e^{-4\bm{\omega}} + \left(K^2 + 3\Lambda\right)\frac{e^{2\bm{\omega}}}{6}=0 \, .
\end{align}
Here $\mathcal{D}$ is the covariant derivative with respect to the boundary conformal metric $[g]_{mn}$, indices are raised with $[g]_{mn}$ and $\mathcal R_\Gamma$ is its corresponding Ricci scalar. Linearizing \eqref{eq: weyl mode eom} around a background with $\bm{\omega}=0$ yields the equation of motion for the Weyl mode perturbation $\delta\bm{\omega}$. In general, this equation couples $\delta\bm{\omega}$ to bulk perturbations via the $T^{mn}\delta T_{mn}$ term.

\subsection{Background solutions}
\label{sec:backgrounds}

The bulk equations $\mathcal{R}_{\mu\nu}=\Lambda g_{\mu\nu}$ derived from \eqref{eq:action} admit both the Schwarzschild-AdS$_4$ black hole and its planar counterpart, the AdS$_4$ black brane, as solutions. In the following, we summarize the main features of these solutions, focusing on the horizon and asymptotic infinity in both cases, as well as on the photon sphere in the Schwarzschild-AdS$_4$ case, as these will play a central role throughout the analysis.

\subsubsection*{Schwarzschild-AdS$_4$ black hole}
We consider the Schwarzschild-AdS$_4$ black hole solution
\begin{align} \label{eq:general-static-background}
  \dd s^2
  &=-f(r)\dd t^2+\frac{\dd r^2}{f(r)}+r^2\dd\Omega_2^2,\qquad
  f(r)=1+\frac{r^2}{\ads^2}-\frac{\mu}{r},
\end{align}
with $\mu>0$ and horizon $r_h$ satisfying 
\begin{equation}
  \mu=r_h\left(1+\frac{r_h^2}{\ads^2}\right).
  \label{eq:horizon-data}
\end{equation}
We write the metric on the $S^2$ as $d\Omega_2^2=d\theta^2+\sin^2\theta d\varphi^2$. The surface gravity and the Hawking temperature are given by
\begin{align}
  \kappa_h=\frac{f'(r_h)}{2}
  =\frac{1+3r_h^2/\ads^2}{2r_h},\qquad T=\frac{\kappa_h}{2\pi}=\frac{1+3r_h^2/\ads^2}{4\pi r_h}
\end{align}
We place a timelike boundary $\Gamma$ at  $r=\wallr>r_h$ and retain the region between the black hole horizon and the boundary, which we refer to as the horizon patch.

We choose the induced metric on the timelike boundary,
\begin{equation}
  \bar g_{mn}\dd x^m\dd x^n \big|_{\Gamma}
  =-f(\wallr)\dd t^2+\wallr^2\dd\Omega_2^2,
  \label{eq:background-induced-metric}
\end{equation}
as the representative of the fixed conformal class. Thus, \eqref{eq:general-static-background} corresponds to $\bm{\omega}=0$.

The outward unit normal to the boundary is
$n^\mu\partial_\mu=\sqrt{f(\wallr)}\,\partial_r$.
With the convention $K_{mn}=\frac12\mathcal L_n g_{mn}$, the mean curvature is given by
\begin{equation}
  K(\wallr)
  =\frac{f'(\wallr)}{2\sqrt{f(\wallr)}}
   +\frac{2\sqrt{f(\wallr)}}{\wallr}.
  \label{eq:K-background}
\end{equation}
Together, the conformal class represented by \eqref{eq:background-induced-metric} and the mean curvature \eqref{eq:K-background} constitute the conformal boundary data.

For the Schwarzschild AdS$_4$ black hole, the components of the conformal stress tensor \eqref{eq:CBY-definition} take the simple form
\begin{equation}
  \bar{T}^m{}_n
  =
  \frac{\Delta(\wallr)}
       {48\pi G_N\wallr\sqrt{f(\wallr)}}
  \operatorname{diag}(-2,1,1),
  \label{eq:CBY-general-f}
\end{equation}
where the entries are ordered as $(t,\theta,\varphi)$ and we introduced
\begin{equation}
  \Delta(\wallr)
  \equiv\wallr f'(\wallr)-2f(\wallr)
  =\frac{3\mu}{\wallr}-2.
  \label{eq:Delta-background}
\end{equation}
Using \eqref{eq:CBY-general-f}, we obtain
\begin{equation}
  (8\pi G_N)^2 \bar{T}_{mn}\bar{T}^{mn}
  =
  \frac{\Delta(\wallr)^2}{6\wallr^2f(\wallr)}
  =
  \frac{2}{3}K^2-\frac{6}{\ads^2}-\frac{2}{\wallr^2}.
  \label{eq:T2-SAdS}
\end{equation}
Since $\mathcal{R}_{\Gamma}=2/\wallr^2$, \eqref{eq:T2-SAdS} implies that \eqref{eq: weyl mode eom} is satisfied for the background geometry \eqref{eq:general-static-background}.

We observe that the conformal stress tensor vanishes at the special location of the boundary where $\Delta(\wallr)=0$, which we denote by
\begin{align}\label{eq: position photon sphere}
  \rph&=\frac{3\mu}{2}.
\end{align}
Computing the temporal and angular components of the extrinsic curvature 
\begin{align}
  K^t_{\;t}|_{\Gamma}&=\frac{f'(\wallr)}{2\sqrt{f(\wallr)}},\qquad K^{\theta}_{\;\theta}|_{\Gamma}=K^{\varphi}_{\;\varphi}|_{\Gamma}=\frac{\sqrt{f(\wallr)}}{\wallr},
\end{align}
we note that $\Delta(\wallr)$ measures their difference 
\begin{align}
    K^t_{\;t}-K^{\theta}_{\;\theta}|_{\Gamma}&=\frac{\Delta(\wallr)}{2\wallr \sqrt{f(\wallr)}}.
\end{align}
At the particular location \eqref{eq: position photon sphere}, the conformal
boundary is therefore umbilic,
\begin{align}\label{eq: umbilic condition general}
  K_{mn}&=\frac{K}{3}g_{mn}.
\end{align}
As is well known, in a static spherically symmetric geometry a static, spherically symmetric timelike surface is umbilic precisely when it is located at the photon sphere radius, defined by the existence of circular null geodesics \cite{ClaudelVirbhadraEllis2001}. Indeed, in a background of the form \eqref{eq:general-static-background}, a circular null geodesic lies at a stationary point of the optical potential \cite{CardosoEtAl2009},
\begin{equation}
  U(r)=\frac{f(r)}{r^2},
  \qquad
  U'(r)=\frac{rf'(r)-2f(r)}{r^3}
       =\frac{\Delta(r)}{r^3}.
  \label{eq:background-U}
\end{equation}
In the case of Schwarzschild-AdS$_4$, the unique finite radius for which $U'(r)=0$ is \eqref{eq: position photon sphere}\footnote{Note that the explicit dependence
on $\ads$ cancels from the condition determining $\rph$. The photon sphere is therefore located at the same radius as for the asymptotically flat Schwarzschild black hole.
}. Since
\begin{equation}
  U''(\rph)=-\frac{2}{\rph^4}<0,
  \label{eq:rph}
\end{equation}
the circular null orbit is unstable.

The conformal stress tensor defines a conserved conformal energy associated with time translations \cite{AnninosAriasGalanteManeerat2025},
\begin{align}
  E_{\rm conf}
  &=
  \frac{\wallr^2}{\sqrt{f(\wallr)}}
  \int_{S^2}\dd\Omega_2\,T_{tt}
  \label{eq:Econf-definition}\\
  &=
  \frac{\wallr\,\Delta(\wallr)}{6G_N}
  =\frac{3\mu-2\wallr}{6G_N}.
  \label{eq:Econf-SAdS}
\end{align}
Since the conformal stress tensor is covariantly conserved, the conformal energy is time-independent. 
It is positive for $r_h<\wallr<\rph$, vanishes at $\rph$, and is negative outside the photon sphere. The vanishing of the conformal energy at this surface was previously observed in the thermodynamics of conformal boundaries in asymptotically flat spacetimes \cite{BanihashemiShaghoulianShashi2025}. In this paper we examine its dynamical significance.

Finally, moving the conformal boundary towards the black hole horizon or towards the asymptotic AdS boundary corresponds to two distinct limits of the mean curvature, which we will examine separately in later sections. Near the horizon,
\begin{align}\label{eq:near the horizon expansion}
   f(r)=2\kappa_h(r-r_h)+\Order((r-r_h)^2),
\end{align}
and inverting \eqref{eq:K-background} yields
\begin{equation}
  \wallr-r_h
  =\frac{\kappa_h}{2K^2}+\Order(K^{-4}).
  \label{eq:R-of-K-leading}
\end{equation}
Thus, moving the conformal boundary towards the horizon, $\wallr\rightarrow r_h$, corresponds to a large-$K$ limit. This is the black hole counterpart of the large-$K$ expansion in \cite{AnninosGalanteGeorgescuManeeratSvesko2026}, and hence we will refer to it as the stretched-horizon limit.

At the other end, $\wallr\to\infty$,
\begin{equation}
  K\ads
  =3+\frac{\ads^2}{2\wallr^2}
   +\Order\!\left(\frac{\ads^4}{\wallr^4}\right).
  \label{eq:K-large-R}
\end{equation}
Removing the finite cut-off therefore sends $K\ads\rightarrow 3^+$. Although the umbilic condition \eqref{eq: umbilic condition general} has a unique solution at finite radius, namely the photon sphere \eqref{eq: position photon sphere}, the conformal boundary becomes asymptotically umbilic as $\wallr\rightarrow\infty$,
\begin{align}
  K^t_{\;t}-K^{\theta}_{\;\theta}|_{\Gamma}&=-\frac{\ads}{\wallr^2} + \mathcal{O}(\wallr^{-3}).
\end{align}

\subsubsection*{AdS$_4$ black brane}
We consider now the AdS$_4$ black brane,
\begin{align}\label{eq: black brane metric}
  \dd s^2
  &=-f_{\rm br}(r)\dd t^2+\frac{\dd r^2}{f_{\rm br}(r)}
  +\frac{r^2}{\ads^2}\left(\dd x^2+\dd y^2\right),
  \qquad
  f_{\rm br}(r)=\frac{r^2}{\ads^2}\left(1-\frac{r_h^3}{r^3}\right),
\end{align}
whose horizon radius we also denote by $r_h$. The corresponding surface gravity and Hawking temperature are given by
\begin{equation}
  \kappa_h=\frac{f'_{\rm br}(r_h)}{2}
  =\frac{3r_h}{2\ads^2},\qquad T=\frac{\kappa_h}{2\pi}=\frac{3r_h}{4\pi\ads^2}.
  \label{eq:surface gr brane}
\end{equation}
We place a timelike boundary at $r=\wallr>r_h$ and choose the induced metric
\begin{equation}
  \bar g_{mn}\dd x^m\dd x^n\big|_\Gamma
  =-f_{\rm br}(\wallr)\dd t^2
  +\frac{\wallr^2}{\ads^2}\left(\dd x^2+\dd y^2\right)
\end{equation}
as the representative of the fixed conformal class. The mean curvature is
again given by \eqref{eq:K-background}, with $f$ replaced by $f_{\rm br}$.

The background conformal stress tensor takes the same form as
\eqref{eq:CBY-general-f},
\begin{equation}
  \bar{T}^m{}_n
  =
  \frac{\Delta_{\rm br}(\wallr)}
       {48\pi G_N\wallr\sqrt{f_{\rm br}(\wallr)}}
  \operatorname{diag}(-2,1,1),
  \qquad
  \Delta_{\rm br}(\wallr)
  \equiv \wallr f'_{\rm br}(\wallr)-2f_{\rm br}(\wallr)
  =\frac{3 r_h^3}{\wallr \ads^2}.
\end{equation}
Since $\Delta_{\rm br}(r)>0$ at every finite radius outside the horizon, there is no finite-radius umbilic conformal boundary. Equivalently, the optical potential
\begin{equation}
  U_{\rm br}(r)=\frac{f_{\rm br}(r)}{r^2}
  =\frac{1}{\ads^2}\bigg(1-\frac{r_h^3}{r^3}\bigg), \qquad U'_{\rm br}(r)=\frac{3 r_h^3}{\ads^2 r^4}>0
\end{equation}
is monotonic, increasing from 0 at the horizon to $1/\ads^2$ at the asymptotic AdS boundary. Hence, the black brane has no photon sphere.

Just as for Schwarzschild-AdS$_4$, we will consider the stretched-horizon limit, in which the conformal boundary approaches the black brane horizon. Near the horizon,
\begin{align}
  \wallr-r_h
  &=\frac{\kappa_h}{2K^2}+\Order(K^{-4}),
\end{align}
so that $\wallr\rightarrow r_h$ again corresponds to a large-$K$ limit. At the other extreme, the asymptotic limit $\wallr\rightarrow\infty$ again corresponds to $K\ads\rightarrow 3^+$,
\begin{align}
  K \ads&= 3+ \frac{3 r_h^6}{8\wallr^6}+ \mathcal{O}(\wallr^{-9}).
\end{align}
The conformal boundary becomes asymptotically umbilic in this limit.

\subsection{Linearized dynamics}
\label{sec:scalar-problem}
\label{sec:vectorial-problem}

We study the dynamics of small perturbations around the Schwarzschild AdS$_4$ black hole and AdS$_4$ black brane that preserve the conformal boundary conditions specified in Section \ref{sec:background-cbc}. Writing
\begin{equation}
    g_{\mu\nu} = \bar{g}_{\mu\nu} + \varepsilon \, h_{\mu\nu} \,, \;\;|\varepsilon| \ll1 \,,
\end{equation}
where $\bar{g}_{\mu\nu}$ is given by either \eqref{eq:general-static-background} or \eqref{eq: black brane metric}, we aim to solve the linearized Einstein's equations for $h_{\mu\nu}$. The maximal symmetry of the two-dimensional spatial sections, spherical for the black hole and planar for the black brane, allows the metric perturbations to be organized into scalar and vector sectors using the Kodama-Ishibashi formalism \cite{KodamaIshibashiSeto2000,KodamaIshibashi2003},
\begin{align}
  h_{\mu\nu}&=h_{\mu\nu}^{(S)}+h_{\mu\nu}^{(V)}.
\end{align}
Throughout the spectral analysis we use the static time in
\eqref{eq:general-static-background},\eqref{eq: black brane metric} and the convention
\begin{equation}
  h_{\mu\nu}^{(S,V)}\propto\e^{-\ii\omega t},
  \qquad
  \operatorname{Im}\omega>0
  \quad\Longleftrightarrow\quad
  \text{exponential growth}.
  \label{eq:time-convention}
\end{equation}
In the following, we present the analysis for the Schwarzschild-AdS$_4$ black hole, where the presence of the photon sphere makes the structure of the perturbation problem richer, while the simpler AdS$_4$ black brane analysis is relegated to Appendix \ref{app:brane-CBC}.

\subsubsection{Scalar sector}

\noindent \textbf{Metric perturbations.}
In the scalar sector, the metric perturbations $h^{(S)}_{\mu\nu}$ are written as
\begin{align}\label{eq:scalar sector KI functions}
  h^{(S)}_{ab}&=f_{ab}\mathbb{S},\qquad h^{(S)}_{ai}=r f_{a}^S \mathbb{S}_i, \qquad h^{(S)}_{ij}=2r^2 H^S \mathbb{S}_{ij}+2r^2 \gamma \sigma_{ij}\mathbb{S},
\end{align}
where $a,b=\{t,r\}$, $i,j=\{\theta,\varphi\}$, $\sigma_{ij}$ is the metric on the unit $S^2$, $\mathbb{S}$ are the eigenfunctions of the unit two-sphere Laplacian with eigenvalues $k^{2}_{S}=l(l+1)\equiv \mathfrak{L}^2$ for $l=0,1,2,...$ and $\mathbb{S}_i=-\frac{1}{\mathfrak{L}}\nabla_i \mathbb{S}$, $\mathbb{S}_{ij}=\frac{1}{\mathfrak{L}^2}\nabla_i\nabla_j \mathbb{S}+\frac{1}{2}\sigma_{ij}\mathbb{S}$, with $\nabla$ the covariant derivative with respect to $\sigma_{ij}$. We focus on the case $l\geq2$, for which the linearized Einstein's equations reduce to
the scalar Kodama-Ishibashi equation for the master field $\Phi^{(S)}(r,t)$
\cite{KodamaIshibashiSeto2000,KodamaIshibashi2003}. The spherically symmetric $l=0$ case is analyzed in Appendix \ref{app:homogeneous-nonlinear}. Upon Fourier decomposition
\begin{align}
  \Phi^{(S)}(r,t)= \operatorname{Re} e^{-i\omega t}\phi^{(S)}(r),
\end{align}
the master field equation becomes the following second order ODE for the radial master field,
  \begin{equation}
  f^2\phi^{(S)''}+ff'\phi^{(S)'}
  +\left(\omega^2-V_S\right)\phi^{(S)}=0,
  \label{eq:master}
\end{equation}
where primes denote radial derivatives and $f(r)$ is given in \eqref{eq:general-static-background}. The scalar potential is
\begin{equation}
  V_S=
  \frac{f}{r^2H^2}
  \left[
    m^3+2m^2+3m^2x+9mx^2+9x^3-18x^2y_{\rm KI}
  \right],
  \label{eq:KI-potential}
\end{equation}
where 
\begin{equation}
  m=\mathfrak{L}^2-2,
  \qquad
  x=\frac{\mu}{r},
  \qquad
  y_{\rm KI}=-\frac{r^2}{\ads^2},
  \qquad
  H=m+3x.
  \label{eq:KI-definitions}
\end{equation}
Of the two linearly independent solutions to \eqref{eq:master}, we select the one which is ingoing at the horizon,
\begin{equation}
  \phi_{\rm in}^{(S)}(r)
  =
  c_S (r-r_h)^{-\ii\omega/(2\kappa_h)}
  \left[1+\Order(r-r_h)\right],
  \label{eq:ingoing}
\end{equation}
where $c_S$ is an overall normalization.

The Kodama-Ishibashi invariants $F$ and $F_{ab}$ are reconstructed from this solution using
the expressions in Appendix \ref{app:reconstruction}.  They completely determine the gauge-invariant content of the scalar perturbations. The functions entering the metric perturbations \eqref{eq:scalar sector KI functions} are determined in terms of these invariants once a gauge is chosen, as explained in \cite{KodamaIshibashiSeto2000,KodamaIshibashi2003}.

\medskip

\noindent \textbf{Imposing boundary conditions.}
The remaining step is to impose the conformal boundary conditions at $r=\wallr$. Together with the ingoing condition at the horizon, these determine the allowed frequencies of the perturbations. For this purpose, we reconstruct the metric in a convenient gauge. The analyses of \cite{AnninosGalanteManeerat2023,AnninosAriasGalanteManeerat2025,AnninosGalanteManeerat2024} used the gauge
\begin{align}\label{eq: usual gauge}
  H^S=0,\qquad f_t^S=0,\qquad f_{tt}=-2 f \gamma, 
\end{align}
which has the advantage that it automatically fixes the conformal class of the induced metric, so that the only boundary condition that remains to be imposed in the scalar sector is $\delta K|_{\Gamma}=0$.

However, the reconstruction of the metric perturbations in this gauge involves division by $\Delta(r)=r f'-2f$, which vanishes at the photon sphere \eqref{eq: position photon sphere}. Hence, for $\wallr\geq\rph$, the photon sphere lies within the horizon patch, and this gauge cannot be used throughout the entire bulk region. Moreover, the resulting expression for
$\delta K |_{\Gamma}$ involves division by $\Delta(\wallr)$ and therefore becomes singular when the conformal boundary itself is placed at the photon sphere. This singularity is a feature of the gauge choice rather than of the underlying perturbation problem: both the bulk master equation and the gauge-invariant Kodama-Ishibashi invariants remain regular at $r=\rph$. To retain a description valid across this radius,
we reconstruct the metric in the Regge-Wheeler gauge,
\begin{align}
  H^S&=0, \qquad f_a^S=0.
\end{align}
In this gauge, the metric perturbations take simple expressions in terms of the Kodama-Ishibashi invariants
\begin{align}
  f_{ab}=F_{ab},\qquad \gamma=F.
\end{align}
The Regge-Wheeler gauge is, however, too restrictive if the conformal boundary is simultaneously kept at the fixed coordinate position $r=\wallr$. In that case, imposing the conformal boundary conditions leaves only the trivial scalar perturbation. We therefore allow the position of the boundary to fluctuate under an infinitesimal diffeomorphism. Let $\delta X^\mu$ denote the vector field generating this displacement, whose normal component at the boundary we denote by
\begin{align}
      \delta X^\mu|_{\Gamma}=\zeta n^\mu.
\end{align}
The total scalar metric perturbation is then
\begin{equation}
    h^{(S){\rm tot}}_{\mu\nu}
    =
    h^{(S)}_{\mu\nu}
    +\mathcal{L}_{\delta X}\bar g_{\mu\nu},
    \label{eq:total-scalar-perturbation}
\end{equation}
where $h^{(S)}_{\mu\nu}$ is the perturbation reconstructed from the Kodama-Ishibashi master field in Regge-Wheeler gauge. The full induced metric at the boundary is the pullback of $h^{(S){\rm tot}}_{\mu\nu}$ to $\Gamma$. Suppressing the common harmonic dependence, it can be written as
\begin{align}
       \delta g_{mn}=h_{mn}+ 2\zeta K_{mn}.
  \label{eq:displaced-wall-metric}
\end{align}
The conformal boundary conditions \eqref{eq:CBC-linear}, imposed on \eqref{eq:displaced-wall-metric} yield the following system of equations 
\begin{align}
  \mathcal I+\alpha\zeta&=0,
  \label{eq:conformal-regular}\\
  \mathcal B+\mathcal J\zeta&=0,
  \label{eq:K-regular}
\end{align}
where here and in the following boundary expressions, all radial quantities are evaluated at $r=\wallr$, and we introduced
\begin{align}
    \mathcal I&\equiv F_{tt}+2fF \\
    \alpha&\equiv \frac{\sqrt{f}}{\wallr}(2f-\wallr f')=-\frac{\sqrt{f}}{\wallr}\Delta(\wallr)\label{eq:alpha-definition}\\ 
    \mathcal B&\equiv
  2\sqrt f\,F'
  -\frac{1}{2\sqrt f}
   \left(F_{tt}'-\frac{f'}{f}F_{tt}\right)
  -\frac{fK}{2}F_{rr}
  -\frac{\ii\omega}{\sqrt f}F_{tr},
  \label{eq:D-definition}\\
  \mathcal J &\equiv
  -\frac{\omega^2}{f}
  +\frac{\mathfrak{L}^2}{\wallr^2}
  -\frac{f'^2}{4f}
  -\frac{2f}{\wallr^2}
  +\frac{3}{\ads^2}.
  \label{eq:J-definition}
\end{align}
The first equation \eqref{eq:conformal-regular} corresponds to fixing the conformal class of the induced metric at the boundary, while the second equation \eqref{eq:K-regular} corresponds to  $\delta K|_{\Gamma}=0$. $\mathcal{I}$ and $\mathcal B$ denote variations of the conformal class and of $K$, respectively, induced by the metric perturbation at fixed embedding, while $\alpha \zeta$ and $\mathcal J \zeta$ are the corresponding variations induced by a normal displacement of the boundary in the background geometry. In general, the two types of contributions compensate each other and neither is required to vanish separately. A nontrivial solution exists when the determinant of the system of equations vanishes,
\begin{equation}
  \mathcal{R}_S(\omega;\wallr,\mathfrak{L})
  \equiv
  \alpha\mathcal B-\mathcal J \mathcal I=0.
  \label{eq:regular-residual}
\end{equation}
Here and from now on $\mathcal I$ and $\mathcal B$ are evaluated on the ingoing solution $\phi^{(S)}_{\rm in}$ of the master field equation and therefore inherit its overall normalization $c_S$. Note that \eqref{eq:regular-residual} involves no division by $\Delta(\wallr)$. Away from the photon sphere, this condition is equivalent to the $\delta K|_{\Gamma}=0$ boundary condition obtained in the fixed-boundary description \eqref{eq: usual gauge}. At $\wallr=\rph$, where $\alpha=0$, the spectral condition factorizes and admits a branch of solutions with vanishing master field and nonzero displacement $\zeta$. We examine this
factorization in Section \ref{sec:exact-umbilic-finite-l}.

Away from special limits, such as the stretched-horizon and asymptotic regimes, the master field equation \eqref{eq:master} does not admit analytic solutions. We therefore integrate \eqref{eq:master} numerically from a point close to the future horizon, using the local ingoing expansion \eqref{eq:ingoing}. The resulting solution is then substituted into the regular conformal boundary condition \eqref{eq:regular-residual}, which determines the allowed frequencies.

\medskip

\noindent \textbf{Weyl mode dynamics.}
The perturbation of the boundary Weyl mode follows directly from  \eqref{eq:displaced-wall-metric} as
\begin{equation}
  \delta\bm{\omega}
  =F+\frac{\sqrt f}{\wallr}\zeta.
  \label{eq:weyl-mode}
\end{equation}
Its dynamics is governed by the linearization of \eqref{eq: weyl mode eom}, which takes the form
\begin{equation}
  \mathcal D_m\mathcal D^m\delta\bm{\omega}
  +
  \left[
    64\pi^2G_N^2\bar T_{mn}\bar T^{mn}
    +\frac{K^2}{3}+\Lambda
  \right]\delta\bm{\omega}
  =
  32\pi^2G_N^2\bar T^{mn}\delta T_{mn},
  \label{eq:linearised-Weyl-equation}
\end{equation}
where $\delta T_{mn}$ is the variation of the conformal stress tensor \eqref{eq:CBY-definition}. The Weyl mode has the same harmonic dependence on the boundary coordinates as the scalar perturbation, $\delta \bm{\omega}=c_{\bomega} e^{-i\omega t}\mathbb{S}$, where $c_{\bomega}$ is constant. Using this decomposition, together with \eqref{eq:K-background} and
\eqref{eq:T2-SAdS}, the differential operator acting on
$\delta\bm{\omega}$ in \eqref{eq:linearised-Weyl-equation} can be
identified with $-\mathcal J$ defined in
\eqref{eq:J-definition},
\begin{equation}\label{eq:JlWeyl}
\mathcal{J}\delta\boldsymbol{\omega}=-32\pi^2G_N^2\,\bar T^{mn}\delta T_{mn}.
\end{equation}
The perturbation of the conformal stress tensor sources the Weyl mode through its contraction with the background conformal stress tensor. At the photon sphere, $\bar T_{mn}=0$, and the source vanishes. The Weyl
mode equation \eqref{eq:linearised-Weyl-equation}  therefore reduces to
\begin{equation}
  \left(
    \mathcal D_m\mathcal D^m+\frac{1}{\rph^2}
  \right)\delta\bm{\omega}=0,
  \qquad
  \wallr=\rph.
  \label{eq:umbilic-Weyl-equation}
\end{equation}
A nonzero Weyl mode perturbation with the harmonic dependence specified above must therefore have the frequency
\begin{equation}
  \omega_{\rm bdy}^2
  =
  \Omph^2\left(\angL^2-1\right),
  \qquad
  \Omph\equiv\frac{\sqrt{f(\rph)}}{\rph}.
  \label{eq:umbilic-boundary-dispersion}
\end{equation}
For $l\geq2$, i.e. $\mathfrak{L}\geq \sqrt{6}$, $\omega_{\rm bdy}$ is real. The subscript `bdy' indicates that, for these modes, the Weyl mode decouples from the bulk dynamics.

On the other hand, modes with $\delta\bm{\omega}=0$ need not satisfy \eqref{eq:umbilic-boundary-dispersion}. Their frequencies are instead determined by the Kodama-Ishibashi master field equation together with the remaining conformal boundary conditions at $\wallr=\rph$.

\subsubsection{Vector sector}
\label{sec:vector-problem}

\noindent \textbf{Metric perturbations.} In the vector sector, the metric perturbations $h^{(V)}_{\mu\nu}$ are written as
\begin{align}
  h^{(V)}_{ai}&=r f^V_a \mathbb{V}_i, \qquad h^{(V)}_{ij}=2r^2 H^V \mathbb{V}_{ij},
  \label{eq:vector perturbations def}
\end{align}
where $a,b=\{t,r\}$, $i,j=\{\theta,\varphi\}$, $\mathbb{V}_i$ are the vector spherical harmonics, i.e. vectorial eigenfunctions of the unit two-sphere Laplacian with eigenvalues $k_V^2=l(l+1)-1=\mathfrak{L}^2-1$ with $l=1,2,...$ and $\mathbb{V}_{ij}=-\frac{1}{2 k_V}(\nabla_i \mathbb{V}_j+\nabla_j \mathbb{V}_i)$. Just like in the scalar sector, we focus on the case $l\geq2$, for which the linearized Einstein's equations reduce to
the vector Kodama-Ishibashi equation for the master field $\Phi^{(V)}(r,t)$
\cite{KodamaIshibashiSeto2000,KodamaIshibashi2003}. Writing
\begin{align}
  \Phi^{(V)}(r,t)= \operatorname{Re} e^{-i\omega t}\phi^{(V)}(r),
\end{align}
the master field equation becomes the following second order ODE for the radial master field,
  \begin{equation}
  f^2 \phi^{(V)''}
  +f f'\phi^{(V)'}
  +\left(\omega^2-V_V\right)\phi^{(V)}=0,
  \label{eq:vector-master}
\end{equation}
where the vector sector potential is given by
\begin{equation}
  V_V(r)  =\frac{f}{r^2}
  \left[\angL^2-\frac{3\mu}{r}\right].
  \label{eq:vector-potential}
\end{equation}
Just like for the scalar sector, we select the solution to \eqref{eq:vector-master} which is ingoing at the horizon,
\begin{equation}
  \phi_{\rm in}^{(V)}(r)
  =
  c_V(r-r_h)^{-\ii\omega/(2\kappa_h)}
  \left[1+\Order(r-r_h)\right],
\end{equation}
where $c_V$ is an overall normalization. The Regge-Wheeler gauge sets
\begin{align}
  H^V=0.
\end{align}
The remaining nontrivial metric perturbations can be reconstructed from the master field as
\begin{align}
  f^V_r&=\frac{1}{f}\,\partial_t\Phi^{(V)}\,,\qquad f^V_t=\frac{f}{r}\,\partial_r\!\left(r\Phi^{(V)}\right).
  \label{eq:vector-RW-reconstruction}
\end{align}
They remain regular throughout the region from the horizon to the asymptotic boundary. In particular, they do not develop any singularity at the photon sphere. The gauge-invariant construction and its relation to the general Kodama-Ishibashi decomposition are given in Appendix \ref{app:reconstruction}. 

\medskip

\noindent \textbf{Imposing boundary conditions.}
The conformal boundary conditions take a simple form in this sector. A normal displacement of the boundary transforms as a scalar on the $S^2$ and therefore does not contribute to vector perturbations, so no additional displacement parameter is required. Moreover, the variation of the mean curvature vanishes identically in this sector. The only nontrivial boundary condition is therefore that the conformal class of the induced metric at the boundary remain fixed. This condition yields
\begin{equation}
  \mathcal R_V(\omega;\wallr,\mathcal{L})
  \equiv
  \phi_{\rm in}^{(V)}(\wallr)
  +\wallr\,
   \phi_{\rm in}^{(V)'}(\wallr)
  =0.
  \label{eq:vector-residual}
\end{equation}
This Robin boundary condition has the same form as the vector conformal boundary condition in \cite{AnninosAriasGalanteManeerat2025,
AnninosGalanteGeorgescuManeeratSvesko2026}.
The black hole geometry enters through the master field satisfying \eqref{eq:vector-master}. The vector boundary condition \eqref{eq:vector-residual} remains regular at $\wallr=\rph$, with no analogue of the scalar sector factorization for $\alpha=0$.

Just as in the scalar sector, away from special limits, the master field equation \eqref{eq:vector-master} does not admit analytic solutions. We therefore integrate it numerically using the ingoing expansion and determine the allowed frequencies by imposing the boundary condition \eqref{eq:vector-residual}.

 \medskip

\noindent \textbf{Weyl mode dynamics.}
A pure vector perturbation does not contribute to the scalar boundary Weyl factor,
\begin{equation}
  \delta{\bm{\omega}}^{(V)}=0.
  \label{eq:vector-Weyl-K}
\end{equation}
The boundary conformal stress tensor is nevertheless nontrivial. However, the vector-harmonic structure implies
\begin{align}
  \bar T^{mn}\delta T^{(V)}_{mn}=0,
\end{align}
consistent with the Weyl mode equation \eqref{eq:linearised-Weyl-equation} and \eqref{eq:vector-Weyl-K}.

\subsection{Umbilic conformal boundaries}
\label{sec:exact-umbilic-finite-l}
We now focus on umbilic conformal boundaries. As explained in Section \ref{sec:backgrounds}, for Schwarzschild-AdS$_4$ the conformal boundary is exactly umbilic when it is placed at the photon sphere \eqref{eq: position photon sphere}, where
\begin{equation}
  K(\rph)=3\Omega_{\rm ph}.
  \label{eq:exact umbilic K}
\end{equation}
For both Schwarzschild-AdS$_4$ and the AdS$_4$ black brane, the conformal boundary also becomes asymptotically umbilic as $\wallr\rightarrow\infty$, with
\begin{equation}
  K\rightarrow \frac{3}{\ads}.
\end{equation}

\subsubsection{The photon sphere: exact factorization of the scalar spectral problem}
We place the conformal boundary at the photon sphere, $\wallr=\rph$, and examine the resulting spectral problem. We focus on the scalar sector, where the umbilic boundary leads to a richer structure of the solutions. Denoting $f_{\rm ph}\equiv f(\rph)$, we have
\begin{align}
   f_{\rm ph}&=\frac{1}{3}+\frac{\rph^2}{\ads^2}, \qquad\rph f'_{\rm ph}=2f_{\rm ph}.
\end{align}
Hence, from \eqref{eq:alpha-definition} we find
\begin{align}
  \alpha(\rph)&=0,
\end{align}
which implies that the spectral condition \eqref{eq:regular-residual} factorizes
\begin{align}\label{eq:exact-umbilic-factorization}
  \mathcal{I}(\rph;\omega,\mathfrak{L}) \mathcal{J}(\rph;\omega,\mathfrak{L})=0.
\end{align}
This gives rise to two branches of solutions, which we call the boundary and bulk branches, for reasons explained below.

\medskip

\noindent \textbf{Boundary branch.} The boundary branch is given by the condition
\begin{align}\label{eq:J boundary branch}
  \mathcal{J}=0.
\end{align}
Using \eqref{eq:conformal-regular} and \eqref{eq:K-regular}, it implies $\mathcal{I}=0=\mathcal{B}$, which forces the scalar master field to vanish. The metric perturbation is, thus, fully given by the boundary displacement $\zeta\neq 0$. At the photon sphere, it is easy to see that
\begin{align}
 \mathcal J(\rph)
  =
  \frac{\omega_{\rm bdy}^2-\omega^2}{f_{\rm ph}}.
  \label{eq:J-umbilic-exact}
\end{align}
Hence, \eqref{eq:J boundary branch} sets the frequency of the perturbation to \eqref{eq:umbilic-boundary-dispersion}. Thus the frequencies predicted by the sourceless Weyl equation are realized by gravitational perturbations with
\begin{equation}
  \phi^{(S)}=0,
  \qquad
  \delta\bm{\omega}_{\rm bdy}=\Omega_{\rm ph}\zeta.
  \label{eq:pure-boundary-weyl}
\end{equation}
These perturbations are generated by diffeomorphisms with a nonzero normal component at the boundary. We call them boundary modes because the Weyl mode decouples from the bulk, while the Kodama-Ishibashi master field vanishes. Moreover, the generating vector field can be chosen to vanish near the black hole horizon and to have support only in a neighbourhood of the boundary.

Moving the boundary slightly away from the photon sphere $\mathfrak r=\rph+\delta r$, with $|\delta r|\ll \rph$, 
\begin{align}
  \alpha(\mathfrak r)=\mathcal{O}(\mathfrak r-\rph),
\end{align}
and therefore the branch with $\zeta\neq 0$ obeys
\begin{equation}
    \frac{\,{\cal I}(\mathfrak r)}{\zeta}=-
    \alpha(\mathfrak r) = \mathcal{O}(\mathfrak r-\rph).
\end{equation}
Thus, away from the photon sphere the metric perturbation contains a nonzero Kodama-Ishibashi master field contribution and is not a pure diffeomorphism. As $\wallr\rightarrow \rph$, this contribution vanishes relative to the boundary displacement, and the mode approaches the pure boundary diffeomorphism with frequency $\omega_{\rm bdy}$. This provides a finite-radius analogue of the ``near-physical diffeomorphisms'' encountered when a conformal boundary approaches the asymptotic boundary of global AdS$_4$ \cite{AnninosAriasGalanteManeerat2025}.

\medskip

\noindent \textbf{Bulk branch.} The bulk branch is given by the condition
\begin{align}\label{eq:I boundary branch}
  \mathcal{I}=0.
\end{align}
Using the Kodama-Ishibashi reconstruction presented in Appendix \ref{app:reconstruction}, this condition can be written as the following frequency-dependent Robin boundary condition
\begin{equation}
  \left[
    \frac{\omega^2 \rph^2}{f_{\rm ph}}
    -(\angL^2-1)
    +\frac{2f_{\rm ph}(\angL^2-2)}{\angL^4}
  \right]\phi^{(S)}(\rph)
  -
  \frac{f_{\rm ph}(\angL^2-2)}{\angL^2}\,
  \rph\,\phi^{(S)'}(\rph)=0.
  \label{eq:exact-umbilic-bulk-Robin}
\end{equation}
Together with \eqref{eq:ingoing}, this defines the spectral problem for this branch, which we analyze in the next sections.

It is interesting to note that, at $\wallr=\rph$, we have
\begin{equation}
  \mathcal B(\rph)
  =
  \frac{\rph}{\sqrt{f_{\rm ph}}}\,
  \mathcal J(\rph)\,F(\rph).
  \label{eq:D-J-bulk-umbilic}
\end{equation}
Then, for $\mathcal J(\rph)\neq0$, the mean-curvature condition fixes the boundary displacement in this branch as 
\begin{align}
  \zeta
  &=-
  \frac{\rph}{\sqrt{f_{\rm ph}}}\,F(\rph).
  \label{eq:zeta-bulk-umbilic}
\end{align}
Using \eqref{eq:weyl-mode}, this implies that the bulk branch has vanishing Weyl mode perturbation
\begin{align}
  \delta\bm{\omega}_{\rm bulk} &=F(\rph)+\Omega_{\rm ph}\zeta  =0.
  \label{eq:bulk-zero-weyl}
\end{align}
The exactly umbilic conformal boundary therefore separates the scalar spectral problem into a boundary branch with vanishing Kodama-Ishibashi master field and a bulk branch with vanishing boundary Weyl factor. This separation follows from the conformal boundary conditions and does not require the perturbed boundary to remain umbilic.

\medskip
\noindent\textbf{Mode stability at finite angular momentum.}
In the boundary branch, the frequencies are real and the corresponding modes are stable. Here we give a stability argument for the modes in the bulk branch, which obey the Robin boundary condition \eqref{eq:exact-umbilic-bulk-Robin}. We use an $S$-deformation of the master equation \cite{Kodama:2003ck,IshibashiKodama2003StabilityVfive}, chosen so that the deformed potential is manifestly positive,
\begin{align}  
D_S&=\partial_{r^*}+S_S,\qquad S_S=f\frac{H'}{H},\qquad
\widetilde V_S=V_S+fS_S'-S_S^2
=\frac{f m\mathfrak L^2}{r^2H}>0.
\label{eq:Strquant}
\end{align}
Here we introduced the tortoise coordinate
$d r^*/d r=f^{-1}$, while $m$ and $H$ are defined in \eqref{eq:KI-definitions}. Primes denote derivatives with respect to $r$. At the photon sphere, $H(\rph)=\mathfrak L^2$ and $S_S(\rph)=-2f_{\rm ph}/(\rph\mathfrak L^2)$. The Robin condition becomes
\begin{align}
  D_S\phi^{(S)}(\rph)
  &=\left(a_S\omega^2-b_S\right)\phi^{(S)}(\rph),
  \label{eq:positive-scalar-robin-v5}
\end{align}
with positive coefficients
\begin{align}
  a_S&=\frac{\rph\mathfrak L^2}{f_{\rm ph}m}>0,\qquad
  b_S=\frac{\mathfrak L^2(\mathfrak L^2-1)}{m\rph}>0.
\end{align}
We now multiply the master field equation by $\phi^{(S)*}$ and integrate over the radial domain,
\begin{align}
\int_{-\infty}^{r^*(\rph)}\!\dd r^*\,
\phi^{(S)*}
\left(-\partial_{r^*}^2+V_S\right)\phi^{(S)}
=
\omega^2
\int_{-\infty}^{r^*(\rph)}\!\dd r^*\,|\phi^{(S)}|^2.
\end{align}
After integrating by parts and rewriting the left-hand side in terms of the $S$-deformed quantities, we obtain
\begin{align}
\int_{-\infty}^{r^*(\rph)}\!\dd r^*\,
\left(
|D_S\phi^{(S)}|^2
+\widetilde V_S|\phi^{(S)}|^2
\right)
-
\left[
\phi^{(S)*}D_S\phi^{(S)}
\right]_{-\infty}^{r^*(\rph)}
=
\omega^2
\int_{-\infty}^{r^*(\rph)}\!\dd r^*\,
|\phi^{(S)}|^2.
\end{align}
Assuming that a nontrivial ingoing solution has $\operatorname{Im}\omega>0$, both the master field and its tortoise derivative decay at the horizon, $r^*\to-\infty$, so the horizon boundary term vanishes. Using the Robin condition \eqref{eq:positive-scalar-robin-v5} to evaluate the remaining boundary term, we obtain
\begin{equation}
\begin{split}
 &\int_{-\infty}^{r^*(\rph)}\!\dd r^*\,
   \left(|D_S\phi^{(S)}|^2+\widetilde V_S|\phi^{(S)}|^2\right)
   +b_S|\phi^{(S)}(\rph)|^2
 \\
 &\qquad=\omega^2\left[
   \int_{-\infty}^{r^*(\rph)}\!\dd r^*\,|\phi^{(S)}|^2
   +a_S|\phi^{(S)}(\rph)|^2\right].
\end{split}
\label{eq:umbilic-positive-energy-v5}
\end{equation}
For $l\geq2$, both the left-hand side and the expression in brackets on the right-hand side are strictly positive. Hence $\omega^2$ must be real and positive, which contradicts the assumption $\operatorname{Im}\omega>0$.

For vector perturbations, the same argument is even simpler. Taking
\begin{align}
 D_V&=\partial_{r^*}+\frac{f}{r},\qquad  \widetilde V_V=V_V+f\left(\frac{f}{r}\right)'
                   -\frac{f^2}{r^2}
                 =\frac{f(\mathfrak L^2-2)}{r^2}>0,
 \label{eq:positive-vector-potential-v5}
\end{align}
the boundary condition \eqref{eq:vector-residual} is $D_V\phi^{(V)}(\wallr)=0$. The boundary term therefore vanishes, leaving a positive bulk integral equal to $\omega^2\int\dd r^*|\phi^{(V)}|^2$. Growing vector modes are therefore excluded for every $l\geq2$ and any finite $\wallr>r_h$, including the photon sphere.

\subsubsection{The asymptotic boundary: asymptotic factorization of the spectral problem}\label{subsection:asy AdS}

\noindent \textbf{Scalar sector.} In the asymptotic limit $\wallr\to\infty$, the master field solution which is ingoing at the horizon has the asymptotic expansion
\begin{equation}
  \phi^{(S)}(r)
  =
  A_\infty+\frac{B_\infty}{r}
  +\Order(r^{-2}),
  \label{eq:ads-master-expansion}
\end{equation}
where $A_\infty$ and $B_\infty$ depend on
$\omega$ and $\mathfrak{L}$ and are determined, up to an overall normalization, by the ingoing condition at the horizon.

From \eqref{eq:J-definition}, it follows that
\begin{align}
  \mathcal J(\wallr)
  &=
  \frac{\angL^2-1-(\omega\ads)^2}{\wallr^2}
  +\Order(\wallr^{-3}).
  \label{eq:ads-J-expansion}
\end{align}
The contribution $\alpha\mathcal B$ to \eqref{eq:regular-residual} vanishes as $\wallr\rightarrow\infty$, whereas $\mathcal J \mathcal I$ has a finite asymptotic limit. The determinant therefore factorizes asymptotically as
\begin{equation}
  \mathcal{R}_S(\omega;\wallr,\mathfrak{L})
  =
  \frac{\angL^2-1-(\omega\ads)^2}{\ads^4}
  \left(
    B_\infty+\frac{3\mu}{\mathfrak{L}^2-2}A_\infty
  \right)
  +\Order(\wallr^{-1}).
  \label{eq:ads-CBC-factorization}
\end{equation}
Similarly to the exactly umbilic case discussed above, we refer to the first factor as the boundary branch and to the second as the bulk branch. We interpret this asymptotic factorization as a consequence of the conformal boundary becoming asymptotically umbilic.

The boundary branch has the asymptotic frequencies
\begin{equation}
  \omega\ads=\pm\sqrt{\angL^2-1},
  \label{eq:ads-boundary-branch}
\end{equation}
which are independent of the black hole geometry. These are precisely the same asymptotic frequencies found when a conformal boundary approaches the asymptotic boundary of global AdS$_4$ in \cite{AnninosAriasGalanteManeerat2025}.

The bulk branch has the asymptotic frequencies fixed by
\begin{equation}
  B_\infty+\frac{3\mu}{\mathfrak{L}^2-2}A_\infty=0,
  \label{eq:ads-scalar-Robin}
\end{equation}
which depends on the black hole parameters. This condition coincides with the gravitational Robin boundary condition imposed in \cite{IshibashiWald2004,MichalogiorgakisPufu2007,DiasSantos2013} to define the standard Schwarzschild-AdS$_4$ quasinormal-mode problem with an undeformed asymptotic boundary metric.

In the case of the AdS$_4$ black brane, a similar factorization holds in the asymptotic limit. The boundary branch is given by
   \begin{align}
       \omega\ads &=\pm k_S,
       \label{eq:boundary branch asy brane}
   \end{align}
where $k_S$, the planar analogue of $\mathfrak{L}$, is the momentum along the brane. The bulk branch satisfies
   \begin{align}
         B_{\infty}+\frac{3 r_h^3}{k_S^2\ads^2} A_{\infty}=0.
         \label{eq:bulk branch asy brane}
   \end{align}
The bulk-branch condition matches the Robin boundary condition imposed in the standard black-brane quasinormal-mode problem \cite{Morgan:2009pn}, which selects perturbations that leave the boundary metric undeformed. As explained in detail in \cite{Morgan:2009pn}, this condition is equivalent to imposing Dirichlet boundary conditions on the Kovtun-Starinets scalar variable and is essential for recovering the holographic sound spectrum
\cite{MirandaMorganZanchin2008}. Here, we recover the same condition directly from the conformal boundary condition $\delta K|_{\Gamma}=0$, in the asymptotic limit $\wallr\rightarrow\infty$.

\medskip
\noindent \textbf{Vector sector.} For completeness, the ingoing vector-sector master field has a similar asymptotic expansion,
\begin{equation}
  \phi^{(V)}(r)
  =A_\infty^{(V)}
   +\frac{B_\infty^{(V)}}{r}
   +\Order(r^{-2}).
  \label{eq:vector-asymptotic-expansion}
\end{equation}
The contribution proportional to $B_\infty^{(V)}$ cancels
between the two terms in \eqref{eq:vector-residual}, leaving
\begin{align}
  \mathcal R_V(\omega;\wallr,\mathfrak{L})
  =A_\infty^{(V)}+\Order(\wallr^{-2}).
\end{align}
Therefore, in the limit $\wallr\to\infty$, vector perturbations obey the
Dirichlet boundary condition
\begin{equation}
  A_\infty^{(V)}=0.
  \label{eq:ads-vector-Dirichlet}
\end{equation}
This coincides with the standard asymptotic condition imposed on the vector Kodama-Ishibashi master field for an undeformed boundary metric \cite{DiasSantos2013}. Together with the ingoing condition at the horizon,
it defines the standard vector gravitational quasinormal-mode problem.

In the case of the AdS$_4$ black brane, the same Dirichlet condition \eqref{eq:ads-vector-Dirichlet} holds asymptotically, and coincides with the Dirichlet boundary condition imposed on the Kovtun-Starinets vector variable in the standard black brane quasinormal-mode problem \cite{Morgan:2009pn}.

\section{The large angular momentum limit}
\label{sec:large L analysis}
\label{sec:fixedR}

In general, the master field equations together with the conformal boundary conditions that determine the allowed frequencies can only be solved numerically. Analytic solutions can nevertheless be obtained in certain limits. In this section, we consider such a limit, namely the large angular momentum regime
$\mathfrak{L}\rightarrow\infty$. This type of analysis was first carried out in Minkowski spacetime with a
spherical conformal boundary in \cite{AnninosGalanteManeerat2023}, and was extended to arbitrary spherically symmetric backgrounds with non-degenerate horizons, for particular locations of the boundary, in Appendix B of \cite{AnninosGalanteGeorgescuManeeratSvesko2026}. For backgrounds with a photon sphere, corresponding at large $\mathfrak L$ to a maximum of the effective master field potential, the analysis of \cite{AnninosGalanteGeorgescuManeeratSvesko2026} considered boundaries located on the horizon side of the photon sphere, so that the maximum of the potential lies outside the spacetime region under consideration. In this case, the scalar sector contains unstable modes with complex frequencies whose positive imaginary parts scale as $\mathfrak{L}^{1/3}$,
\begin{equation}
    \frac{\omega \wallr}{\sqrt{f(\wallr)}} = \mathfrak{L} + \nu \left(48 \pi^2 G_N^2\wallr^2 \bar{T}_{mn}\bar{T}^{mn}\right)^{1/3}\mathfrak{L}^{1/3} + \mathcal{O}(\mathfrak{L}^{-1/3}) \, , \qquad \text{Im}\;\nu>0.
    \label{eq:large L general}
\end{equation}
Their radial profiles are localized near the conformal boundary. In \cite{AnninosAriasGalanteManeerat2025}, these instabilities were shown to be suppressed as the conformal boundary approaches the asymptotic boundary of AdS$_4$, while \cite{AnninosGalanteGeorgescuManeeratSvesko2026} showed that they are likewise suppressed in the stretched-horizon limit.

Here we carry out a more complete analysis of the large-$\mathfrak L$ spectrum of gravitational perturbations around Schwarzschild-AdS$_4$,
allowing the conformal boundary to lie on either side of, or precisely at, the photon sphere. When the conformal boundary is exactly umbilic, we find no unstable large-$\mathfrak L$ modes; in particular, the
$\mathfrak L^{1/3}$ instability is absent. Likewise, instabilities are suppressed near the asymptotic boundary of AdS.

The presence of the photon sphere organizes the scalar spectrum according to the position of the conformal boundary relative to it. We first consider boundaries that remain a finite distance from the photon sphere in the large-$\mathfrak L$ limit, $|\rph-\wallr|=\mathcal{O}(1)$. We refer to the problem as interior when the boundary lies on the horizon side of the photon sphere, $\wallr<\rph$, and as exterior when it lies on the asymptotic side, $\wallr>\rph$. In both cases, the spacetime region under consideration extends from the black hole horizon to the conformal boundary. The structure of the large-$\mathfrak L$ solution is determined by the turning points of the effective potential. We focus on frequencies below the maximum of the potential, for which at least one turning point exists. Depending on the frequency, a second turning point may appear on the exterior side of the photon sphere \footnote{More precisely, two finite turning points exist only when
$\omega^2/\mathfrak L^2>\ads^{-2}$. For $\omega^2/\mathfrak L^2<\ads^{-2}$, the outer turning point is absent because the potential approaches $\ads^{-2}$ as $r\to\infty$.}. Depending on the position of the boundary, however, one of them may lie outside the spacetime region under consideration. When both turning points lie within the physical region, the radial solution consists schematically of a WKB region extending from the horizon, an Airy region around the first turning point, a second WKB region between the turning points, and another Airy region around the second turning point. The details of the solutions for the master field using the WKB approximation can be found in Appendix \ref{app:wkb}. The conformal boundary condition then selects the allowed frequencies. We focus on the subset of modes for which a turning point approaches the conformal boundary in the large-$\mathfrak L$ limit. Equivalently, the boundary lies within the corresponding Airy region. These modes are therefore localized near, and controlled by, the conformal boundary.

Second, we consider the case in which the conformal boundary is placed at the photon sphere, $\wallr=\rph$, and hence is exactly umbilic. We analyze the large-$\mathfrak{L}$ spectrum for the boundary and bulk branches discussed in Section \ref{sec:exact-umbilic-finite-l}. Since the boundary now coincides with the maximum of the effective potential, its slope vanishes and the Airy approximation for the master field used away from the photon sphere is no longer applicable. Instead, the near-maximum region is described by a parabolic-cylinder master field equation. We again focus on frequencies for which turning points approach the boundary in the large-$\mathfrak L$ limit, so that the corresponding turning-point region is contained within the parabolic-cylinder region. We also comment on the near-umbilic case, in which the boundary does not lie exactly at the photon sphere but approaches it in the large-$\mathfrak L$ limit.

Finally, as the conformal boundary approaches the asymptotic AdS boundary, the spectral problem factorizes asymptotically into a real boundary branch and a bulk branch corresponding to the standard Schwarzschild-AdS$_4$ quasinormal-mode spectrum. In this limit, the large-$\mathfrak L$ instabilities are suppressed, as in global AdS$_4$ \cite{AnninosAriasGalanteManeerat2025}.

Throughout this section, we focus on the Schwarzschild-AdS$_4$ black hole, for which the presence of the photon sphere leads to the richer large-$\mathfrak L$ structure discussed above. We return to the AdS$_4$ black brane in the following sections.

\subsection{Conformal boundary away from the photon sphere}
\label{sec:airy}\label{sec:pole-black-hole}\label{sec:exterior}
We first consider the case in which the boundary remains away from the photon sphere in the large-\(\mathfrak L\) limit, namely $|\rph-\wallr|=\mathcal{O}(1)$.

\begin{figure}[t]
  \centering
  \paperfigure[width=0.78\textwidth]
    {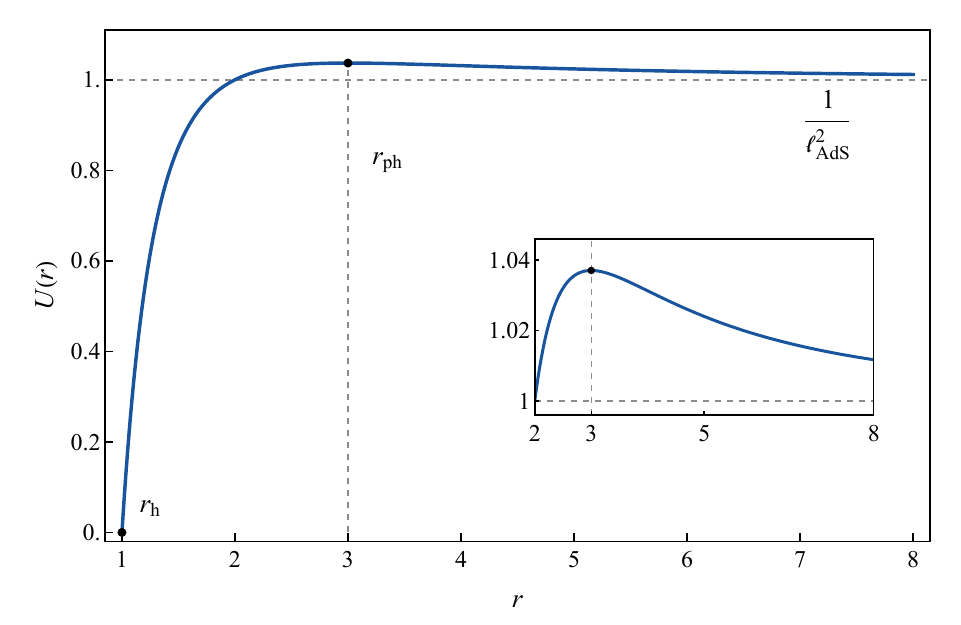}
  \caption{
Optical potential $U=f/r^2$ for $r_h=\ads=1$.
The potential vanishes at the horizon, reaches its maximum at
$\rph=3$, and approaches the AdS value $1/\ads^2$ from above.
The inset resolves the shallow maximum and the exterior side
of the barrier; the horizontal dashed line marks the AdS
asymptote. For $\wallr<\rph$, the potential is monotone
between the horizon and the conformal boundary. For $\wallr>\rph$, the photon-sphere
barrier lies between the conformal boundary and the horizon.
}
  \label{fig:eikonal-potential}
\end{figure}

Using the tortoise coordinate, the master field equation
\eqref{eq:master} takes, in the large angular momentum limit, the form of a one-dimensional Schrödinger equation 
\begin{equation}
  \partial_{r^*}^2 \phi^{(S)}
  +\left[
    \omega^2-\angL^2U(r^*)+\Order(\angL^0)
  \right]\phi^{(S)}=0,
  \label{eq:eikonal-equation}
\end{equation}
where we recall that $\angL=\sqrt{l(l+1)}$ and $U$ is the optical potential defined in \eqref{eq:background-U}, expressed now in terms of the tortoise coordinate. The optical potential is shown in Figure \ref{fig:eikonal-potential}, as a function of $r$. For the numerical examples in this section we set
\begin{equation}
  r_h=\ads=1 \quad \Rightarrow \quad \mu=2, \;\rph=3.
  \label{eq:fiducial}
\end{equation}
Although we focus on the scalar sector, the same leading large-$\mathfrak{L}$ equation holds in the vector sector, since the two effective potentials coincide at leading order in $\mathfrak{L}$. We return to the vector sector at the end of this subsection.

The master field equation \eqref{eq:eikonal-equation} can be solved by standard WKB matching from the near-horizon region to asymptotic infinity, as detailed in Appendix \ref{app:wkb}. The conformal boundary condition then determines the spectrum.

As discussed at the beginning of this section, we refer to the cases $\wallr<\rph$ and $\wallr>\rph$ as the interior and exterior problems, respectively. In the former, the conformal boundary truncates the potential before its maximum, while in the latter the photon sphere barrier lies within the physical domain. In both cases, the physical domain extends from the horizon to the boundary, and the solution to the master field equation \eqref{eq:eikonal-equation} must satisfy the ingoing condition \eqref{eq:ingoing}.

We focus on the subset of modes for which a turning point of the effective potential, $r_t$, approaches the conformal boundary in the large-$\mathfrak L$ limit as $|\wallr^*-r_t^*|=\mathcal O(\mathfrak L^{-2/3})$.

For these modes, the master field equation implies, at leading order in the $\mathfrak L\rightarrow\infty$ limit
\begin{align} \label{eq:omega-scales}
    \frac{\omega}{\mathfrak{L}}\rightarrow \Omega_{\wallr},\qquad \Omega_{\wallr}\equiv \sqrt{U(\wallr)}=\frac{\sqrt{f(\wallr)}}{\wallr}.
\end{align}
In the vicinity of the turning point, the WKB approximation for the master field breaks down within a layer of tortoise coordinate width  $\Delta r^*\propto\angL^{-2/3}$. Zooming into this layer, the master field equation reduces, at leading nontrivial order in the large-$\mathfrak L$ expansion, to an Airy equation \footnote{Note that this statement relies on the fact that $U'(\wallr)\neq0$ away from the photon sphere and therefore we can expand the optical potential linearly in the vicinity of the turning point.}. The Airy solution is selected by matching to the ingoing WKB solution near the horizon. The frequency expansion for these modes takes the form
\begin{align}
  \omega
  &=
  \angL\OmR+C_{\wallr}\nu\,\angL^{1/3}
  +\Order(\angL^{-1/3}),
  \label{eq:airy-frequency}
  \end{align}
where the coefficient of the $\mathfrak L^{1/3}$ term is given by
\begin{align} 
 C_{\wallr}
  &=
  \frac{|fU'|^{2/3}}{2\OmR}
  =
  \OmR\left[\frac{\Delta^2}{8f}\right]^{1/3},
  \label{eq:CR}
\end{align}
with all the background quantities evaluated at $r=\wallr$, and $\Delta$ defined
in \eqref{eq:Delta-background}. As remarked in \cite{AnninosGalanteGeorgescuManeeratSvesko2026}, $C_{\wallr}$ can also be expressed in terms of the background conformal stress tensor, using \eqref{eq:T2-SAdS}, and takes the form \eqref{eq:large L general}.

The conformal boundary is chosen to lie within the Airy layer and the conformal boundary conditions evaluated on the selected Airy solution fix the dimensionless number $\nu$.

As explained in Appendix \ref{app:wkb}, for the interior problem the ingoing condition at the horizon, implemented by matching to the near-horizon WKB solution, selects the rotated Airy function $ \operatorname{Ai}\!\big(-\e^{2\pi i/3}\nu\big)$. In the large-$\mathfrak L$ limit, the conformal boundary coincides with the turning point located on the horizon side of the photon sphere. The scalar conformal boundary condition then yields
\begin{equation}
    \e^{2\pi\ii/3}
    \operatorname{Ai}'\!\left(-\e^{2\pi i/3}\nu\right)
    -4\nu^2
    \operatorname{Ai}\!\left(-\e^{2\pi i/3}\nu\right)
    =0.
  \label{eq:black-hole-airy}
\end{equation}
As noted in \cite{LiuReallSantosWiseman2025,AnninosGalanteGeorgescuManeeratSvesko2026}, this equation admits a root with positive imaginary part,\footnote{The equation admits additional roots, but among those found numerically, only the one displayed here has a positive imaginary part.}
\begin{align}\label{cosmic}
   \nu_{\rm int}
    =-0.4042886500+0.1555613292 i.
\end{align}
Since $C_{\wallr}>0$ away from the photon sphere, the corresponding mode is unstable and grows exponentially in time in the $\mathfrak L\rightarrow\infty$ limit.

For the exterior problem, the ingoing condition selects the Airy function $\operatorname{Ai}(-\nu)$\footnote{Exact matching with the WKB solution which is ingoing at the horizon also introduces a $\operatorname{Bi}$ component, which is exponentially suppressed in the large-$\mathfrak L$ limit. We provide the details of this mixture in Appendix \ref{app:wkb}.}. In the large-$\mathfrak L$ limit, the conformal boundary coincides with the turning point on the asymptotic side of the photon sphere. The scalar conformal boundary condition then yields
\begin{equation}
    \operatorname{Ai}'(-\nu)
    -4\nu^2\operatorname{Ai}(-\nu)=0.
  \label{eq:pole-airy}
\end{equation}
The equation admits the growing root 
\begin{align}\label{pole}
   \nu_{\mathrm{ext}} =0.0674242621+0.4279049060 i.
\end{align} 
These growing roots are familiar in the context of gravitational dynamics with conformal boundaries. The root \eqref{pole} was previously found in the pole patch of the de Sitter static patch \cite{AnninosGalanteGeorgescuManeeratSvesko2026} and in the interior spherical cavity in Minkowski spacetime \cite{LiuReallSantosWiseman2025}. Similarly, the root \eqref{cosmic} was found in the cosmic patch of the de Sitter static patch \cite{AnninosGalanteGeorgescuManeeratSvesko2026} and in the exterior spherical cavity in Minkowski spacetime \cite{LiuReallSantosWiseman2025}.

The solutions found in both the interior and exterior problems have a non-vanishing master field and Weyl factor, related by \eqref{eq:weyl-mode}. Away from the Airy region, the exterior solution decays into the photon-sphere barrier, whereas the interior solution has an oscillatory ingoing continuation toward the horizon. Among these solutions, both the interior and exterior problems admit unstable modes of the form \eqref{eq:large L general}, whose growth rates become unbounded as \(\mathfrak L\to\infty\), for fixed \(\wallr\) away from the photon sphere.

\begin{figure}[t]
  \centering
  \paperfigure[width=\textwidth]
    {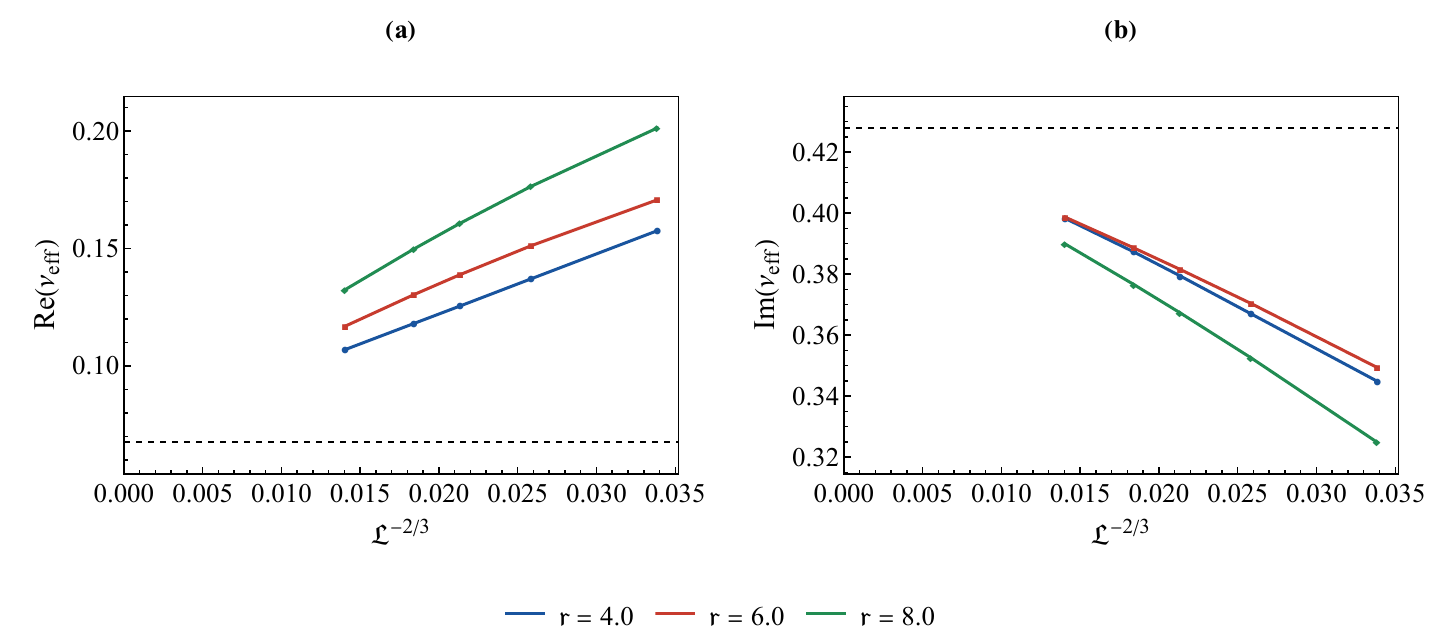}

  \par\medskip

  \paperfigure[width=\textwidth]
    {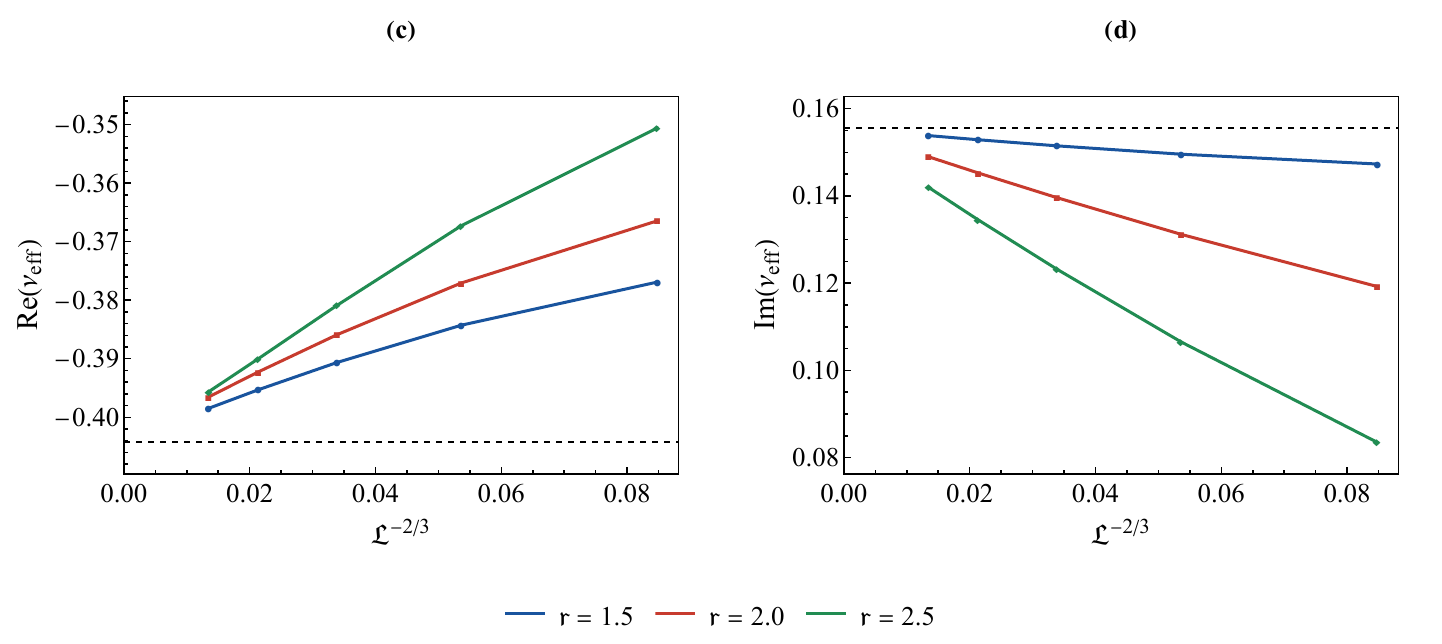}
  \caption{
Scalar sector numerically determined frequencies compared with the analytic Airy predictions. Panels $(a,b)$ show the exterior problem at $\wallr>\rph$,
and panels $(c,d)$ the interior problem at
$r_h<\wallr<\rph$. The real and imaginary parts of $\nu_{\mathrm{eff}}$ are
plotted against $\angL^{-2/3}$ for the indicated boundary radii. The horizontal dashed lines mark the corresponding growing Airy roots in \eqref{cosmic} and \eqref{pole}.
The large angular momentum limit lies at the left edge of
each panel.
}
  \label{fig:wall-airy}
  \label{fig:exterior-airy}
  \label{fig:interior-airy}
\end{figure}

To compare the frequencies corresponding to \eqref{cosmic} and \eqref{pole}, obtained from the Airy analysis, with the numerical results, we introduce
\begin{equation}
  \nu_{\mathrm{eff}}
  \equiv
  \frac{\omega-\mathfrak{L}\Omega_{\wallr}}
       {C_{\wallr}\mathfrak{L}^{1/3}}
  =
  \nu+\Order(\mathfrak{L}^{-2/3}).
  \label{eq:effective-airy-parameter}
\end{equation}
At fixed boundary position, the numerically determined $\nu_{\mathrm{eff}}$ approaches the corresponding analytic Airy root, \eqref{cosmic} or \eqref{pole}, as $\mathfrak L\to\infty$, as shown in Figure \ref{fig:wall-airy}. The agreement becomes less accurate as the boundary approaches the photon sphere, where the local Airy approximation breaks down. It also worsens towards the asymptotic AdS boundary in the exterior problem\footnote{For the exterior radii in Figure~\ref{fig:wall-airy}, the asymptotic effect dominates. Boundaries closer to the photon sphere, such as $\wallr=3.5$, display the expected degradation near the photon sphere.}. In the interior problem, it instead improves towards the horizon, until $|Kr_h|\sim\mathfrak L$, where the large-$\mathfrak L$ expansion itself ceases to apply.

Indeed, the width of the Airy layer depends on the position of the conformal boundary as
\begin{equation}
  \Delta r_{\mathrm{Airy}}
  \sim
  \left(
    \frac{f(\wallr)^2}
    {\mathfrak{L}^2|U'(\wallr)|}
  \right)^{1/3}.
  \label{eq:airy-width}
\end{equation}
Near the photon sphere, this scales as
$\Delta r_{\mathrm{Airy}}\sim
\mathfrak{L}^{-2/3}|\wallr-\rph|^{-1/3}$,
so that the Airy layer broadens as $\wallr\to\rph$. At the same time, $C_{\wallr}\to0$. Once the width of the Airy layer becomes comparable to the distance
between the conformal boundary and the photon sphere, the linear approximation to the potential used to derive \eqref{cosmic} and \eqref{pole} ceases to be valid.

\medskip
\noindent\textbf{Growth rate near umbilicity.}
We can now quantify how the large-$\mathfrak{L}$ instabilities are suppressed as the conformal boundary approaches the photon sphere. Let $K_{\rm ph}=K(\rph)=3\Omega_{\rm ph}$ and define the dimensionless departure from umbilicity
\begin{align}
 \epsilon_K&\equiv\rph\left[K(\wallr)-K_{\rm ph}\right]
 =-\frac{\wallr-\rph}{\rph\sqrt{f_{\rm ph}}}
   +\Order\!\left(\frac{(\wallr-\rph)^2}{\rph^2}\right).
 \label{eq:departure-umbilicity-v5}
\end{align}
Near the photon sphere, the coefficient $C_{\wallr}$ \eqref{eq:CR} controlling the $\mathfrak{L}^{1/3}$ term behaves as
\begin{equation}
 \frac{C_{\wallr}}{\Omega_{\rm ph}}
 =\frac{|\epsilon_K|^{2/3}}{2^{1/3}}
  \left[1+\Order(|\epsilon_K|)\right],
 \label{eq:near-umbilic-coefficient-v5}
\end{equation}
so that the unstable interior and exterior branches have growth rate
\begin{equation}
 \frac{\operatorname{Im}\omega}{\Omega_{\rm ph}}
 \sim\operatorname{Im}\nu_{\rm int,ext}
       \left(\frac{\mathfrak L\epsilon_K^2}{2}\right)^{1/3},
 \label{eq:near-umbilic-growth-v5}
\end{equation}
provided that $\mathfrak L\epsilon_K^2\gg1.$ This condition corresponds to the regime of validity of the Airy approximation, for which the width \eqref{eq:airy-width} must remain much smaller than $|\wallr-\rph|$. Equivalently, the angular momentum scale at which the Airy regime becomes valid grows as
\begin{equation}
 \mathfrak L_{\rm Airy}\sim|\epsilon_K|^{-2}.
 \label{eq:umbilic-angular-crossover-v5}
\end{equation}
Thus, as the boundary approaches the umbilic surface, increasingly large angular momentum is required to recover the Airy growth law. The crossover occurs for $\mathfrak L\epsilon_K^2=\Order(1)$, precisely the near-photon-sphere scaling studied in Section \ref{sec:uniform}.

Most of the analysis in this section so far also applies to the vector sector. In particular, the frequency expansion \eqref{eq:airy-frequency} and the Airy solutions for the master field in the interior and exterior problems remain unchanged. The only difference lies in the conformal boundary condition that determines $\nu$. In the vector sector, the analogues of \eqref{eq:black-hole-airy} and \eqref{eq:pole-airy} are
\begin{align}
\mathrm{Ai}'(-e^{2\pi i/3}\nu)=0,
\qquad
\mathrm{Ai}'(-\nu)=0,
\end{align}
respectively. The former admits only roots with negative imaginary parts, while the latter has only real roots. Thus, unlike in the scalar sector, no unstable modes arise in the large-angular-momentum expansion. As in the scalar sector, the Airy approximation breaks down as the conformal boundary approaches the photon sphere, where the slope of the effective potential vanishes. We now turn to this near-photon-sphere regime, including the exactly umbilic case $\wallr=\rph$.

\subsection{Conformal boundary near and at the photon sphere}
\label{sec:uniform}
We now consider the regime in which the conformal boundary approaches the photon sphere in the large-$\mathfrak L$ limit, namely $|\wallr-\rph|\to0$ as $\mathfrak L\to\infty$. We focus on frequencies approaching the top of the effective potential, for which the two turning points coalesce near the photon sphere. Their Airy regions are then no longer parametrically separated, and the linear expansion of the potential around each turning point ceases to be valid. Instead, we retain the quadratic expansion of the potential about its maximum. Balancing this quadratic variation against the radial-derivative term in \eqref{eq:eikonal-equation} gives a characteristic width $\Delta r\sim\mathfrak L^{-1/2}$, over which the two contributions enter at the same order. This region is the near-maximum analogue of the Airy layer around a simple turning point. As detailed in Appendix \ref{app:wkb}, within it, the WKB approximation breaks down.\footnote{Indeed, if $|r^*-\rph^*|=\mathcal{O}(\mathfrak{L}^{\gamma})$, the quadratic variation of the potential contributes to \eqref{eq:eikonal-equation} at order $\mathcal{O}(\mathfrak{L}^{2+2\gamma})$, while the radial-derivative term is of order $\mathcal{O}(\mathfrak{L}^{-2\gamma})$. Balancing the two gives $\gamma=-1/2$. By contrast, near a simple turning point the linear variation of the potential contributes at order $\mathcal{O}(\mathfrak{L}^{2+\gamma})$, which balances the derivative term for $\gamma=-2/3$, reproducing the Airy scaling encountered above.} We focus on modes for which both turning points lie within this near-maximum region. We zoom into this region by introducing the rescaled radial coordinate
\begin{equation}
  r=\rph+\frac{y}{\sqrt{\mathfrak L}}.
  \label{eq:uniform-scaling}
\end{equation}
We choose the conformal boundary to lie within the same region,
\begin{align}
   \wallr=\rph+\frac{\mathfrak{y}}{\sqrt{\mathfrak L}}, \qquad \mathfrak{y}=\mathcal O(1).
   \label{eq:position boundary PC}
\end{align}
Expanding the master field equation in the large-$\mathfrak L$ limit then requires the frequency to take the form
\begin{align}\label{eq:form of frequency largel}
  \omega=\Omega_{\rm ph}\mathfrak L+W,
\end{align}
with $W=\mathcal O(1)$. Unlike in the Airy regime, there is therefore no $\mathfrak L^{1/3}$ term. Using \eqref{eq:form of frequency largel}, the master field equation reduces in the large-$\mathfrak{L}$ limit to an inverted harmonic-oscillator equation,
\begin{equation}
  \phi^{(S)\prime\prime}(y)+q(y)\phi^{(S)}(y)=0,
  \qquad
  q(y)=a y^2+\frac{2\Omega_{\rm ph}}{f_{\rm ph}^2}W,
  \label{eq:uniform-master}
\end{equation}
where primes now denote derivatives with respect to $y$, and
\begin{equation}
  a=-\frac{U''(\rph)}{2f_{\rm ph}^2}
   =\frac{1}{\rph^4f_{\rm ph}^2}.
  \label{eq:a-q}
\end{equation}
Expanding the conformal boundary condition \eqref{eq:regular-residual} in the same near-maximum scaling, we obtain
\begin{equation}
  \mathfrak{y} \phi^{(S)'}(\mathfrak{y})+\frac{2}{a}q(\mathfrak{y})^2\phi^{(S)}(\mathfrak{y})=0.
  \label{eq:uniform-CBC}
\end{equation}
Keeping both terms in \eqref{eq:regular-residual} is essential. Although the normal-displacement coefficient $\alpha$ vanishes as the conformal boundary approaches the potential maximum, the corresponding fixed-embedding contribution to the mean-curvature variation $\mathcal{B}$ grows on the same scale. Their product therefore remains finite and contributes to \eqref{eq:uniform-CBC}.

\medskip

\noindent \textbf{Solution to the master field equation.} The solutions to the inverted harmonic-oscillator problem are parabolic-cylinder functions. The same functions arise in the WKB treatment of black-hole resonances \cite{SchutzWill1985,IyerWill1987,Olver1974}, as well as in the near-photon sphere analysis of \cite{HadarEtAl2022,Hashimoto:2023buz}. Here, we select those that are ingoing at the horizon and satisfy the conformal boundary conditions at $y=\mathfrak{y}$. Introducing
\begin{equation}
  \nu_{\mathrm{PC}}
  =-\frac12+\frac{\ii W}{\Omega_{\rm ph}},
  \label{eq:parabolic-variables}
\end{equation}
the solution to \eqref{eq:uniform-master} selected by the ingoing horizon
condition is \footnote{On the horizon side, the behaviour of \eqref{eq:parabolic-solution} for large negative $y$ is, up to a $y$-independent factor,
\begin{equation*}
  \phi^{(S)}_{\mathrm{in}}(y)
  \sim
  |y|^{-1/2+\ii W/\Omega_{\mathrm{ph}}}
  \exp\left(\frac{\ii \sqrt a}{2}y^2\right).
  \label{eq:parabolic-ingoing-asymptotic}
\end{equation*}
For $y<0$, the derivative of the leading radial phase is negative, selecting propagation towards smaller radii with the time convention \eqref{eq:time-convention}. See Appendix \ref{app:wkb} for details.}

\begin{equation}
  \phi^{(S)}_{\mathrm{in}}(y)
  =D_{\nu_{\mathrm{PC}}}(-\e^{-i \pi/4}(4a)^{1/4}y).
  \label{eq:parabolic-solution}
\end{equation}
Substituting \eqref{eq:parabolic-solution} into the
conformal boundary condition \eqref{eq:uniform-CBC} gives the following equation for the allowed frequencies
\begin{equation}
  \mathfrak{y}
  \partial_y D_{\nu_{\mathrm{PC}}}(-\e^{-i \pi/4}(4a)^{1/4}\mathfrak{y})
  +\frac{2}{a}q(\mathfrak{y})^2
  D_{\nu_{\mathrm{PC}}}(-\e^{-i \pi/4}(4a)^{1/4}\mathfrak{y})
  =0.
  \label{eq:uniform-frequency}
\end{equation}

\medskip

\noindent \textbf{Matching to the Airy regime.} As the conformal boundary moves away from the near-maximum region, the parabolic-cylinder description should match onto the Airy analysis of Section \ref{sec:airy}. To make this connection explicit, we examine the large-$|\mathfrak{y}|$ limit of the master field equation \eqref{eq:uniform-master} and the conformal boundary condition \eqref{eq:uniform-CBC}. Both the interior and exterior problems can be treated by introducing
\begin{equation}
  \kappa=(2a|\mathfrak{y}|)^{1/3},
  \qquad
  \xi=\operatorname{sgn}(\mathfrak{y})\,\kappa(\mathfrak{y}-y),
  \qquad
  \nu=\frac{q(\mathfrak{y})}{\kappa^2}.
  \label{eq:PC-Airy-variables}
\end{equation}
The conformal boundary is located at $\xi=0$. In these variables, equations \eqref{eq:uniform-master} and \eqref{eq:uniform-CBC} become
\begin{align}
  \partial_{\xi}^2\phi^{(S)}
  +\left(
    \nu-\xi+\frac{a}{\kappa^4}\xi^2
  \right)\phi^{(S)}  &=0,
  \label{eq:outer-Airy-overlap-equation}\\
  \left(4\nu^2-\partial_\xi\right)\phi^{(S)}(0) &=0.
  \label{eq:overlap-Airy-CBC}
\end{align}
Since $a/\kappa^4=\Order(|\mathfrak{y}|^{-4/3})$, the quadratic term is suppressed at large $|\mathfrak{y}|$, and the master-field equation reduces to Airy form. Which Airy solution is selected depends on the side of the potential maximum on which the boundary lies. For $\mathfrak{y}>0$, corresponding to the exterior problem, moving inward from the boundary toward the horizon gives $\xi>0$ and enters the classically forbidden photon-sphere barrier. As in the exterior Airy analysis of Section \ref{sec:airy}, the local solution is selected by matching to the decaying WKB branch. For $\mathfrak{y}<0$, corresponding to the interior problem, the horizon side has $\xi<0$ and lies in the oscillatory region, where the local solution is instead fixed by matching to the ingoing WKB branch. Accordingly, the large-$|\mathfrak{y}|$ limits of \eqref{eq:parabolic-solution} are
\begin{equation}
  \phi^{(S)}(\xi)\propto
  \begin{cases}
    \operatorname{Ai}(\xi-\nu),
      &  \mathfrak{y}>0,\\[2mm]
    \operatorname{Ai}\!\left(
      \e^{2\pi i/3}(\xi-\nu)
    \right),
      & \mathfrak{y}<0.
  \end{cases}
  \label{eq:PC-Airy-continuations}
\end{equation}
Substituting these expressions into \eqref{eq:overlap-Airy-CBC} reproduces the exterior and interior Airy conditions, leading to the coefficients \eqref{pole} and \eqref{cosmic}, respectively. The same matching also reproduces the corresponding large-$\mathfrak L$ frequency expansion. Using $q(\mathfrak{y})=\nu\kappa^2$, we find
\begin{equation}
  W  =
  \frac{U''(\rph)}{4\Omega_{\rm ph}}\mathfrak{y}^2
  +
  \Omega_{\rm ph}
  \left(\frac{\mathfrak{y}^2}{2 \rph^2 f_{\rm ph}}\right)^{1/3}\nu
  +
  \Order(|\mathfrak{y}|^{-2/3}),
  \label{eq:uniform-large-y-frequency}
\end{equation}
with $\nu=\nu_{\rm int}$ or $\nu_{\rm ext}$ for the growing branches. The two terms reproduce, respectively, the expansions of $\mathfrak L(\Omega_{\wallr}-\Omega_{\rm ph})$ and $C_{\wallr}\nu\mathfrak L^{1/3}$ near the photon sphere.

\begin{figure}[t]
  \centering
  \paperfigure[width=\textwidth]{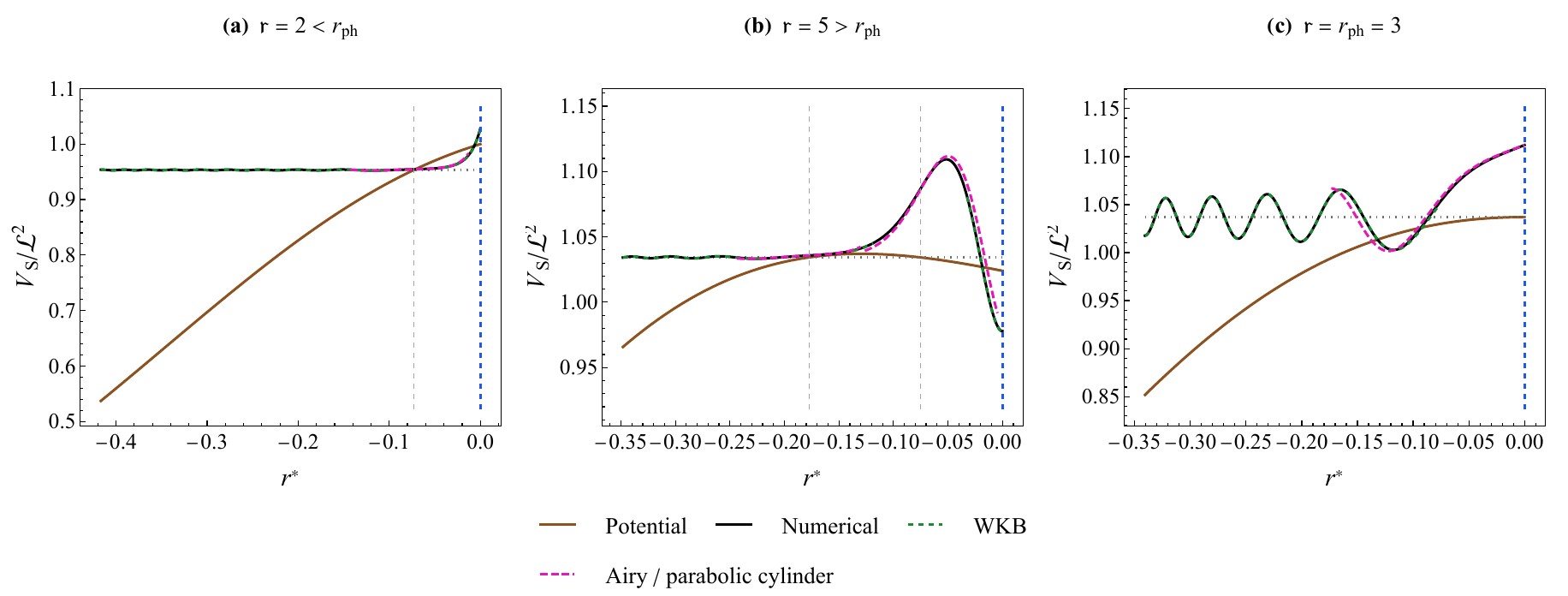}
  \caption{
  Radial solutions of the scalar KI equation and their
  asymptotic descriptions for three positions of the boundary, with
  $r_h=\ads=1$.
  The brown curve is $V_S/\angL^2$.
  Black curves show the real part of the numerical solution,
  while green and magenta dashed curves show the WKB and
  local uniform approximations, respectively.
  Profiles share a common normalization within each panel
  and are shifted and rescaled for display about the
  horizontal energy line $\omega^2/\angL^2$.
  The blue vertical line marks the conformal boundary, where the additive
  constant of $r^*$ is fixed by $r^*(\wallr)=0$.
  (a) An isolated simple turning point for $\wallr<\rph$,
  with $l=400$.
  (b) Two separated turning points enclosing the exterior
  barrier for $\wallr>\rph$, with $l=800$.
  (c) The photon-sphere region at the barrier-top energy
  of the leading uniform problem, with $l=400$, where
  the local description uses parabolic-cylinder functions.
  Real frequencies are chosen to display the turning-point
  geometry; no CBC eigenvalue condition is imposed.
  The numerical integration starts with ingoing WKB data
  in the allowed region, and the relative coefficients of
  the asymptotic solutions follow from the connection
  formulae.
  }
  \label{fig:profile-matching}
\end{figure}

The matching of the radial master field profiles is illustrated in Figure \ref{fig:profile-matching}. Away from the turning points, the master field is described by the WKB approximation. An isolated simple turning point is described locally by an Airy solution, whereas near the photon-sphere maximum the two nearby turning points must be treated together using the parabolic-cylinder solution \eqref{eq:parabolic-solution}.

\medskip

\begin{figure}[t]
  \centering
  \paperfigure[width=\textwidth]
    {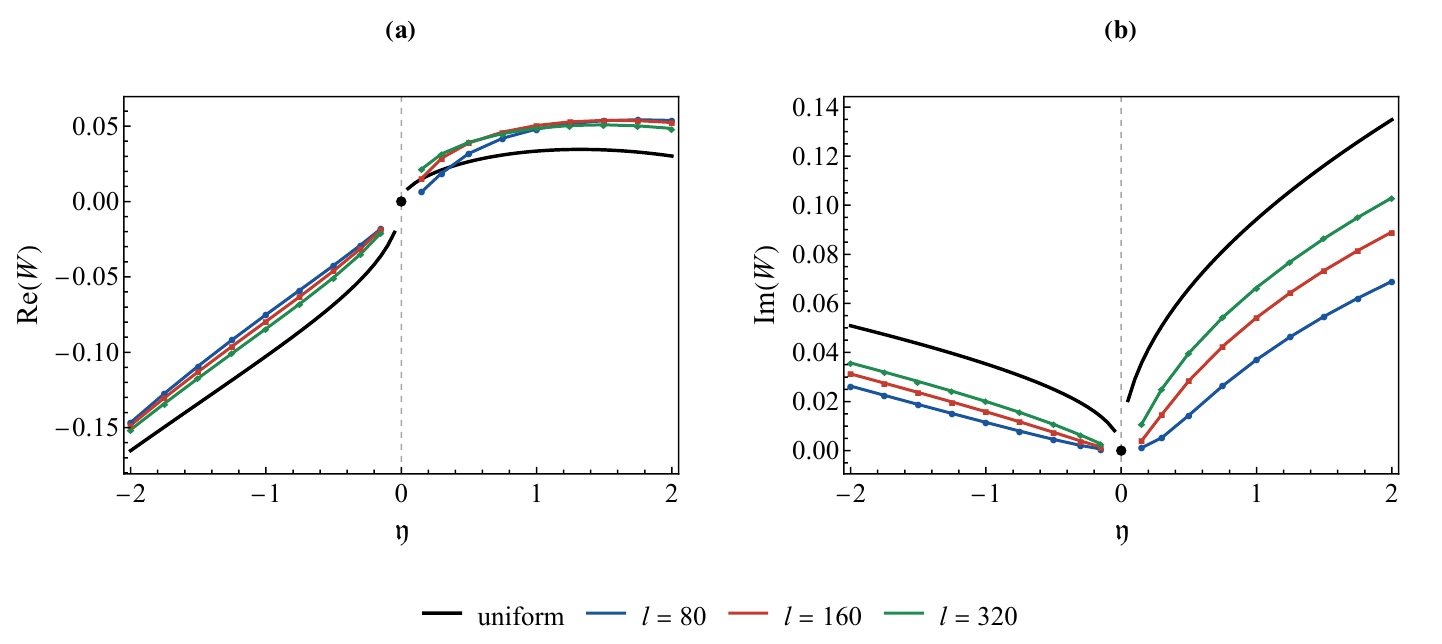}
  \caption{
Frequency crossover in the photon-sphere scaling
\eqref{eq:position boundary PC}.
Panels (a) and (b) show the real and imaginary parts of $W$, respectively. The black curves give the leading uniform prediction from \eqref{eq:uniform-frequency}, while the colored curves show the full scalar CBC roots at the indicated angular momenta.
Negative and positive $\mathfrak{y}$ correspond to conformal boundaries inside and
outside the photon sphere, respectively.
The vertical dashed line marks the umbilic boundary.
The black point at $\mathfrak{y}=W=0$ belongs to the leading uniform problem. As $\mathfrak{y}\rightarrow 0$, the positive imaginary parts of the frequencies vanish.
}
  \label{fig:photon-crossover}
\end{figure}

\noindent \textbf{Allowed frequencies near the photon sphere.}
We now turn to solving the condition \eqref{eq:uniform-frequency} for the allowed frequencies. For a generic boundary position $\mathfrak{y}=\mathcal{O}(1)$, it must be solved numerically. We have done so for a range of boundary positions, and representative results are shown in Figure \ref{fig:photon-crossover}. One of the resulting branches is compared there with the roots of the full conformal boundary condition determinant \eqref{eq:regular-residual} at finite angular momentum. As $\mathfrak{L}$ increases, the frequencies of the full problem approach those predicted by \eqref{eq:uniform-frequency}. The parabolic-cylinder description therefore follows the spectrum through the region in which the separate Airy approximations cease to apply.

When the conformal boundary approaches the photon sphere, $\mathfrak{y}\rightarrow 0$, \eqref{eq:uniform-frequency} gives
\begin{equation}
  W^2
  =
  -\mathfrak{y}\,
  \frac{\Omega_{\mathrm{ph}}^2(4a)^{1/4}}
       {4\sqrt2}
  \e^{-i\pi/4}
  \frac{\Gamma(3/4)}{\Gamma(1/4)}
  +\Order(|\mathfrak{y}|^{3/2}).
  \label{eq:Puiseux}
\end{equation}
Thus, $W=\mathcal O(|\mathfrak{y}|^{1/2})$, corresponding to a square-root splitting of two nearby roots. At $\mathfrak{y}=0$, the two roots coalesce at $W=0$ at leading order in the near-photon-sphere approximation. In the following, we identify them with the bulk and boundary branches of the exact factorization \eqref{eq:exact-umbilic-factorization}.

\medskip

\noindent \textbf{Exactly umbilic case.}
In the particular case $\mathfrak{y}=0$, corresponding to an exactly umbilic conformal boundary, the allowed frequencies can be determined analytically in the large-$\mathfrak{L}$ expansion. Taking the strict $\mathfrak{y}\rightarrow 0$ limit of \eqref{eq:uniform-frequency}, we obtain
\begin{align}
     0&=W^2 D_{\nu_{\rm PC}}(0)
  =
  W^2\,
  \frac{2^{\nu_{\rm PC}/2}\sqrt{\pi}}
       {\Gamma\!\left[(1-\nu_{\rm PC})/2\right]}.
       \label{eq:umbilic-gamma-factor}
\end{align}
We therefore find two classes of solutions: (i) a double zero at $W=0$, anticipated by \eqref{eq:Puiseux}, and (ii) an infinite tower arising from the poles of the Gamma function in the denominator. 

\begin{figure}[t]
  \centering
  \paperfigure[width=\textwidth]
    {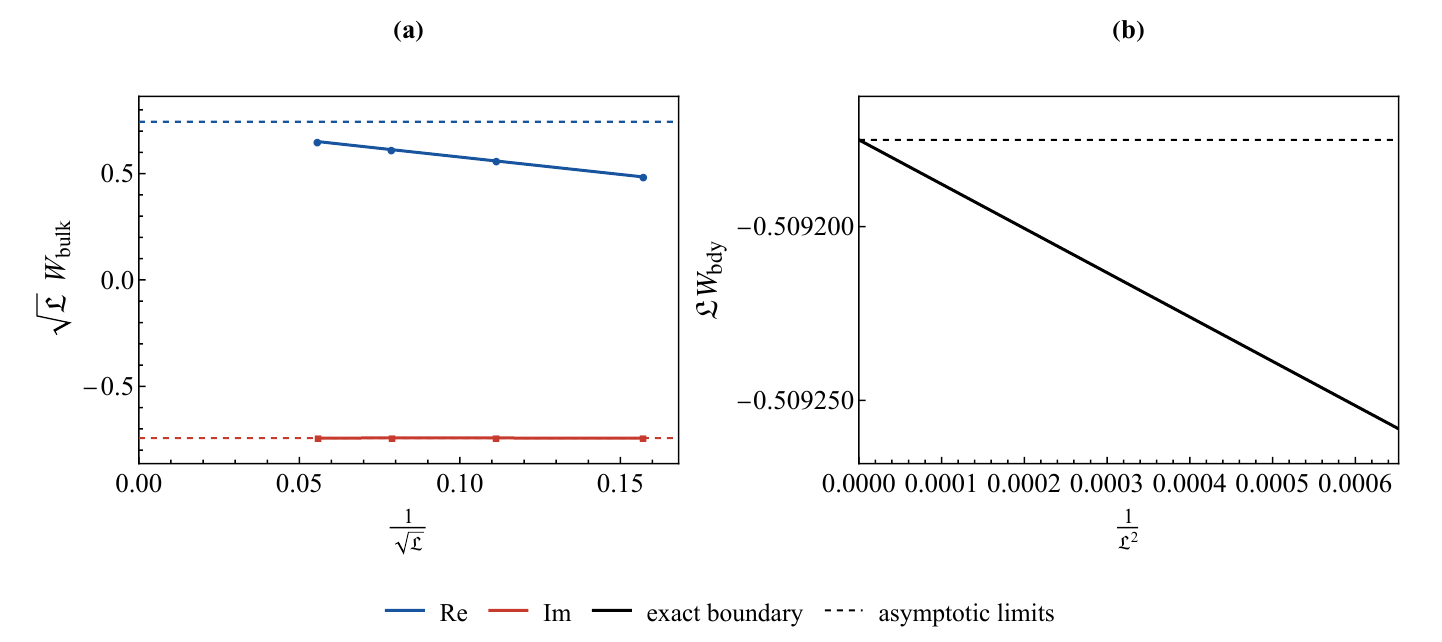}
  \caption{
Boundary and bulk frequency shifts at the umbilic conformal boundary, for $r_h=\ads=1$.
(a) Real and imaginary parts of
$\sqrt{\angL}\,W_{\rm bulk}$, obtained from the exact bulk
Robin condition for $l=40,80,160,320$.
The dashed lines mark the corresponding components of $w_1$
in \eqref{eq:w1-bulk-umbilic}.
(b) The rescaled boundary shift $\angL W_{\rm bdy}$,
evaluated from the exact boundary dispersion relation.
The dashed line marks its limiting value $-\Omega_{\rm ph}/2$.
The different rescalings resolve the two sectors that
coincide at $W=0$ in the leading uniform description.
}
  \label{fig:umbilic-splitting}
\end{figure}

\medskip
\noindent \textbf{(i) Double zero at $W=0$.} One of the two zeros at $W=0$ corresponds to the boundary branch \eqref{eq:umbilic-boundary-dispersion}. Expanding the exact dispersion relation at large $\mathfrak{L}$ gives
\begin{align}
    W_{\rm bdy}
  =
  -\frac{\Omega_{\rm ph}}{2\mathfrak{L}}
  +\Order(\mathfrak{L}^{-3}).
\end{align}
The other zero belongs to the bulk branch and can be obtained by expanding the exact Robin condition \eqref{eq:exact-umbilic-bulk-Robin}, evaluated on the parabolic-cylinder solution. The appropriate scaling is
\begin{equation}
  W_{\rm bulk}
  =
  \frac{w_1}{\sqrt{\mathfrak{L}}}
  +\mathcal{O}(\mathfrak{L}^{-1}),
  \label{eq:umbilic-largeL-dispersion}
\end{equation}
with
\begin{equation}
  w_1  =
  \frac{f_{\rm ph}}{\rph}
  \frac{\Gamma(3/4)}{\Gamma(1/4)}
  \e^{-i\pi/4}\approx 0.238994(1-i)\frac{f_{\rm ph}}{\rph}.
  \label{eq:w1-bulk-umbilic}
\end{equation}
Both the boundary and bulk roots discussed above are therefore stable. Their shifts away from $\Omega_{\rm ph}\mathfrak{L}$ vanish in the large-$\mathfrak{L}$ limit, but on different scales, see Figure \ref{fig:umbilic-splitting}.

Away from the umbilic boundary, the normal displacement term in the conformal boundary condition \eqref{eq:regular-residual} couples the two sectors, producing the square-root splitting \eqref{eq:Puiseux}.

\medskip
\noindent \textbf{(ii) Infinite tower.} The other class of solutions to \eqref{eq:umbilic-gamma-factor} is the infinite tower
\begin{equation}
  W_n
  =
  -i \Omega_{\rm ph}
  \left(2n+1+\frac12\right),
  \qquad
  n=0,1,2,\ldots,
  \label{eq:umbilic-gamma-tower}
\end{equation}
which follows from imposing a Dirichlet condition on the parabolic-cylinder master field. These modes are also stable and belong to the bulk branch. Their origin lies in the inverted harmonic-oscillator problem.\footnote{We note that a similar tower was identified in \cite{HadarEtAl2022} for asymptotically flat Schwarzschild, though different boundary conditions were used; indeed, that set-up had no boundary but rather outgoing conditions were imposed at spatial infinity. In our case, the conformal boundary condition, which reduces to a Dirichlet condition at $\mathfrak{y}=0$, removes the even members of their tower. As explained in \cite{HadarEtAl2022}, the inverted harmonic oscillator wave equation leads to an $\mathfrak{sl}(2,\mathbb{R})$ symmetry of the spectrum, where the two towers (odd and even contributions) correspond to two different irreducible representations. In our case, only one of the irreducible representations, the odd one, is allowed by the boundary conditions in the scalar sector. A similar tower was also identified in \cite{Hashimoto:2023buz} for the Schwarzschild-AdS black hole, without a finite boundary.}

To verify that the roots \eqref{eq:umbilic-gamma-tower} persist at finite $\mathfrak{L}$, we followed the $n=0$ mode numerically in the exact problem. Its first subleading large-$\mathfrak{L}$ correction can also be obtained analytically, allowing for a more precise comparison with the numerical roots,
\begin{equation}
  W_{0}
  =
  -\frac{3 i}{2}\Omega_{\rm ph}
  +
  \sqrt{\frac{2}{\pi}}\,
  \frac{f_{\rm ph}}{\rph}\,
  \frac{1-i}{\sqrt{\mathfrak{L}}}
  +\Order(\mathfrak{L}^{-1}).
  \label{eq:first-gamma-root}
\end{equation}

\begin{figure}[t]
  \centering
  \paperfigure[width=\textwidth]{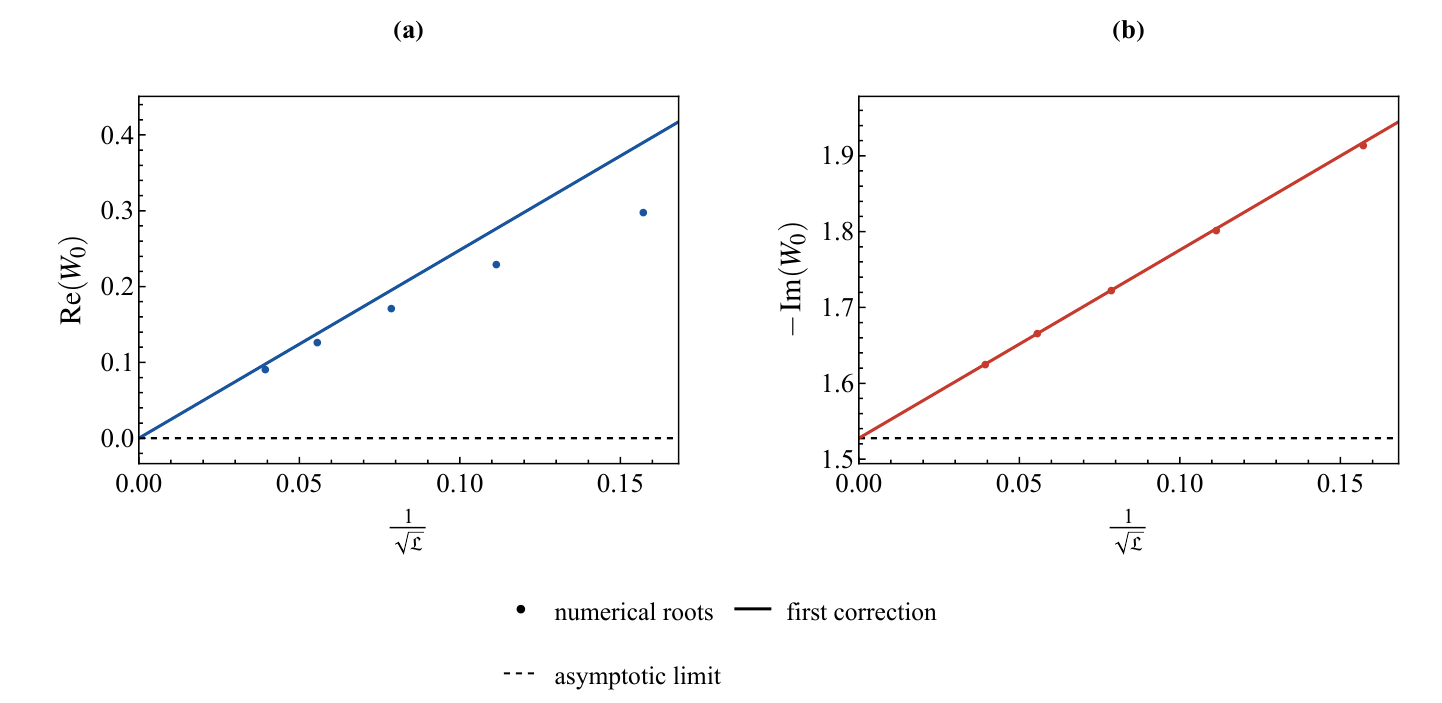}
  \caption{
  The first bulk root of the tower at the umbilic boundary, for
  $r_h=\ads=1$.
   Points show the real part (a) and damping (b) of the frequency
  shift $W_{0}$ obtained from the exact bulk Robin condition
  at $l=40,80,160,320,640$.
  Solid lines include the first large-angular-momentum correction;
  dashed lines mark the limiting value
  $W_0=-3 i\Omega_{\rm ph}/2$.
  Unlike the two branches approaching $W=0$, this mode retains
  finite damping in the uniform limit.
  }
  \label{fig:umbilic-bulk-tower}
\end{figure}

The exact roots at $l=40,80,160,320,640$ approach \eqref{eq:first-gamma-root}, as shown in Figure \ref{fig:umbilic-bulk-tower}. After subtracting the terms displayed above, the remainder scales as $\Order(\mathfrak{L}^{-1})$. We will not pursue the higher members of the tower \eqref{eq:umbilic-gamma-tower} here.

We conclude that the large-$\mathfrak{L}$ growing modes \eqref{eq:large L general} are suppressed as the conformal boundary approaches the photon sphere and are absent in the exactly umbilic case. The resulting spectrum is stable and separates into bulk and boundary sectors. We will find an analogous suppression of the large-$\mathfrak{L}$ instability in the asymptotic AdS limit.

\medskip

\noindent\textbf{Vector sector.} Most of the analysis above also applies to the vector sector. Using the same rescaled coordinate $y$, the master field equation reduces to the same inverted harmonic-oscillator problem, with the ingoing solution given by \eqref{eq:parabolic-solution}. The difference lies in the conformal boundary condition \eqref{eq:vector-residual}, which at leading order in the near-maximum scaling reduces to a Neumann boundary condition,
\begin{align}
\partial_y\phi^{(V)}(\mathfrak y)=0.
\end{align}
Substituting the ingoing parabolic-cylinder solution, we obtain at the exactly umbilic boundary
\begin{align}
  D'_{\nu_{\rm PC}}(0)=0,
\end{align}
where $\nu_{\rm PC}$ is given by \eqref{eq:parabolic-variables}, just like in the scalar sector. The corresponding solutions form an infinite tower
\begin{align}
W_n^{(V)}=-i\Omega_{\rm ph}\left(2n+\frac12\right),
\qquad n=0,1,2,\ldots,
\end{align}
which is precisely the even counterpart of the scalar sector tower \eqref{eq:umbilic-gamma-tower}. All these vector sector modes are therefore damped.

\subsection{Conformal boundary towards the asymptotic AdS boundary}
\label{sec:ads-limit}
\label{sec:ads-qnm-connection}
Finally, we consider the limit in which the conformal boundary approaches the asymptotic AdS boundary and becomes asymptotically umbilic. In this regime, the large-$\mathfrak{L}$ instabilities are attenuated, as first observed for global AdS$_4$ in \cite{AnninosAriasGalanteManeerat2025} and further analyzed in \cite{AnninosGalanteGeorgescuManeeratSvesko2026}. Since the analysis closely parallels that of these works, we keep the discussion brief.

Let us introduce the parameter
\begin{equation}
  \chi\equiv
  \frac{\mathfrak{L}\ads^2}{\wallr^2}.
  \label{eq:eta-ads}
\end{equation}
Using \eqref{eq:K-large-R}, this can equivalently be written as
\begin{align}
  \chi
  =
  2\mathfrak{L}(K\ads-3)+o(1),
\end{align}
in the joint large-$\wallr$, large-$\mathfrak{L}$ limit with $\chi$ held fixed. Taking the asymptotic limit, $K\ads\rightarrow 3$ , of the large-$\mathfrak{L}$ frequency expansion for the exterior problem \eqref{eq:airy-frequency}, we obtain
\begin{equation}
  \omega \ads
  =
  \mathfrak{L}+\frac{\chi}{2}+
  \nu_{\mathrm{ext}}
  \left(\frac{\chi}{2}\right)^{1/3}
  +\mathcal{O}(\chi^{-1/3}).
  \label{eq:ads-airy}
\end{equation}
This expansion is valid for
\begin{align}
  \chi\gg 1.
\end{align}
Its regime of validity can also be understood from the size of the Airy layer. Using \eqref{eq:airy-width}, we find
\begin{align}
  \frac{\Delta r_{\mathrm{Airy}}}{\wallr}
  \sim
  \frac{1}{2^{1/3}\chi^{2/3}}.
\end{align}
Thus, the Airy layer remains parametrically small compared with the position of the conformal boundary only for $\chi\gg1$.

The parameter $\chi$ therefore distinguishes the two possible orderings of the large-$\mathfrak{L}$ and asymptotic AdS limits. For $\chi\gg1$, the large-$\mathfrak{L}$ limit can be taken first and the exterior Airy branch persists, with a growth rate that decreases as the conformal boundary is moved outward. When $\chi=\mathcal{O}(1)$, the Airy approximation ceases to be valid. In the opposite regime, $\chi\ll1$, the asymptotic AdS limit is instead taken first. As discussed in Section \ref{subsection:asy AdS}, the conformal boundary conditions then factorize into stable boundary and bulk sectors. The $\mathfrak{L}^{1/3}$ Airy instability \eqref{eq:large L general} is therefore absent in the strict asymptotic limit, as in global AdS$_4$ \cite{AnninosAriasGalanteManeerat2025}. We interpret this stability as a consequence of the conformal boundary becoming asymptotically umbilic.

\section{The stretched-horizon limit}
\label{sec:str-horizon}

So far, we have identified two regions in which the large-$\mathfrak L$ instabilities \eqref{eq:large L general} are suppressed and the spectral problem becomes analytically tractable: near the photon sphere and near the asymptotic AdS boundary. A third such regime, already mentioned in Section \ref{sec:backgrounds}, is the stretched-horizon limit. In \cite{AnninosGalanteGeorgescuManeeratSvesko2026}, this limit was proposed as a way to suppress the large-$\mathfrak L$ instabilities near non-degenerate horizons. It was further argued that the resulting near-horizon spectrum has a universal structure, already captured by Rindler geometry. More precisely, the Rindler problem reproduces the stretched-horizon spectrum in the joint large-$|K r_h|$, large-$\mathfrak L$ limit, with their ratio held fixed. An earlier fluid description of the de Sitter horizon was obtained
in \cite{Anninos:2011zn}, where a near-horizon expansion led to
incompressible Navier--Stokes dynamics, including a nonlinear
construction with fixed conformal boundary data. The relativistic
conformal-fluid sector considered here is the one identified in
\cite{AnninosGalanteGeorgescuManeeratSvesko2026}. In this section, we first briefly review these Rindler results. We then derive the near-horizon spectra for Schwarzschild-AdS$_4$ and the AdS$_4$ black brane directly.

\subsection{Review of the Rindler stretched-horizon spectrum}

Starting from \eqref{eq:general-static-background} or \eqref{eq: black brane metric}, we introduce a rescaled radial coordinate that zooms into the near-horizon region,
\begin{equation}
  r-r_h=\frac{\kappa_h z^2}{2(K r_h)^2},
  \qquad
  z_0=\kappa_h^{-1}.
  \label{eq:near-horizon-rindler-map}
\end{equation}
At leading order in the large-$|K r_h|$ expansion, the metric takes the Rindler form
\begin{equation}
  \dd s^2\simeq
  \frac{1}{(K r_h)^2}
  \left[
  -\frac{z^2}{z_0^2}\dd t^2+\dd z^2
  +r_h^2(K r_h)^2\dd \hat{s}^2
  \right],
  \label{eq:near-horizon-rindler-metric}
\end{equation}
where
\begin{equation}
  \dd \hat{s}^2=
  \begin{cases}
    \dd\Omega_2^2,
      & \text{Schwarzschild-AdS}_4,\\[1mm]
    \ell_{\rm AdS}^{-2}(\dd x^2+\dd y^2),
      & \text{AdS}_4\text{ black brane}.
  \end{cases}
\end{equation}
In the Schwarzschild-AdS$_4$ case, the $S^2$ factor in \eqref{eq:near-horizon-rindler-metric} has an effective radius $r_h|K r_h|$, which diverges in the stretched-horizon limit. The sphere can therefore be approximated locally by two-dimensional flat space. As explained in \cite{AnninosGalanteGeorgescuManeeratSvesko2026}, the spherical harmonics are then replaced locally by planar waves with momentum
\begin{equation}
  k_{\mathcal R}=\frac{\mathfrak L}{r_h|K r_h|}.
\end{equation}
Keeping a finite nonzero momentum in the Rindler limit therefore requires taking $\mathfrak L\rightarrow\infty$ together with $|K r_h|\rightarrow\infty$, with their ratio held fixed. It is useful to introduce\footnote{In the notation of \cite{AnninosGalanteGeorgescuManeeratSvesko2026}, our $q_{\mathcal{R}}$ corresponds to their $\kappa$.} 
\begin{equation}
  q_{\mathcal{R}}\equiv
  \frac{\mathfrak L}{|K r_h|}
  \frac{\mathfrak z}{r_h}
  \simeq \mathfrak{z} k_{\mathcal{R}},
  \qquad
  \widehat\omega\equiv\omega z_0=\frac{\omega}{\kappa_h}.
  \label{eq:q-X}
\end{equation}
In the case of the black brane, the transverse geometry is already planar. However, because of the Rindler scaling, keeping a finite momentum in the Rindler limit requires taking $|Kr_h|\to\infty$ together with $k_{S,V}\to\infty$, while keeping their ratio fixed. We therefore define the analogues of (4.4) and (4.5) by replacing $\mathfrak L$ with $k_{S,V}$, the dimensionless momentum parameters in the scalar and vector sectors, respectively. We use the same notation $k_R$ and $q_R$ in both cases, since the Rindler analysis below is identical.

As argued in \cite{AnninosGalanteGeorgescuManeeratSvesko2026}, in the joint near-horizon limits described above, gravitational perturbations of Schwarzschild-AdS$_4$ and the AdS$_4$ black brane reduce to the same Rindler problem, which is particularly simple.

We consider linearized gravitational perturbations around
\begin{align}
    \dd s^2
    &=
    -\frac{z^2}{z_0^2}\dd t^2+\dd z^2+ \dd x^2+\dd y^2,
    \label{eq:Rindler geometry}
\end{align}
with a conformal boundary at $z=\mathfrak z$. We retain the region between the conformal boundary and the Rindler horizon at $z=0$. In both the scalar and vector sectors, the master field equation reduces to the same Bessel equation. The solution that is ingoing at the Rindler horizon is
\begin{align}
    \phi^{(S,V)}_{\rm in}(z)
    &=
    J_{-i\widehat\omega}(-ik_{\mathcal{R}}z).
\end{align}
Substituting these solutions into the conformal boundary conditions gives the equations determining the allowed frequencies. 

\medskip

\noindent \textbf{Scalar sector.}
In the scalar sector, the conformal boundary condition takes the form
\begin{align}
  \left[
    2q_{\mathcal{R}}^4+2\widehat\omega^2(1+\widehat\omega^2)
    +q_{\mathcal{R}}^2(-1+\widehat\omega(i-4\widehat\omega))
  \right]J_{-i\widehat\omega}(-i q_{\mathcal{R}})
  -i q_{\mathcal{R}}^3J_{1-i\widehat\omega}(-i q_{\mathcal{R}})=0.
  \label{eq:rindler-determinant}
\end{align}
In the $q_{\mathcal{R}}\rightarrow0$ limit, the scalar spectrum organizes into three families:
\begin{itemize}
    \item \textit{Sound modes.} These are gapless fluid-dynamical roots,
    \begin{equation}
      \widehat\omega_{\mathrm{sound}}
      =\pm\frac{q_{\mathcal{R}}}{\sqrt2}
      -\frac{i}{4}q_{\mathcal{R}}^2+\Order(q_{\mathcal{R}}^3).
      \label{eq:rindler-smallq}
    \end{equation}
    The real part describes sound propagation with the conformal-fluid speed $c_s=1/\sqrt2$, while the negative imaginary part gives the leading attenuation.

    \item \textit{Soft modes.} These remain finite as $q_{\mathcal{R}}\rightarrow0$ and are given by
      \begin{align}
    \widehat\omega_{+}
  &=i-\frac{i}{2}q_{\mathcal{R}}^2+\Order(q_{\mathcal{R}}^4),
  \label{eq:soft-plus-rindler}\\
  \widehat\omega_{-}^{(\pm)}
  &=-i+\frac{i}{8}(3\pm\sqrt{17})q_{\mathcal{R}}^2
  +\Order(q_{\mathcal{R}}^4).
  \label{eq:soft-minus-rindler}
\end{align}

    \item \textit{QNM-like modes.} These form an infinite tower of dissipative modes,
    \begin{equation}
  \widehat\omega_n^{\mathrm{QNM}}
  =-i n
  +\frac{i (-1)^n q_{\mathcal{R}}^{2n}}
  {2^{2n}\Gamma(n+1)\Gamma(n)}
  +\Order(q_{\mathcal{R}}^{2n+2}),
  \qquad n\geq2.
  \label{eq:qnm-type-rindler}
\end{equation}

\end{itemize}

In the opposite limit, $q_{\mathcal{R}}\rightarrow\infty$, the solutions of \eqref{eq:rindler-determinant} also contain a growing Airy branch,
\begin{equation}
  \widehat\omega
  =q_{\mathcal{R}}+\nu_{\rm int}
  \left(\frac{q_{\mathcal{R}}}{2}\right)^{1/3}
  +\Order(q_{\mathcal{R}}^{-1/3}),
  \label{eq:stretched-airy}
\end{equation}
which reproduces the Rindler limit of
\eqref{eq:airy-frequency},\eqref{eq:black-hole-airy} and connects the Rindler spectrum with the large angular momentum analysis of the previous section.

\medskip

\noindent \textbf{Vector sector.} In the vector sector, the conformal boundary condition takes the form
\begin{equation}
  J_{-1-i\widehat\omega}(-i q_{\mathcal{R}})
  -J_{1-i\widehat\omega}(-i q_{\mathcal{R}})=0.
  \label{eq:vector-rindler-determinant}
\end{equation}
In the small-$q_{\mathcal{R}}$ limit, the vector spectrum organizes into two families:
\begin{itemize}
    \item \textit{Shear modes.} These are gapless fluid-dynamical roots,
    \begin{equation}
      \widehat\omega_{\mathrm{shear}}
      =-\frac{i}{2}q_{\mathcal{R}}^2+\Order(q_{\mathcal{R}}^4).
      \label{eq:vector-rindler-shear}
    \end{equation}
    Their purely imaginary frequency describes diffusive relaxation of transverse momentum, with no propagating part at this order.

    \item \textit{QNM-like modes.} These form an infinite tower of dissipative modes, similar to the one from the scalar sector
    \begin{equation}
  \widehat\omega_n^{\mathrm{QNM}}
  =-i n
  +\frac{i (-1)^{n+1} q_{\mathcal{R}}^{2n}}
  {2^{2n}\Gamma(n+1)\Gamma(n)}
  +\Order(q_{\mathcal{R}}^{2n+2}),
  \qquad n\geq 1.
  \label{eq:qnm-type-rindler}
\end{equation}

\end{itemize}

In the large-$q_{\mathcal{R}}$ limit, the solutions to \eqref{eq:vector-rindler-determinant} are determined by the zeros of the derivative of the Airy function, 
\begin{equation}
  \widehat\omega_j
  =q_{\mathcal{R}}+\e^{-2\pi i/3}a_j
  \left(\frac{q_{\mathcal{R}}}{2}\right)^{1/3}
  +\Order(q_{\mathcal{R}}^{-1/3}),
  \qquad
  \operatorname{Ai}'(-a_j)=0,
  \quad a_j>0.
  \label{eq:vector-rindler-airy}
\end{equation}
Unlike in the scalar sector, these large-$q_{\mathcal{R}}$ roots have negative imaginary parts and are therefore stable.

\subsection{The stretched-horizon spectrum for Schwarzschild AdS$_4$}
We now determine the stretched-horizon spectrum of Schwarzschild-AdS$_4$ at finite $\mathfrak{L}$. The procedure is straightforward. We introduce a radial coordinate that zooms into the near-horizon region while keeping the conformal boundary within it,
\begin{align}
  r&=r_h+\delta\rho,
  \qquad
  \mathfrak{r}=r_h+\delta,\qquad \delta/r_h\ll 1,
  \label{eq:near horizon expansion}
\end{align}
so that the boundary is located at $\rho=1$. We then solve the master field equation order by order in the small-$\delta$ expansion and impose the conformal boundary conditions. Finally, we trade $\delta$ for $|K r_h|$ using \eqref{eq:R-of-K-leading}. We present the derivation explicitly for the sound mode and only quote the results for the remaining modes.

\medskip

\noindent \textbf{Scalar sector.}
We expand the master-field equation \eqref{eq:master} at small $\delta$, writing the master field as
\begin{align}
  \phi^{(S)}(\rho)
  &=c_S \rho^{-i\omega/(2\kappa_h)}
  \left[
    \phi_0(\rho)
    +\delta \phi_1(\rho)
    +\mathcal{O}(\delta^2)
  \right].
\end{align}
The prefactor implements the ingoing near-horizon behaviour \eqref{eq:ingoing}. At leading order, the master field equation gives
\begin{align}
  \kappa_h \rho\,\phi_0''(\rho)
  +(\kappa_h-i\omega)\phi_0'(\rho)=0.
\end{align}
The ingoing solution sets $\phi_0=1$, where the remaining overall normalization has been absorbed into $c_S$. At the next order, we find
\begin{align}
  \phi_1(\rho)
  &=
  \left[
  -\frac{i \omega}{4r_h^2\kappa_h^2}
  +
  \frac{
  \left(\mathfrak{L}^4-2\mathfrak{L}^2+3\right)\ads^4
  +12\ads^2 r_h^2
  +9r_h^4
  }{
  2\ads^4 r_h^2
  \left(\mathfrak{L}^2+2r_h\kappa_h\right)
  (\kappa_h-i \omega)
  }
  \right]\rho.
\end{align}
Higher-order corrections can be obtained recursively in the same way.

Substituting the near-horizon expansion into the conformal boundary conditions, we find a pair of solutions with frequency expansion
\begin{align}
  \omega
  &=\omega_0\sqrt{\delta}
  +\omega_1\delta
  +\mathcal{O}(\delta^{3/2}).
\end{align}
Imposing the boundary conditions order by order fixes
\begin{align}
  \omega_0
  &=\pm \frac{\mathfrak{L}\sqrt{\kappa_h}}{r_h},
  \qquad
  \omega_1
  =-\frac{i(\mathfrak{L}^2-2)}{2r_h^2}.
\end{align}
These coefficients are determined entirely by $\phi_0$ and $\phi_1$, so no higher-order corrections to the master field are required at this order. The procedure can be continued systematically to higher orders in $\delta$. Using \eqref{eq:R-of-K-leading} to trade $\delta$ for $|K r_h|$, we obtain the stretched-horizon sound modes of Schwarzschild-AdS$_4$,
\begin{align}
  \omega_{\rm sound}
  &= \pm\frac{\mathfrak{L}\kappa_h}{\sqrt{2}\,|K r_h|}
  -\frac{i(\mathfrak{L}^2-2)\kappa_h}{4|K r_h|^2}
  +\mathcal{O}(|K r_h|^{-3}).
  \label{eq:sound modes full BH}
\end{align}
The real part of the frequency vanishes linearly in $|K r_h|^{-1}$, while the attenuation is quadratic. Dividing by $\kappa_h$ and using the identification $q_{\mathcal{R}}=\frac{\mathfrak L}{|K r_h|}$, where we used $\mathfrak z=r_h$ at leading order in the near-horizon scaling, we obtain
\begin{align}
  \widehat\omega_{\rm sound}
  &=
  \pm\frac{q_{\mathcal{R}}}{\sqrt{2}}
  -\frac{i q_{\mathcal{R}}^2}{4}
  +\frac{i}{2|K r_h|^2}
  +\cdots.
\end{align}
Taking the simultaneous large-$\mathfrak L$, large-$|K r_h|$ limit with $q_{\mathcal{R}}$ fixed, the last term vanishes and we recover precisely the Rindler sound mode \eqref{eq:rindler-smallq}.

It is useful to also note that the frequencies \eqref{eq:sound modes full BH} are defined with respect to the Schwarzschild time $t$ and vanish as the boundary approaches the horizon. This behavior is a consequence of the near-horizon redshift, as can be seen by considering the frequency measured in the local frame of the boundary. A static observer at $r=\wallr$ measures
\begin{align}
    \omega_{\rm loc}
    =\frac{\omega}{\sqrt{f(\wallr)}},
    \qquad
    k_{\rm loc}
    =\frac{\mathfrak L}{\wallr}.
\end{align}
Near the horizon, $\sqrt{f(\wallr)}\approx \frac{\kappa_h}{|K|}$, and therefore \eqref{eq:sound modes full BH} gives
\begin{align}
    \frac{\rm{Re}\;\omega_{\rm loc}}{k_{\rm loc}}
    \longrightarrow \frac{1}{\sqrt{2}},
\end{align}
while the local attenuation vanishes as $|K r_h|^{-1}$. Thus, in the local boundary frame the mode approaches a sound wave with sound speed $c_s=1/\sqrt{2}$.

The numerical identification of the near-horizon sound modes is shown in Figure \ref{fig:nearhorizon-sound}. The upper row displays the appropriately rescaled frequencies at fixed $l$, and hence fixed $\mathfrak L$, retaining the finite-angular-momentum corrections. As $|K r_h|\to\infty$, they approach the corresponding coefficients predicted by \eqref{eq:sound modes full BH}. The lower row compares the spherical sound roots with the corresponding Rindler roots as $\mathfrak L$ and $|K r_h|$ are taken large together at fixed $q_{\mathcal{R}}$. In this joint limit, the finite-angular-momentum corrections disappear and the universal planar spectrum is recovered.

\begin{figure}[t!]
  \centering
  \paperfigure[width=\textwidth]
  {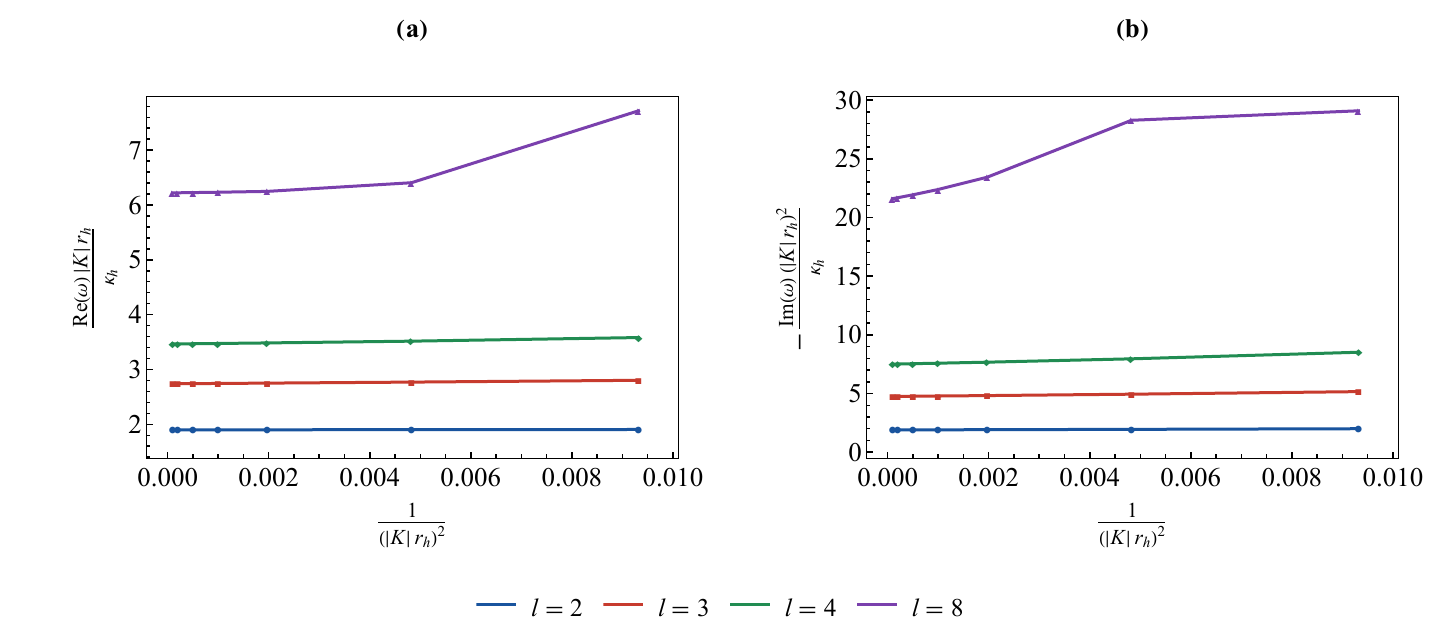}
  \par\medskip
  \paperfigure[width=\textwidth]
  {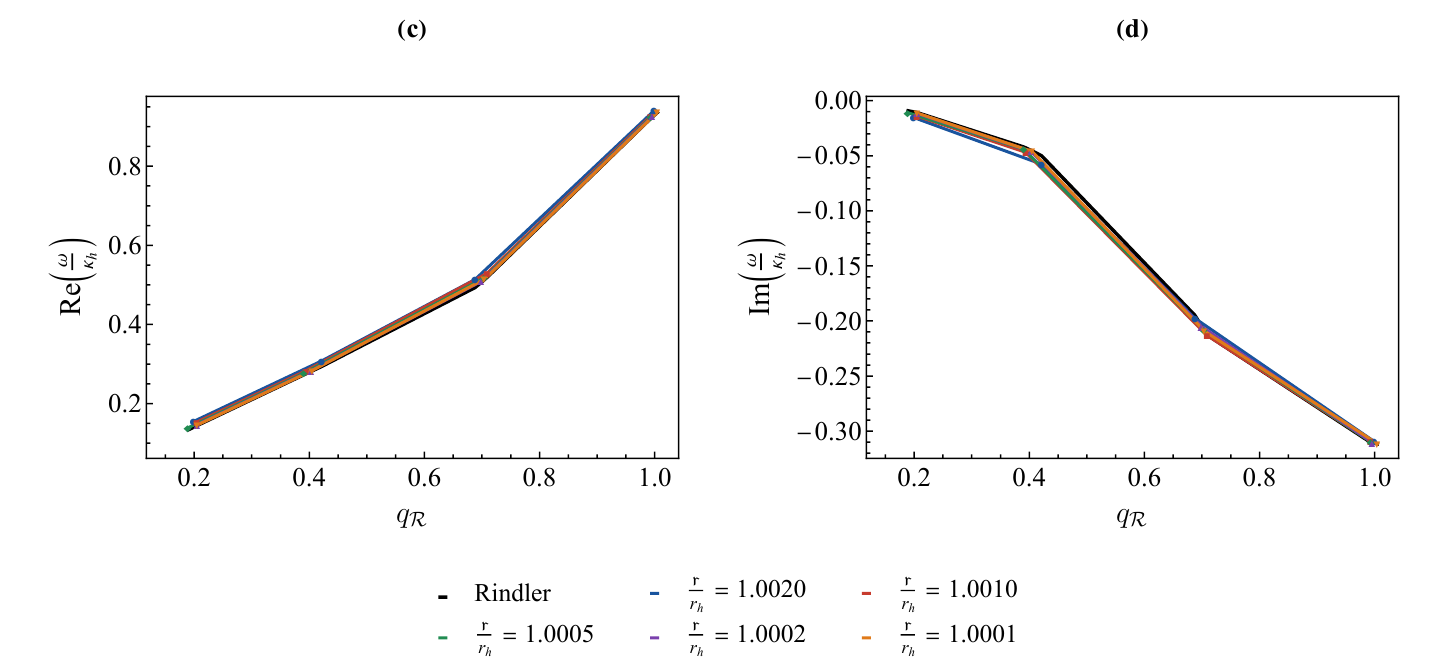}
  \caption{
Numerical checks of the stretched-horizon sound mode for $r_h=\ads=1$. $(a,b)$ At fixed $l$, the rescaled real and imaginary parts of the frequency approach the corresponding coefficients in \eqref{eq:sound modes full BH} as $|K r_h|^{-2}\to0$.
$(c,d)$ The spherical sound frequencies at successively smaller boundary--horizon separations, compared with the corresponding Rindler roots \eqref{eq:rindler-smallq} shown in black. The frequencies are plotted as functions of $q_{\mathcal{R}}=\mathfrak L/|K r_h|$, and lines connect the calculated points. 
}
  \label{fig:nearhorizon-sound}
  \label{fig:fixedell-horizon}
  \label{fig:spherical-rindler}
\end{figure}

For the soft modes, the same procedure gives
\begin{align}
  \omega_+
  &=i\kappa_h
  -\frac{i\kappa_h\left(\mathfrak L^2+2r_h\kappa_h\right)}
  {2|K r_h|^2}
  +\mathcal{O}(|K r_h|^{-4}),
  \\
  \omega_-^{(\pm)}
  &=-i\kappa_h
  +\frac{i\kappa_h}{8|K r_h|^2}
  \bigg[
  3\mathfrak L^2+2r_h\kappa_h-2
  \nonumber\\
  &\pm
  \sqrt{
  17\mathfrak L^4
  +4\left(3r_h\kappa_h-7\right)\mathfrak L^2
  +\left(6r_h\kappa_h+2\right)^2
  }
  \bigg]
  +\mathcal{O}(|K r_h|^{-4}).
\end{align}
The tower of QNM-like modes can also be recovered order by order in the stretched-horizon expansion. At leading order, their frequencies are given by\footnote{Although we have not carried the expansion to sufficiently high $n$ to determine the fate of the tower, it may truncate at finite $\mathfrak{L}$, similarly to the corresponding tower in de Sitter. This contrasts with the Rindler problem, where the tower is infinite \cite{AnninosGalanteGeorgescuManeeratSvesko2026}.}
\begin{align}
  \omega_n^{\mathrm{QNM}}
  &=-i n\kappa_h
  +\mathcal{O}(|K r_h|^{-2n}),
  \qquad n\geq 2.
  \label{eq:qnm Sch}
\end{align}

\medskip

\noindent \textbf{Vector sector.} 
Similarly, solving the vector sector master field equation and conformal boundary conditions order by order in the stretched-horizon expansion gives the shear mode
\begin{align}
\omega_{\rm shear}
&=
-\frac{i\left(\mathfrak L^2-2\right)\kappa_h}
{2|K r_h|^2}
-\frac{
3i\kappa_h\left(\mathfrak L^2-2\right)
\left(\ads^2\mathfrak L^2+12r_h^2\right)
}{
16\ads^2|K r_h|^4
}
+\mathcal{O}(|K r_h|^{-6}).
\label{eq:vector-fixedell-shear}
\end{align}
Thus, at finite angular momentum, the curvature correction already enters the leading damping rate. It becomes negligible relative to the $\mathfrak L^2$ contribution at large-$\mathfrak L$, where \eqref{eq:vector-fixedell-shear} matches the small-$q_{\mathcal{R}}$ Rindler dispersion relation for the shear mode \eqref{eq:vector-rindler-shear}, using the same identifications as for the sound mode. For every $l\geq2$, the leading frequency in
\eqref{eq:vector-fixedell-shear} lies on the negative imaginary
axis and vanishes quadratically as $|K r_h|\to\infty$.

Finally, we also recover the vector sector tower of QNM-like modes. Its leading order frequencies coincide with those of the scalar tower \eqref{eq:qnm Sch}, except that in the vector sector the tower starts at $n=1$.

\subsection{The stretched-horizon spectrum for AdS$_4$ black brane}
The stretched-horizon analysis proceeds similarly to the Schwarzschild case and gives the following spectra for the AdS$_4$ black brane.

\medskip

\noindent \textbf{Scalar sector.} The stretched-horizon scalar frequencies of the AdS$_4$ black brane are
\begin{align}
  \omega_{\rm sound}
  &=\pm \frac{k_S\kappa_h}{\sqrt{2}\,|K r_h|}
  -\frac{i k_S^2\kappa_h}{4|K r_h|^2}
  +\mathcal{O}(|K r_h|^{-3}),\label{sound modes brane}\\
  \omega_+
  &=i\kappa_h
  -\frac{i\kappa_h\left(k_S^2+2r_h\kappa_h\right)}
  {2|K r_h|^2}
  +\mathcal{O}(|K r_h|^{-4}),\\
  \omega_-^{(\pm)}
  &=-i\kappa_h
  +\frac{i\kappa_h}{8|K r_h|^2}
  \bigg[
  3k_S^2+2r_h\kappa_h
  \pm
  \sqrt{
  17k_S^4
  +12r_h\kappa_h k_S^2
  +\left(6r_h\kappa_h\right)^2
  }
  \bigg]
  +\mathcal{O}(|K r_h|^{-4}),\\
  \omega_n^{\mathrm{QNM}}
  &=-i n\kappa_h
  +\mathcal{O}(|K r_h|^{-2n}),
  \qquad n\geq 2.
  \label{eq:brane largeK mode scalar}
\end{align}
Here, $\kappa_h$ is given in \eqref{eq:surface gr brane}, while $k_S$ denotes the momentum along the brane in the scalar sector.

\medskip
\noindent \textbf{Vector sector.} The stretched-horizon vector frequencies of the AdS$_4$ black brane are
\begin{align}
  \omega_{\rm shear}
  &=
  -\frac{i k_V^2\kappa_h}{2|K r_h|^2}
  -\frac{
  3i k_V^2\kappa_h
  \left(\ads^2 k_V^2+12r_h^2\right)
  }{
  16\ads^2|K r_h|^4
  }
  +\mathcal{O}(|K r_h|^{-6}), \label{eq:shear mode near horizon brane}
  \\
  \omega_n^{\rm QNM}
  &=
  -i n\kappa_h
  +\mathcal{O}(|K r_h|^{-2n}),
  \qquad n\geq 1.
  \label{eq:brane largeK mode vector}
\end{align}
Here, $k_V$ denotes the momentum along the brane in the vector sector.

\section{Global continuation of hydrodynamic modes}
\label{sec:global-sound}
A particular sector of the universal stretched-horizon spectrum consists of hydrodynamic sound and shear modes. In \cite{AnninosGalanteGeorgescuManeeratSvesko2026}, a nonlinear completion of this fluid sector was obtained in the stretched-horizon limit, and the resulting dynamics were identified with those of a relativistic conformal fluid with shear viscosity over entropy density $\eta/s=1/(4\pi)$. In Section \ref{sec:str-horizon}, we identified the corresponding near-horizon sound and shear modes for both Schwarzschild-AdS$_4$ and the AdS$_4$ black brane.

In an asymptotically AdS setting, a second fluid description is well-known. When the finite boundary is removed and the conformal boundary is taken to the asymptotic AdS boundary, the long-wavelength dynamics are described by the fluid/gravity correspondence, which arises as the hydrodynamic limit of AdS/CFT. For the AdS$_4$ black brane, the hydrodynamic regime is obtained directly in the long-wavelength limit. For the AdS$_4$ black hole, one must in addition take the large-black-hole limit for the hydrodynamic description to apply.

The goal of this section is to show that these two fluid descriptions are continuously connected as the conformal boundary is moved from the near-horizon region out to the asymptotic AdS boundary.

We begin by reviewing the standard fluid/gravity sound and shear modes of the dual CFT$_3$. We then present the numerical continuation of the near-horizon hydrodynamic modes towards the asymptotic AdS boundary. We first consider the black brane, for which the numerical analysis is simpler owing to the absence of a photon sphere and because no additional large-black-hole limit is required. We then turn to the Schwarzschild-AdS$_4$ black hole.

\subsection{Review of the CFT$_3$ sound and shear modes}
\label{sec:review fluid gravity}

\medskip

\noindent \textbf{Planar case.}
The asymptotic boundary of the AdS$_4$ black brane is $\mathbb R^{2,1}$, with
\begin{align}
    g^{(0)}_{mn}dx^m dx^n=-dt^2+dx^2+dy^2.
\end{align}
For a neutral conformal fluid in $2+1$ dimensions, the stress tensor takes the form, to first order in the derivative expansion,
\begin{align}
    \mathcal{T}_{mn}
    &=p\left(g^{(0)}_{mn}+3u_m u_n\right)
    -\eta\,\sigma_{mn},
\end{align}
where $p$ is the pressure, $u^m$ is the fluid velocity, normalized as $u^m u_m=-1$, and $\eta$ is the shear viscosity. The shear tensor is given by 
\begin{align}
    \sigma_{mn}
    &=
    P_m{}^p P_n{}^q
    \left(
    \nabla_p u_q+\nabla_q u_p-P_{pq}\nabla_k u^k
    \right),
    \qquad
    P_{mn}=g^{(0)}_{mn}+u_m u_n,
\end{align}
where $\nabla$ is the covariant derivative with respect to $g^{(0)}$. The bulk viscosity vanishes as a consequence of conformal invariance. The fluid equations are given by conservation of the stress tensor,
\begin{align}
    \nabla_m \mathcal{T}^{mn}=0.
\end{align}
We linearize these equations around a homogeneous static solution,
\begin{align}
  p&=p_0+\delta p\,e^{-i\omega t}\mathbb{S},\qquad
  u^m=(1,0,0)+\delta u\,e^{-i\omega t}\left(0,\nabla^i\mathbb S\right),
\end{align}
where $\mathbb{S}$ is an eigenfunction of the Laplacian on $\mathbb{R}^2$, $\nabla^2\mathbb{S}=-k^2\mathbb{S}$. The velocity perturbation is chosen to lie in the scalar sector. The linearized conservation equations then reduce to
\begin{align}
    2i\omega\,\delta p+3p_0k^2\,\delta u&=0,\nonumber\\
    \delta p+\left(\eta k^2-3i\omega p_0\right)\delta u&=0.
    \label{eq:fluidline}
\end{align}
For the neutral conformal fluid, $3p_0=sT$, where $s$ denotes the entropy density and $T$ the temperature. Using the holographic result $\eta/s=1/(4\pi)$ \cite{KovtunSonStarinets2005}, we find that \eqref{eq:fluidline} reduces to
\begin{align}
  \omega^2+i\frac{k^2\omega}{4\pi T}-\frac{k^2}{2}=0.
  \label{eq:sound derivation cft3}
\end{align}
In the hydrodynamic regime $k/T\ll1$, we obtain the sound modes
\begin{align}
  \omega_{\rm sound}
  &=\pm\frac{k}{\sqrt{2}}-i\frac{k^2}{8\pi T}
  +\mathcal{O}(k^3/T^2).
  \label{eq:sound mode brane asy}
\end{align}
The leading real frequency measures sound propagation on the two-dimensional flat space, while the imaginary part describes viscous attenuation. This dispersion relation was obtained holographically from the poles of the stress-tensor correlator of the M2-brane theory in \cite{Herzog2003Sound}, and directly from the gravitational QNM problem in \cite{MirandaMorganZanchin2008}.

If instead we consider a velocity perturbation in the vector sector,
\begin{align}
    u^m=(1,0,0)+\delta u\,e^{-i\omega t}(0,\mathbb V^i),
\end{align}
where $\mathbb V^i$ is a transverse vector eigenfunction of the Laplacian on $\mathbb{R}^2$, $\nabla^2\mathbb V^i=-k^2\mathbb V^i$, the linearized energy conservation equation, $\nabla_m T^{mt}=0$, is identically satisfied, while momentum conservation, $\nabla_m T^{mi}=0$, reduces at this order to
\begin{align}
    \left(\eta k^2-3i\omega p_0\right)\delta u=0.
    \label{eq:derivation shear cft3}
\end{align}
Using the same relations as above, we obtain the leading-order shear mode
\begin{align}
    \omega_{\rm shear}=-i\frac{k^2}{4\pi T}.
    \label{eq:vector shear derivation}
\end{align}
Note that the physical momentum $k$ is related to the dimensionless harmonic momentum parameters used in the gravitational problem by
\begin{align}
k=\frac{k_{S,V}}{\ell_{\mathrm{AdS}}},
\label{eq:definiton k}
\end{align}
where $k_S$ and $k_V$ refer to the scalar and vector sectors, respectively.

\medskip

\noindent \textbf{Spherical case.}
The asymptotic boundary of the Schwarzschild-AdS$_4$ black hole is $\mathbb R\times S^2$, with the radius of the two-sphere given by $\ads$. The fluid analysis proceeds as in the planar case, with additional curvature corrections. For a neutral conformal fluid in $2+1$ dimensions, the analogue of \eqref{eq:sound derivation cft3} in the scalar sector is
\begin{equation}
  (\omega\ads)^2
  +i\frac{\mathfrak L^2-2}{4\pi T\ads}(\omega\ads)
  -\frac{\mathfrak L^2}{2}=0.
  \label{eq:spherical-hydro-quadratic}
\end{equation}
In the hydrodynamic regime $\mathfrak L/(T\ads)\ll1$, we obtain the sound modes
\begin{equation}
 \omega_{\rm sound}\ads
  =\pm \frac{\mathfrak L}{\sqrt{2}}
  -i \frac{\mathfrak L^2-2}{8\pi T\ads}
  +\mathcal O\left((T\ads)^{-2}\right).
  \label{eq:sound cft3 again}
\end{equation}
On the other hand, in the vector sector, the analogue of \eqref{eq:derivation shear cft3} gives the leading-order shear mode
\begin{equation}
  \omega_{\mathrm{shear}}\ads
  =-i\,\frac{\mathfrak{L}^2-2}{4\pi T \ads}.
  \label{eq:cft3 shear again}
\end{equation}

\medskip

\noindent \textbf{Gravity analysis.}
The sound and shear modes reviewed above can also be obtained from a linearized gravitational analysis, similar to those considered in the previous sections. For the AdS$_4$ black brane, this is particularly simple, and the analysis is presented in Appendix \ref{app:BB}. For the Schwarzschild-AdS$_4$ black hole, reproducing the results above requires an additional large-black-hole limit. To make this explicit, we parametrize the black-hole size by
\begin{equation}
  h\equiv\frac{r_h}{\ads}.
  \label{eq:h-definition}
\end{equation}
The Hawking temperature can then be written as
\begin{equation}
  4\pi T\ads=\frac{1+3h^2}{h}.
  \label{eq:temperature-h}
\end{equation}
At fixed $\mathfrak{L}$ and for $h\gg1$, the low-lying modes of the gravitational perturbation problem have $\ads\omega=\Order(1)$, so both $\omega/T$ and $\mathfrak{L}/(T\ads)$ are of order $h^{-1}$. Although the angular wavelength remains of order $\ads/\mathfrak L$, and hence fixed relative to the size of the boundary sphere, the thermal length $T^{-1}$ decreases as $h^{-1}$. Thus the mode becomes long-wavelength compared with the thermal scale, which is the regime in which hydrodynamics applies.

In this regime, the gravitational sound and shear modes found in \cite{Siopsis2007,MichalogiorgakisPufu2007} reproduce the fluid results above and are given by
\begin{align}
    \omega_{\rm sound}\ads
    &=\pm\frac{\mathfrak L}{\sqrt{2}}
    -i\frac{\mathfrak L^2-2}{6h}
    +\Order(h^{-2}),
    \label{eq:large-h-KI-sound}\\
    \omega_{\rm shear}\ads
    &=-i\frac{\mathfrak L^2-2}{3h}
    +\Order(h^{-2}).
    \label{eq:global-vector-hydro-frequency}
\end{align}

\subsection{Global continuation of the AdS$_4$ black brane stretched-horizon hydrodynamic modes}
\label{sec:global brane}

We now present the numerical continuation of the near-horizon hydrodynamic modes towards the asymptotic AdS boundary for the AdS$_4$ black brane.

As explained in Section \ref{sec:setup}, $K\ads\rightarrow\infty$ as the conformal boundary approaches the horizon, while $K\ads\rightarrow3$ as the boundary is removed. For each value of $K\ads$, or equivalently each boundary position $\wallr$, we have a well-defined linearized gravitational problem with conformal boundary conditions imposed at $\wallr$. Varying $K\ads$ from infinity to $3$ therefore defines a one-parameter family of static linearized problems interpolating between the stretched-horizon and asymptotic AdS regimes. While the two endpoint regimes were treated analytically in the previous sections, the interpolation between them must be obtained numerically.

We emphasize that the stretched-horizon sound and shear modes are hydrodynamic for any fixed finite $k/T$, for $k$ given in \eqref{eq:definiton k}, but this need not remain true as the boundary is moved outwards. Indeed, the Tolman temperature at the boundary is $T_{\rm loc}=T/\sqrt{f_{\rm br}(\wallr)}$, so that
\begin{equation}
  \frac{k_{\rm loc}}{T_{\rm loc}}
  =
  \frac{k}{T}
  \sqrt{1-\frac{r_h^3}{\wallr^3}},
  \label{eq:brane-local-thermal-ratio}
\end{equation}
where the local momentum is $k_{\rm loc}=k\ads/\wallr$. This ratio vanishes at the horizon and approaches $k/T$ as the boundary is removed. Thus, for generic fixed $k/T$, the modes can leave the hydrodynamic regime along the continuation, so their evolution must be followed using the full gravitational problem.

The radial master field equations and conformal boundary conditions in the scalar and vector sectors are given in Appendix \ref{app:BB}. For each value of $K\ads$, we solve the corresponding master field equation numerically with the ingoing condition at the horizon. Evaluating the conformal boundary conditions on this ingoing solution then gives an equation whose roots determine the allowed complex frequencies.

Throughout this subsection, we use the dimensionless frequency and momentum parameters introduced in Appendix \ref{app:BB},
\begin{equation}
    \varpi=\frac{\omega\ads^2}{r_h},
    \qquad
    q=\frac{k\ads^2}{r_h}
    =\frac{k_{S,V}\ads}{r_h},
    \label{eq:planar-vector-frequency-momentum}
\end{equation}
where $k$ is the physical planar momentum, while $k_S$ and $k_V$ are the corresponding dimensionless momentum parameters in the scalar and vector sectors, respectively. For the numerical examples presented below, we set $r_h=\ads=1$, so that $\varpi=\omega$ and $q=k_{S,V}$ in these units. We will therefore use these notations interchangeably when discussing the numerical results. 

\medskip

\noindent \textbf{Scalar sector.} 
Close to the horizon, the sound roots have the expansion \eqref{sound modes brane}. We start from the positive-frequency sound root and follow it numerically as the boundary is moved outwards, keeping $r_h$, $\ads$, and $k_S$ fixed. Figure \ref{fig:black-brane-sound-flow} shows the full trajectory of this mode for $r_h=\ads=1$ and $k_S=1/2$. Along the continuation, both the real and imaginary parts of the frequency first increase in magnitude, before turning around and approaching finite asymptotic values as the boundary is removed.

\begin{figure}[t]
  \centering
  \paperfigure[width=0.82\textwidth]
    {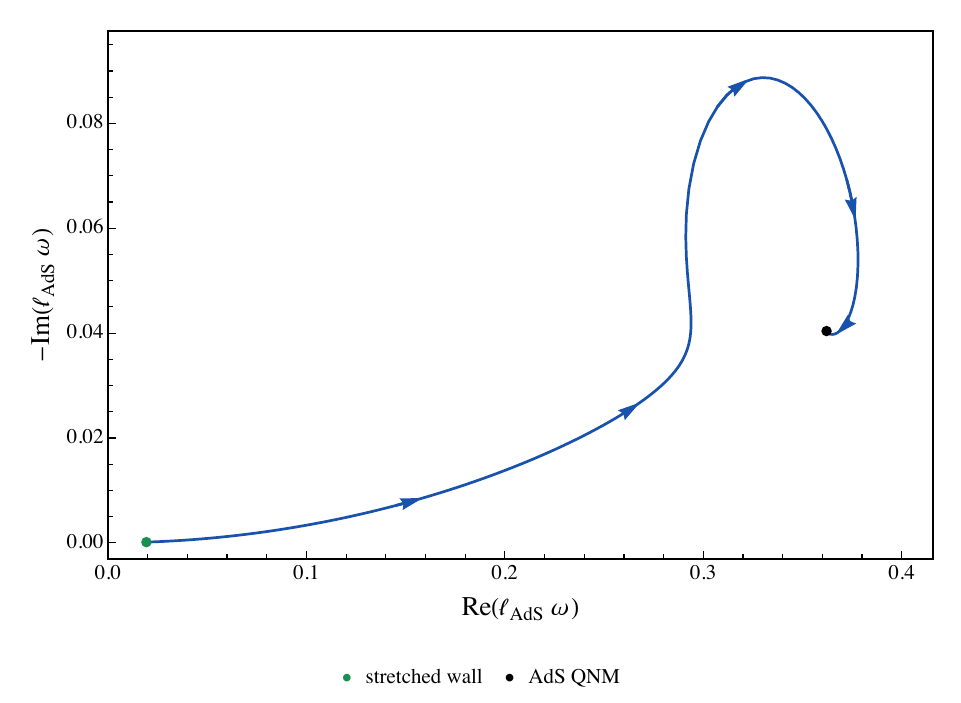}
  \caption{
  Continuation of the positive-frequency scalar sound mode for the planar AdS$_4$ black brane, with $r_h=\ads=1$ and $k_S=1/2$. Arrows indicate the direction in which the conformal boundary is moved towards infinity. The green point marks the stretched-horizon starting point at $\mathfrak{u}\equiv1-r_h/\wallr=10^{-3}$, while the black point denotes the independently computed asymptotic bulk QNM. The branch connects these two regimes without passing through a finite-radius umbilic boundary.
  }
  \label{fig:black-brane-sound-flow}
\end{figure}

According to the asymptotic factorization discussed in Section \ref{sec:exact-umbilic-finite-l}, the endpoint belongs to the bulk branch \eqref{eq:bulk branch asy brane}. For the parameters used in Figure \ref{fig:black-brane-sound-flow}, the numerically continued endpoint agrees with an independent computation of the asymptotic black-brane QNM, 
\begin{equation}
  \varpi_{\rm QNM}
  \simeq 0.36211333-0.04038667\,i.
  \label{eq:brane-sound-QNM-endpoint}
\end{equation}
Thus, the finite-boundary mode reaches the scalar black-brane QNM, whose holographic interpretation is a pole of the retarded stress-tensor correlator
\cite{KovtunStarinets2005,MirandaMorganZanchin2008}.

At small asymptotic momentum, this QNM reduces to the sound pole \eqref{eq:sound mode brane asy}. For comparison, evaluating \eqref{eq:sound mode brane asy} for the parameters used in Figure \ref{fig:black-brane-sound-flow} gives
\begin{align}
   \varpi_{\rm hydro}
   \simeq 0.35355339-0.04166667\,i.
\end{align}
The difference between the two values arises because the momentum used in the numerical continuation is finite, so \eqref{eq:brane-sound-QNM-endpoint} contains corrections beyond the first-order hydrodynamic sound dispersion.

As an additional check, we repeat the numerical continuation at the larger dimensionless momentum $q=3.213$, where the attenuation of the asymptotic black-brane sound QNM reaches its maximum \cite{MirandaMorganZanchin2008,Morgan:2009pn}. The continued branch approaches
\begin{align}
\varpi_{\mathrm{QNM}}\simeq 3.142521-0.414508 i,
\end{align}
in agreement with the damping rate listed in Table 6 of \cite{Morgan:2009pn}.\footnote{For $d=4$ the variables $\mathfrak w,\mathfrak q$ in \cite{Morgan:2009pn} coincide with our $\varpi,q$.} We therefore note that the stretched-horizon sound branch connects to the standard black-brane quasinormal-mode spectrum beyond the hydrodynamic regime.

\medskip 

\begin{figure}[t]
  \centering
  \paperfigure[width=\textwidth]{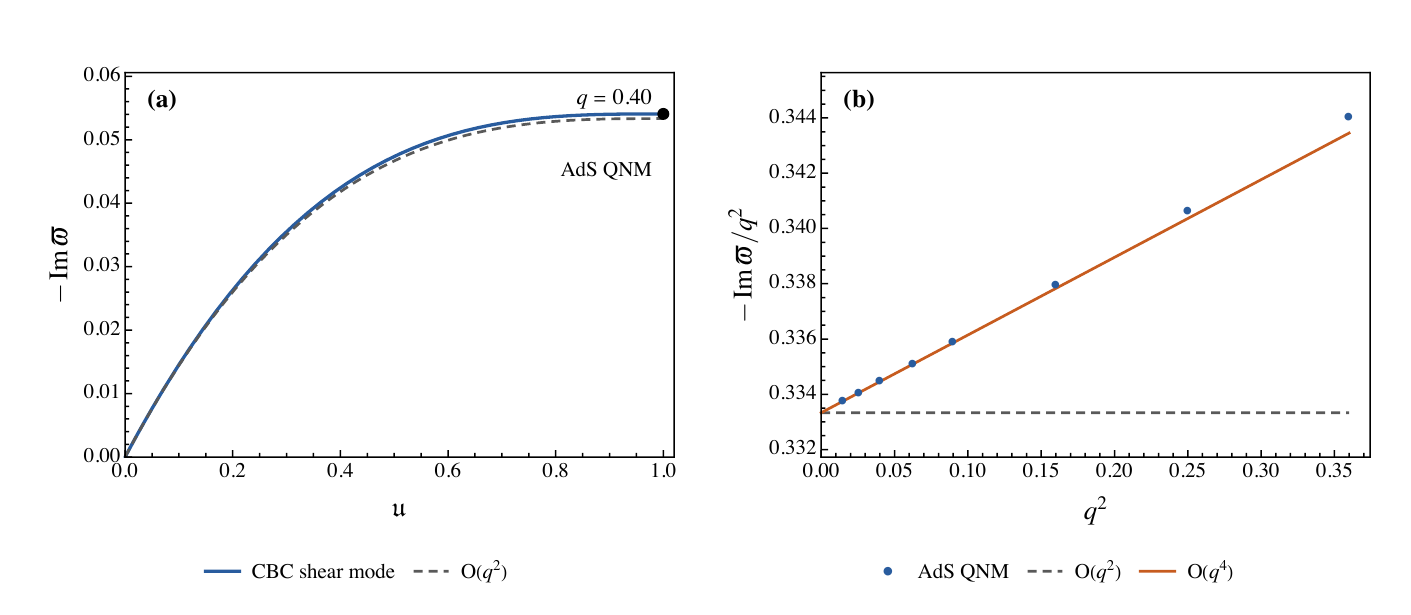}
  \caption{
  Vector shear mode of the AdS$_4$ black brane.
  We use $\varpi=\omega\ads^2/r_h$,
  $q=k\ads^2/r_h=k_V\ads/r_h$, and
  $\mathfrak{u}\equiv 1-r_h/\wallr$.
  (a) Radial continuation at $q=0.4$.
  The blue curve follows the numerical  root from the stretched horizon to the asymptotic AdS boundary.
  The dashed curve is the leading hydrodynamic prediction,
  $-\operatorname{Im}\varpi=q^2(1-(1-\mathfrak{u})^3)/3$,
  and the black point marks the asymptotic QNM.
  (b) Dispersion of the asymptotic shear branch.
  Numerical QNMs are compared with the leading diffusive
  value $-\operatorname{Im}\varpi/q^2=1/3$ and the analytic
  correction through order $q^4$ in the frequency.}
  \label{fig:black-brane-vector-shear}
  \label{fig:planar-vector-shear}
\end{figure}

\noindent \textbf{Vector sector.}
Close to the horizon, the shear root has the expansion \eqref{eq:shear mode near horizon brane}. We start from this root and follow it numerically as the boundary is moved outwards, as in the scalar sector. Using the dimensionless frequency and momentum \eqref{eq:planar-vector-frequency-momentum}, we present in Panel~(a) of Figure \ref{fig:planar-vector-shear} the full trajectory of the mode's attenuation for $q=0.4$. The branch remains purely damped along the computed trajectory and approaches the independently computed asymptotic black-brane QNM,
\begin{equation}
    \varpi_{\mathrm{QNM}}
    \simeq -0.05407056\,i.
    \label{eq:planar-vector-qnm-example}
\end{equation}
At small asymptotic momentum, this QNM reduces to the shear pole, reproduced in Appendix \ref{app:BB}. For comparison, evaluating \eqref{eq:planar-vector-asymptotic-dispersion} at $q=0.4$ gives
\begin{align}
    \varpi_{\rm hydro}
    \simeq -0.054053\,i.
\end{align} 
The leading finite-boundary expression \eqref{eq: hydro shear mode from gravity}, obtained analytically in the hydrodynamic regime, closely follows the numerical trajectory. The small discrepancy visible towards the AdS boundary reflects higher-order corrections in the momentum, which are neglected in the leading hydrodynamic expression but retained in the full numerical eigenfrequency.

Panel~(b) of Figure \ref{fig:planar-vector-shear} displays the numerically computed quantity $-\operatorname{Im}\varpi/q^2$ as a function of $q^2$. In the low-momentum hydrodynamic regime, this quantity approaches the limiting value $1/3$ as $q\to0$. In terms of the physical frequency and momentum, this limiting value corresponds to the diffusion constant $D=1/(4\pi T)$, or equivalently encodes the Einstein-gravity shear-viscosity ratio $\eta/s=1/(4\pi)$ \eqref{eq:planar-vector-diffusion-viscosity}. This is the same value that characterizes the stretched-horizon fluid \cite{AnninosGalanteGeorgescuManeeratSvesko2026}, so the finite-boundary continuation connects the shear transport coefficients at the two endpoints.

At finite momentum, the departure from this constant is well approximated by the analytic quartic correction in \eqref{eq:planar-vector-asymptotic-dispersion}. This comparison identifies the asymptotic endpoint with the hydrodynamic shear branch, including its first higher-order correction in the momentum.

\medskip

\noindent\textbf{Conclusion.} We conclude that both stretched-horizon sound and shear modes can be continuously tracked to the corresponding gravitational QNMs at the asymptotic AdS boundary for the momenta considered here, whose small-momentum limits reproduce the hydrodynamic sound and shear poles. If the continuation is performed at sufficiently small $k/T$, the asymptotic endpoints lie in the CFT$_3$ hydrodynamic regime. This establishes a continuous connection, as the conformal boundary is moved outwards, between the stretched-horizon fluid modes and the hydrodynamic modes of the asymptotic CFT$_3$.

\subsection{Global continuation of the Schwarzschild AdS$_4$ stretched-horizon hydrodynamic modes}
\label{sec:global black hole}
We now turn to the numerical continuation of the near-horizon hydrodynamic modes towards the asymptotic AdS boundary for the Schwarzschild-AdS$_4$ black hole. As anticipated at the beginning of this section, there are two complications relative to the black brane case. The first is the presence of the photon sphere, which provides a distinguished point along the continuation and makes the numerical tracking of the mode more involved. At this radius, the exact factorization discussed in Section \ref{sec:exact-umbilic-finite-l} identifies whether the continued mode belongs to the boundary or bulk sector. Beyond the photon sphere, a second complication arises when comparing the asymptotic endpoint with the CFT$_3$ hydrodynamic results reviewed above. Unlike in the black brane case, reaching the hydrodynamic regime requires an additional large black hole limit. Thus, after continuing the mode to the asymptotic AdS boundary, one must further take 
\begin{align}
    h=\frac{r_h}{\ads}\gg1,
\end{align}
as explained in Section \ref{sec:review fluid gravity}. The radial continuation and the large black hole limit are distinct: the former moves the conformal boundary at fixed background geometry, while the latter changes the background itself.

In the following, we solve numerically the master field equations with the ingoing condition at the horizon and the conformal boundary conditions presented in Section \ref{sec:scalar-problem} for various positions of the boundary, finding the allowed frequencies from the stretched-horizon regime to the asymptotic regime.

\subsubsection{Sound branch}
\label{sec:global-scalar}
We first select the positive-frequency root of the stretched-horizon sound mode \eqref{eq:sound modes full BH}, which describes a long-wavelength excitation of the near-horizon fluid, and continue it by solving the conformal boundary condition \eqref{eq:regular-residual} as the boundary is moved from the near-horizon region towards the asymptotic AdS boundary. Throughout the continuation, we keep the angular momentum $l$, or equivalently $\mathfrak{L}$, and the black-hole size parameter $h$ defined in \eqref{eq:h-definition} fixed.

\medskip

\noindent  \textbf{From the horizon to the photon sphere.}
\label{sec:global-to-photon}
At a generic position of the boundary, both the bulk perturbation and the boundary Weyl factor are nonzero and coupled. At the photon sphere, however, the exact factorization in \eqref{eq:exact-umbilic-factorization} separates the problem into two branches: one with vanishing Weyl factor and nonzero bulk perturbation, and one with nonzero Weyl factor and vanishing bulk perturbation. In all cases reported in Table \ref{tab:global-photon-crossing}, the continued stretched-horizon sound mode reaches the bulk branch, equivalently satisfying the Robin condition \eqref{eq:exact-umbilic-bulk-Robin}. Its boundary Weyl factor therefore vanishes by \eqref{eq:bulk-zero-weyl}, while the Kodama-Ishibashi master field remains nonzero.

\begin{table}[t]
  \centering
  \begin{tabular}{ccc}
    \toprule
    $h$ & $l$ & $\omega\ads(\wallr=\rph)$ \\
    \midrule
    $1$ & $2$ & $2.165354-0.277339\,i$ \\
    $1$ & $3$ & $3.359898-0.351904\,i$ \\
    $2$ & $2$ & $1.893375-0.264546\,i$ \\
    $3$ & $2$ & $1.809619-0.200847\,i$ \\
    $5$ & $2$ & $1.761079-0.128626\,i$ \\
    \bottomrule
  \end{tabular}
  \caption{
  Frequencies of the continued scalar sound branch at the
  photon sphere for several black-hole sizes and angular
  momenta. Each root belongs to the bulk branch selected by
  \eqref{eq:exact-umbilic-bulk-Robin} and has vanishing boundary
  Weyl factor.
  }
  \label{tab:global-photon-crossing}
\end{table}

The branch identification can also be verified by reversing the continuation. For $h=10$ and $l=2$, starting from the bulk root at the photon sphere and moving the boundary inwards recovers the independently determined sound root at $\wallr=1.01r_h$. This confirms the connection between the near-horizon sound mode and the bulk photon-sphere root without relying solely on their proximity in the complex-frequency plane.

\medskip
\noindent \textbf{From the photon sphere to the asymptotic AdS boundary.}
\label{sec:global-to-ads}
Moving past the photon sphere, the Weyl factor becomes nonzero once more and, at a generic finite position of the boundary, is coupled to the bulk perturbation. Near the asymptotic endpoint, the problem factorizes according to \eqref{eq:ads-CBC-factorization}, separating the spectrum into boundary frequencies and bulk QNMs of the Robin problem \eqref{eq:ads-scalar-Robin}. For $h=1,2,3$ in Table \ref{tab:asymptotic-sound-qnms}, the continued stretched-horizon sound branch approaches a zero of the asymptotic Robin problem, corresponding to a perturbation that leaves the asymptotic boundary metric undeformed \cite{MichalogiorgakisPufu2007,DiasSantos2013}, and agrees with an independently computed asymptotic QNM. Thus, removing the boundary connects the stretched-horizon excitation to a gravitational resonance of the asymptotic black hole geometry. 

\begin{table}[t!]
  \centering
  \begin{tabular}{cc}
  \toprule
  $h$ & $\omega_{\rm QNM}\ads$ \\
  \midrule
  $1$  & $2.15571516-0.28549650\,\ii$ \\
  $2$  & $1.89250893-0.26519327\,\ii$ \\
  $3$  & $1.80950041-0.20090736\,\ii$ \\
  $5$  & $1.76107138-0.12862866\,\ii$ \\
  $10$ & $1.73942383-0.06607419\,\ii$ \\
  \bottomrule
\end{tabular}
  \caption{
  Scalar gravitational QNM frequencies for $l=2$, obtained
  independently from the ingoing horizon condition and the
  asymptotic Robin condition \eqref{eq:ads-scalar-Robin}.
  For $h=1,2,3$, these frequencies are also reached by the
  finite-boundary continuation of the near-horizon sound branch.
  The larger-$h$ roots are included for comparison with the
  large-black-hole limit considered below.
  }
  \label{tab:asymptotic-sound-qnms}
\end{table}

\begin{figure}[t!]
  \centering
  \paperfigure[width=0.68\textwidth]
  {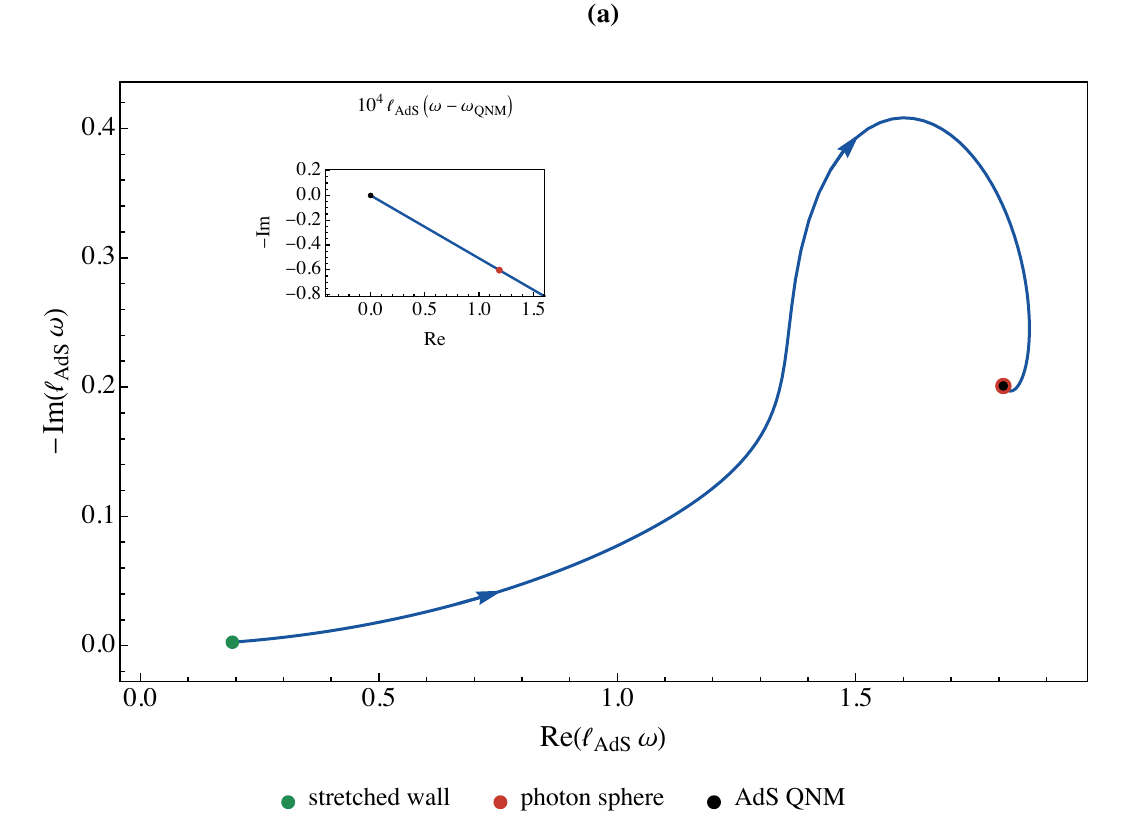}
  \par\medskip
  \paperfigure[width=\textwidth]
  {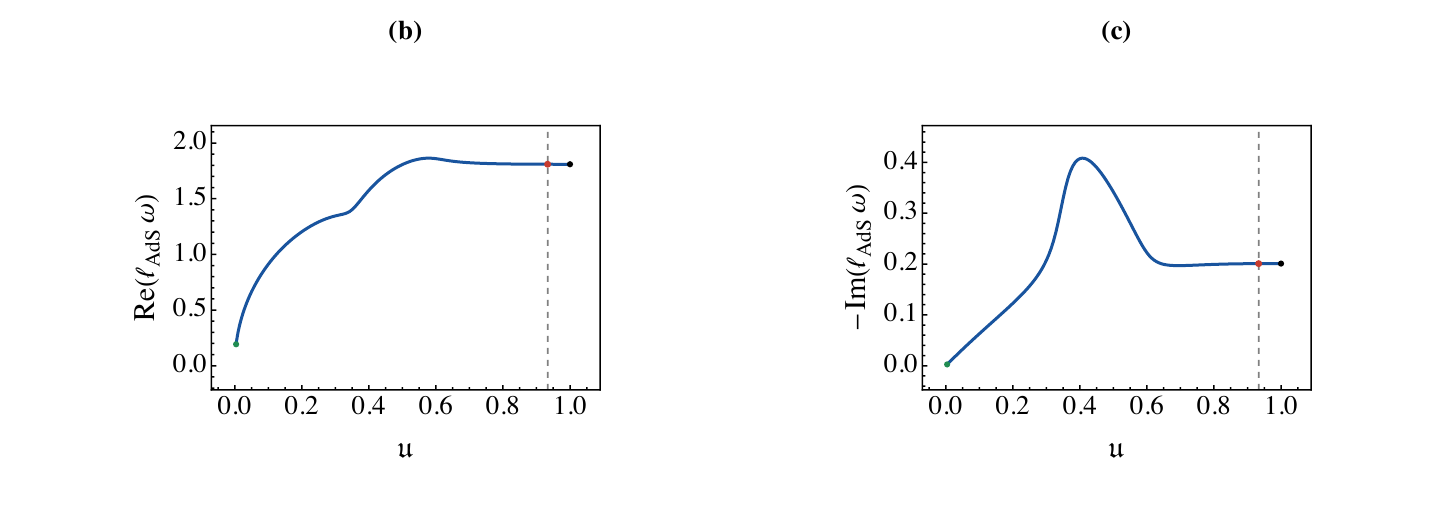}
  \caption{Global continuation of the positive-frequency scalar sound branch for $h=3$ and $l=2$.
(a) Trajectory in the complex-frequency plane as the conformal boundary is moved outwards, with arrows indicating increasing $\wallr$. The highlighted points mark the initial stretched-horizon sound mode, the bulk root at the umbilic photon-sphere boundary, and the independently computed asymptotic Schwarzschild-AdS$_4$ QNM. The inset shows $10^4(\omega-\omega_{\rm QNM})\ads$, resolving the small separation between the photon-sphere frequency and the asymptotic endpoint.
(b,c) Real and imaginary parts of the same trajectory as functions of $\mathfrak u=1-r_h/\wallr$. The vertical dashed line marks the photon sphere, while the black point at $\mathfrak u=1$ denotes the asymptotic QNM. Lines connect the numerically computed finite-boundary frequencies. 
}
  \label{fig:global-cbc-branch}
  \label{fig:wall-flow}
\end{figure}

Figure \ref{fig:global-cbc-branch} shows the full trajectory of the mode in the complex-frequency plane for $h=3$ and $l=2$, from the near-horizon to the asymptotic region. To parameterize the position of the boundary, we use
\begin{equation}
  \mathfrak{u}\equiv 1-\frac{r_h}{\wallr},
  \label{eq:wall-compact-coordinate}
\end{equation}
for which the horizon and asymptotic AdS boundary lie at $\mathfrak{u}=0$ and $\mathfrak{u}=1$, respectively, while the photon sphere is located at
\begin{equation}
  \mathfrak{u}_{\rm ph}
  =1-\frac{r_h}{\rph}
  =1-\frac{2}{3(1+h^2)}.
  \label{eq:uph-global}
\end{equation}
The complex-frequency trajectory shows the evolution of the mode through the spectrum, while plotting its real and imaginary parts as functions of $\mathfrak{u}$ makes the passage through the photon sphere explicit.

Finally, recovering the hydrodynamic dispersion relation of the boundary theory requires a further limit. Having identified the asymptotic QNM branch through the finite-boundary continuation, we now study its behavior as the black-hole size increases. For this purpose, we compute the asymptotic gravitational QNMs independently at larger values of $h$ and examine their approach to the hydrodynamic regime.

\medskip

\begin{figure}[t]
  \centering
  \paperfigure[width=\textwidth]{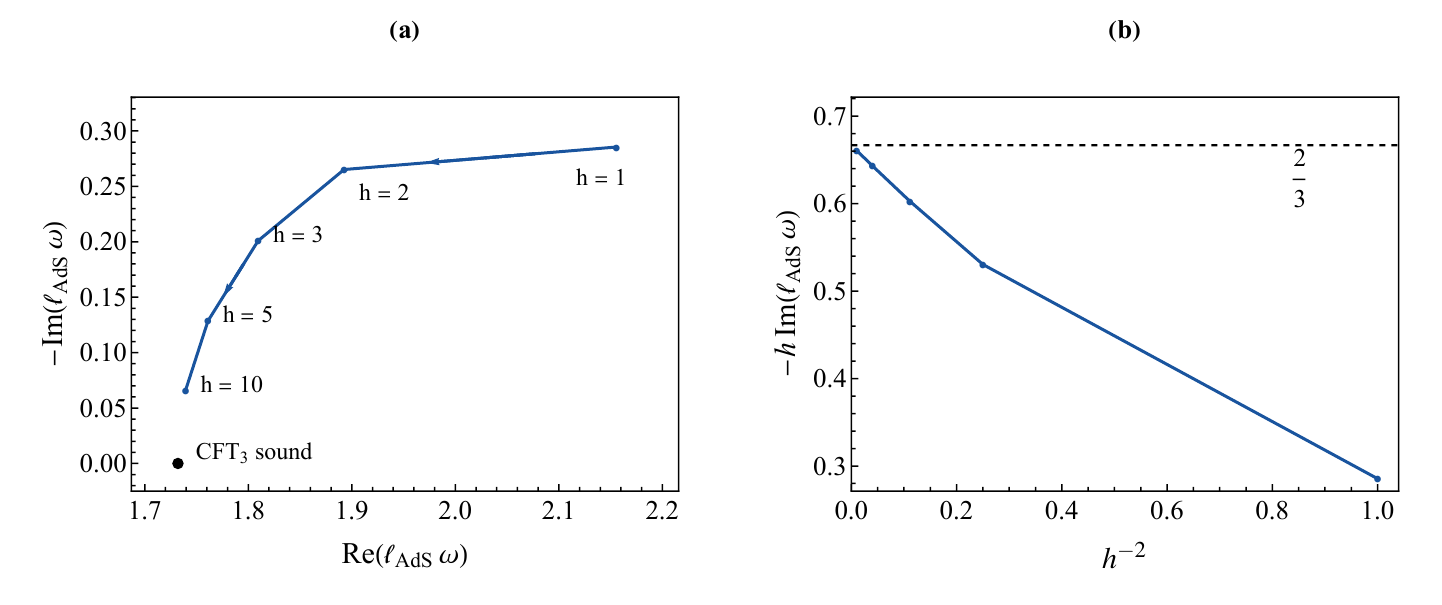}
  \caption{
Approach of the $l=2$ scalar gravitational QNM to the
CFT$_3$ sound mode as the black hole becomes large.
(a) Frequencies for $h=1,2,3,5,10$ in the complex plane,
with arrows indicating increasing $h$.
The black point marks the limiting sound frequency
$(\operatorname{Re}(\omega\ads),-\operatorname{Im}(\omega\ads))
=(\sqrt3,0)$.
(b) The rescaled attenuation $-h\operatorname{Im}(\omega\ads)$
as a function of $h^{-2}$.
The dashed line marks the hydrodynamic limit $2/3$. Lines connect the calculated QNMs.
}
  \label{fig:sound-limit}
\end{figure}

\noindent \textbf{Large black holes and the CFT$_3$ sound mode.}
\label{sec:large-h-sound}
Figure \ref{fig:sound-limit} shows, for the $l=2$ mode, how the asymptotic QNM identified above approaches the hydrodynamic sound dispersion relation \eqref{eq:large-h-KI-sound} in the complex-frequency plane as the black-hole size $h$ is increased. Combined with the radial continuation described above, this establishes a continuous connection between the stretched-horizon sound mode and the sound pole of the asymptotic CFT$_3$ \cite{MichalogiorgakisPufu2007,Siopsis2007}.

\medskip

To sum up, the stretched-horizon sound mode passes through the bulk branch at the photon sphere and approaches a standard scalar gravitational QNM when the finite boundary is removed. Taking the additional large-black-hole limit of this asymptotic endpoint then recovers the sound pole of the dual CFT$_3$.

\subsubsection{Shear branch}
\label{sec:global-shear}
We now select the stretched-horizon shear mode \eqref{eq:vector-fixedell-shear}, and continue it by solving the conformal boundary condition \eqref{eq:vector-residual} as the boundary is moved from the near-horizon region towards the asymptotic AdS boundary. Just like in the scalar sector, we keep the angular momentum $l$ and the black hole size parameter $h$ fixed.

\begin{figure}[t]
  \centering
  \paperfigure[width=\textwidth]{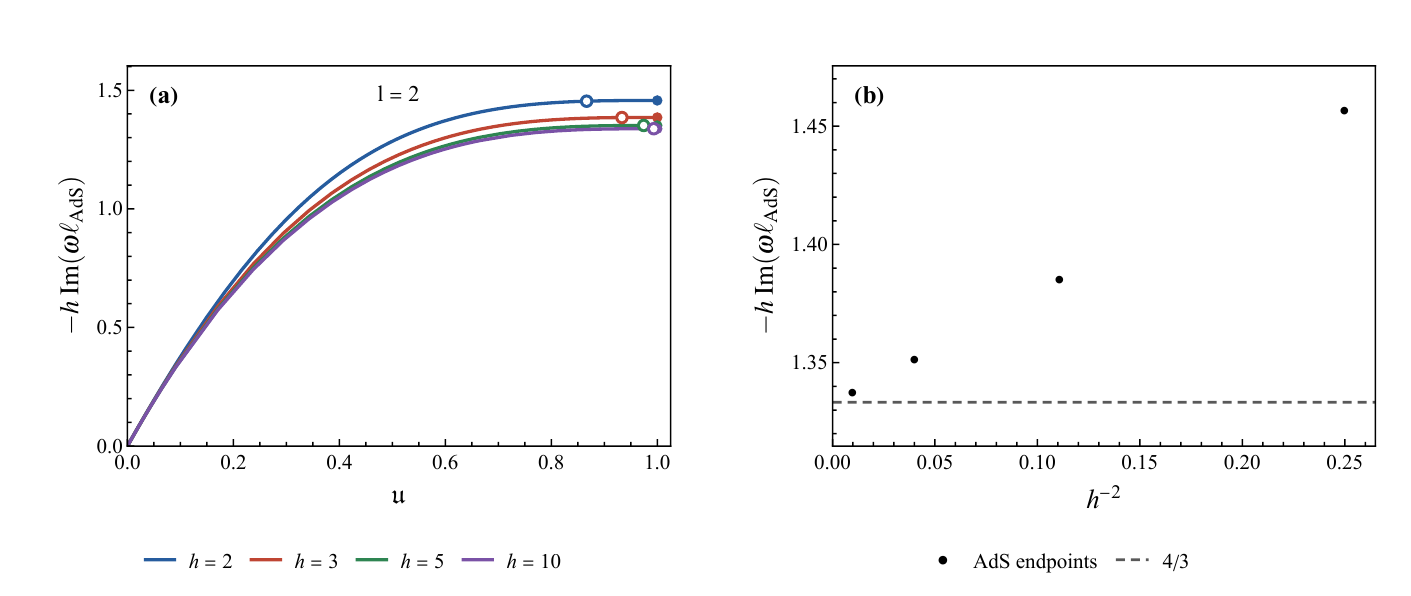}
  \caption{Global continuation and hydrodynamic limit of the spherical vector shear branch for $l=2$.
(a) Rescaled damping rate $-h\,\operatorname{Im}(\ads\omega)$ as a function of $\mathfrak{u}=1-r_h/\wallr$, for $h=2,3,5,10$. Open circles mark the photon-sphere boundary along each trajectory, through which the branches remain smooth. Filled circles at $\mathfrak{u}=1$ denote the asymptotic frequencies obtained by extrapolating the finite-boundary data.
  (b) The same asymptotic endpoints plotted against $h^{-2}$.
  The dashed line marks the large-black-hole shear prediction
  $-h\,\operatorname{Im}(\omega\ads)\to
  (l(l+1)-2)/3=4/3$.
  }
  \label{fig:vector-shear}
  \label{fig:global-vector-shear}
\end{figure}

Figure~\ref{fig:global-vector-shear}(a) shows the numerical continuation for $l=2$ and $h=2,3,5,10$, using the boundary-position variable \eqref{eq:wall-compact-coordinate}. In all these cases, the continued root remains on the negative imaginary axis. Its damping increases as the boundary is moved outwards from the horizon and approaches a finite asymptotic value when the finite boundary is removed. Each trajectory passes smoothly through the photon sphere. Although the optical potential still reaches its maximum there, the vector sector does not exhibit the scalar-sector factorization.

In all cases studied, the continued stretched-horizon shear mode approaches a root of the standard vector gravitational QNM problem with an undeformed asymptotic boundary metric \cite{DiasSantos2013}. This asymptotic frequency can be computed independently by solving the vector perturbation problem with the boundary condition \eqref{eq:ads-vector-Dirichlet}.

As in the scalar sector, recovering the hydrodynamic dispersion relation of the boundary theory requires taking the large-black-hole limit of the asymptotic endpoints. Figure~\ref{fig:global-vector-shear}(b) shows the extrapolated endpoints as a function of $h^{-2}$. For $l=2$, the points approach the value $ -h\,\operatorname{Im}(\ads\omega)\to\frac{4}{3}$, predicted by \eqref{eq:global-vector-hydro-frequency} as $h$ increases, consistently with the identification of the endpoint with the CFT$_3$ shear pole.

\medskip 

To sum up, as in the scalar sector, the stretched-horizon shear mode approaches a standard vector gravitational QNM when the finite boundary is removed. Taking the additional large-black-hole limit of this endpoint then recovers the shear pole of the dual CFT$_3$. We therefore conclude that the stretched-horizon fluid modes are continuously connected to the hydrodynamic poles of the asymptotic boundary theory as the conformal boundary is moved outwards.

\section{Conclusions}
\label{sec:conclusions}
\label{sec:discussion}

Conformal boundary conditions allow the scale of a finite gravitational boundary to become dynamical. The resulting excitations are sensitive both to the local geometry of the boundary and to the propagation of gravitational perturbations in the enclosed region. Schwarzschild-AdS$_4$ black holes and AdS$_4$ black branes provide  settings in which to explore this interplay. Both possess a horizon and an asymptotic AdS region, while Schwarzschild-AdS$_4$ additionally features a photon-sphere barrier, which plays a distinguished role in the dynamics. This special role becomes manifest when the conformal boundary is placed precisely at the photon sphere. At this radius, the boundary is exactly umbilic and the background conformal stress tensor vanishes. We have shown that the scalar conformal boundary conditions then factorize into two independent branches: a boundary branch with vanishing gravitational master field and a bulk branch with vanishing boundary Weyl factor. The former describes pure boundary displacements with exactly real frequencies for $l\geq 2$, while the latter describes bulk gravitational perturbations decoupled from the boundary Weyl mode.

The factorization of the conformal boundary conditions at the photon sphere has important consequences for the large-angular-momentum spectrum. While generic finite-radius conformal boundaries admit unstable modes whose growth rates scale as $\mathfrak L^{1/3}$, these instabilities are suppressed as the boundary approaches the photon sphere and are absent at the exactly umbilic surface. In this regime, the master-field equation reduces to an inverted harmonic-oscillator problem, with no growing modes in the expansion considered.

Turning to the near-horizon regime of Schwarzschild-AdS$_4$ black holes and AdS$_4$ black branes, we recovered the universal stretched-horizon spectra of \cite{AnninosGalanteGeorgescuManeeratSvesko2026}, including the sound and shear modes of an effective relativistic conformal fluid.

Finally, as the conformal boundary approaches the asymptotic AdS boundary, it becomes asymptotically umbilic. In this limit, the scalar conformal boundary conditions again factorize into a boundary branch and a bulk branch, the latter reproducing the standard gravitational quasinormal-mode problem with an undeformed asymptotic boundary metric. The vector sector likewise reduces to the usual quasinormal-mode problem. In the appropriate hydrodynamic regime, the low-lying quasinormal modes reproduce the sound and shear poles of the dual CFT$_3$.

The emergence of fluid modes both near the stretched horizon and at the asymptotic AdS boundary naturally raises the question of whether these two fluid descriptions are continuously connected. To address this, we numerically tracked the sound and shear modes as the conformal boundary is moved from the near-horizon region towards the asymptotic AdS boundary. For both Schwarzschild-AdS$_4$ black holes and AdS$_4$ black branes, we found that these modes continuously approach standard gravitational quasinormal modes, which reproduce the sound and shear poles of the dual CFT$_3$ in the appropriate hydrodynamic regime. In the Schwarzschild-AdS$_4$ case, the continued sound mode passes through the bulk branch at the photon sphere, while the shear mode crosses this radius without an analogous factorization.

\section*{Acknowledgments}

The authors thank Damian Galante for suggesting this problem and Dionysios Anninos for helpful discussions. R.A. and V.D.P. are supported by CONICET and UNLP, Argentina. S.G. is funded by a Royal Society Newton International Fellowship NIF/R1/241888.

\appendix

\section{Kodama-Ishibashi reconstruction}
\label{app:reconstruction}
In this appendix, we collect the details of the Kodama-Ishibashi reconstruction of the perturbed metric around the Schwarzschild AdS$_4$ black hole in the Regge-Wheeler gauge in terms of the bulk master fields $\phi^{(S,V)}$. All expressions below are valid for $l\geq2$.

\subsection{Scalar sector}
In the scalar sector, the metric perturbations take the form \eqref{eq:scalar sector KI functions}. In the Regge-Wheeler gauge, $\gamma=F$, $f_{ab}=F_{ab}$, so the Kodama-Ishibashi invariants directly determine the nonvanishing components of the metric perturbation as
\begin{equation}
    h^{(S)}_{\mu\nu}\dd x^\mu\dd x^\nu
    =
    \e^{-i\omega t}\mathbb S
    \left(
        F_{tt}\dd t^2
        +2F_{tr}\dd t\,\dd r
        +F_{rr}\dd r^2
        +2r^2F\dd\Omega_2^2
    \right),
    \label{eq:KI-RW-metric}
\end{equation}
where, as throughout, taking the real part of the Fourier decomposition is understood. It is convenient to express the reconstruction through three radial
functions $X$, $Y$, and $Z$:
\begin{align}
  X(r)
  &=
  \frac{1}{r}
  \left[
    \left(
      \frac{\omega^2r^2}{f}
      -\frac{P_X}{16H^2}
    \right)\phi^{(S)}
    +\frac{Q_X}{4H}r\phi^{(S)'}
  \right],
  \notag\\
  Y(r)
  &=
  \frac{1}{r}
  \left[
    \left(
      -\frac{\omega^2r^2}{f}
      -\frac{P_Y}{16H^2}
    \right)\phi^{(S)}
    +\frac{Q_Y}{4H}r\phi^{(S)'}
  \right],
  \notag\\
  Z(r)
  &=
  \ii\omega
  \left[
    \frac{P_Z}{4H}\phi^{(S)}-fr\phi^{(S)'}
  \right].
  \label{eq:KI-XYZ-reconstruction}
\end{align}
Primes here denote radial derivatives. In terms of the variables
defined in \eqref{eq:KI-definitions}, the coefficient polynomials are
\begin{equation}
  \begin{aligned}
    P_X={}&
    48mxy_{\rm KI}+216x^3+192mx^2
    +72m^2x+48mx+16m^3+32m^2,\\
    Q_X={}&
    4my_{\rm KI}+18x^2+2(5m-6)x-8m,\\
    P_Y={}&
    (288x^2+48mx)y_{\rm KI}+72x^3+48mx^2
    +24m^2x+48mx,\\
    Q_Y={}&
    (24x+4m)y_{\rm KI}+6x^2-2mx-12x,\\
    P_Z={}&
    -12xy_{\rm KI}+6x^2+6mx-4m.
  \end{aligned}
  \label{eq:KI-reconstruction-polynomials}
\end{equation}
The invariants appearing in
\eqref{eq:KI-RW-metric} are then given by
\begin{equation}
  \begin{aligned}
    F&=-\frac{X+Y}{4},
    &\qquad
    F_{tt}&=-\frac{f}{2}(X-Y),\\
    F_{rr}&=\frac{-X+Y}{2f},
    &
    F_{tr}&=\frac{Z}{f}.
  \end{aligned}
  \label{eq:invariant-reconstruction}
\end{equation}
This reconstruction introduces no singular denominator at the photon sphere. It captures the bulk part of the perturbation and must be supplemented by the normal displacement, as explained in Section \ref{sec:scalar-problem}.

\subsection{Vector sector}
\label{app:vector-reconstruction}
In the vector sector, the metric perturbations take the form \eqref{eq:vector perturbations def}. The Regge--Wheeler gauge sets $H^V=0$, so the Kodama--Ishibashi invariants are directly related to the metric functions,
\begin{equation}
    F_a^{V}=f_a^{V},
    \qquad a=t,r.
    \label{eq:app-vector-invariant}
\end{equation}
In terms of the vector master field, they are given by
\begin{equation}
    r F_a^{V}
    =\epsilon_{ab}D^b\left(r\Phi^{(V)}\right),
    \label{eq:app-vector-master-normalization}
\end{equation}
where $D$ is the covariant derivative on the two-dimensional orbit space $(t,r)$. Choosing the convention $\epsilon_{tr}=+1$, we recover precisely the reconstruction \eqref{eq:vector-RW-reconstruction}.

The master-field equation is
\begin{equation}
    D^aD_a\Phi^{(V)}
    -\frac{k_V^2-1+2f-rf'}{r^2}\Phi^{(V)}=0.
    \label{eq:app-vector-covariant-master}
\end{equation}
For Schwarzschild-AdS$_4$,
\begin{equation}
    2f-rf'=2-\frac{3\mu}{r},
\end{equation}
which reproduces the vector potential \eqref{eq:vector-potential} and, after Fourier decomposition, gives the master-field equation \eqref{eq:vector-master}.

\medskip

\noindent\textbf{The mean-curvature condition.} We now show that the mean curvature is automatically preserved by metric perturbations from the vector sector.  
We use the outward unit normal
\begin{align}
    n^\mu\partial_\mu=\sqrt{f}\,\partial_r,
\end{align}
together with $K_{mn}=\frac{1}{2}\mathcal L_n g_{mn}$. In the Regge-Wheeler gauge, the vector perturbation has only mixed time-angular and radial-angular components. The variation of the inverse induced metric therefore gives no contribution to $\delta K^{(V)}$, since the background extrinsic curvature is diagonal and isotropic on the sphere. We then find
\begin{align}
    \delta K^{(V)}
    &=
    -\left.
    \frac{\sqrt f}{r^2}\,
    \widehat D^i h_{ri}^{(V)}
    \right|_{r=\wallr}
    =
    -\left.
    \frac{\partial_t\Phi^{(V)}}{r\sqrt f}\,
    \widehat D^i\mathbb V_i
    \right|_{r=\wallr}
    =0,
    \label{eq:app-vector-mean-curvature}
\end{align}
where $\hat{D}$ is the covariant derivative on the two-sphere. In the second line we used the reconstruction \eqref{eq:vector-RW-reconstruction}. Thus, the mean-curvature condition is automatically satisfied by transversality,
$\widehat D^i\mathbb V_i=0$, and supplies no additional radial boundary condition.

\medskip

\noindent\textbf{The conformal class of the induced metric condition.}
A Weyl perturbation transforms as a scalar on the sphere, so the vector part of the induced metric perturbation must vanish. Before choosing a gauge, this requires $H^{V}=0=f_t^{V}$ at the boundary. In the Regge-Wheeler gauge, the condition becomes
\begin{equation}
    F_t^{V}\big|_{r=\wallr}=0
    \qquad\Longleftrightarrow\qquad
    \partial_r\!\left(r\Phi^{(V)}\right)\big|_{r=\wallr}=0,
    \label{eq:app-vector-metric-CBC}
\end{equation}
which precisely reproduces the boundary condition \eqref{eq:vector-residual}.

\medskip

\noindent\textbf{The conformal stress tensor.}  Setting the mixed metric component to zero at the boundary does not constrain its normal derivative, which is the origin of the nontrivial vector response of the conformal stress tensor. With the convention \eqref{eq:CBY-definition}, the boundary conditions imply
\begin{equation}
    8\pi G_N\,\delta T_{mn}^{(V)}
    =-\delta K_{mn}^{(V)}.
    \label{eq:app-vector-stress-variation}
\end{equation}
Using the master-field equation together with the conformal boundary condition, \eqref{eq:app-vector-stress-variation} gives
\begin{equation}
  \begin{aligned}
  8\pi G_N\,\delta T_{ti}^{(V)}
  &=
  -\frac{\sqrt{f(\wallr)}}{2\wallr}
   (k_V^2-1)\Phi^{(V)}(t,\wallr)\mathbb V_i,
  \\
  8\pi G_N\,\delta T_{ij}^{(V)}
  &=
  \frac{\wallr}{2\sqrt{f(\wallr)}}
  \partial_t\Phi^{(V)}(t,\wallr)
  \left(
    \widehat D_i\mathbb V_j+\widehat D_j\mathbb V_i
  \right),
  \\
  \delta T_{tt}^{(V)}&=0.
  \end{aligned}
  \label{eq:app-vector-boundary-stress}
\end{equation}
These expressions have the same structure as the vector boundary response found in
\cite{AnninosGalanteGeorgescuManeeratSvesko2026}, and explicitly exhibit the momentum density and shear stress carried by the master field.

\section{Nonlinear homogeneous dynamics at the photon sphere}
\label{app:homogeneous-nonlinear}

In the main text, we have focused on the modes with $l\geq2$. In this appendix, we consider the spherically symmetric $l=0$ sector for Schwarzschild AdS$_4$, which allows us to study the dynamics of the boundary Weyl mode beyond linear order. By Birkhoff's theorem, the bulk geometry is fixed locally, and the dynamics reduces to that of a moving timelike boundary whose trajectory is determined by the prescribed conformal class and mean curvature.

The general nonlinear equation governing the $l=0$ Weyl mode in this setting was derived in Appendix C.2 of \cite{AnninosAriasGalanteManeerat2025}. Here, we recast the dynamics in terms of the conserved conformal energy. At the photon sphere, the sector with vanishing conformal energy can be solved in closed form. In this sector, the boundary remains totally umbilic throughout its nonlinear evolution and admits a simple interpretation as a moving photon surface.

\subsection{Moving conformal boundary and conserved conformal energy}
\label{app:homogeneous-energy}
We use the same Schwarzschild-AdS$_4$ geometry as in the main text, but for the purposes of this appendix we denote the static time coordinate by $T$,
\begin{align}
  d\bar{s}^2
  &=
  -f(r)\,dT^2+\frac{dr^2}{f(r)}+r^2 d\Omega_2^2,
  \qquad
  f(r)=1+\frac{r^2}{\ads^2}-\frac{\mu}{r}.
\end{align}
Let $R$ be a fixed reference radius. We choose the boundary data
\begin{equation}
  g_{mn}\dd x^m\dd x^n
  =
  e^{2\bm{\omega}(t)}
  \left[-f_R\dd t^2+R^2\dd\Omega_2^2\right],
  \qquad
  K=K_R,
  \label{eq:hom-CBC}
\end{equation}
where $t$ is the boundary time coordinate and
\begin{equation}
  f_R\equiv f(R),
  \qquad
  K_R\equiv \frac{f'(R)}{2\sqrt{f_R}}
      +\frac{2\sqrt{f_R}}{R}.
  \label{eq:hom-KR}
\end{equation}
These boundary data admit the static solution $\bm{\omega}=0$. Allowing the Weyl factor to evolve changes the physical areal radius of the boundary according to
\begin{equation}
  a(t)=R\e^{\bm{\omega}(t)},
  \label{eq:hom-radius}
\end{equation}
while the reference radius $R$ and the prescribed mean curvature $K=K_R$ are kept fixed.

The distinction between the different time coordinates is useful here. For the static solution, the boundary coordinate $t$ coincides with the bulk static time $T$, but this is no longer true along a general moving trajectory. We first introduce the dimensionless conformal time
\begin{equation}
  s=\Omega_R t,
  \qquad
  \Omega_R=\frac{\sqrt{f_R}}{R}.
  \label{eq:hom-conformal-time}
\end{equation}
Using $a=R e^{\bm{\omega}}$, the boundary metric \eqref{eq:hom-CBC} then takes the form
\begin{equation}
  g_{mn}\dd x^m\dd x^n
  =
  a^2(s)\left(-\dd s^2+\dd\Omega_2^2\right)= -d\tau^2+a^2(s)d\Omega_2^2,
\end{equation}
where in the second equality we introduced the proper time $\tau$ of an observer comoving with the boundary,
\begin{equation}
  \dd\tau=a\,\dd s.
  \label{eq:hom-proper-time}
\end{equation}
In the following, dots denote $\tau$-derivatives, while primes on $\bm{\omega}$ denote $s$-derivatives. Primes on $f$ continue to denote derivatives with respect to the radial coordinate $r$.

Writing the trajectory of the boundary as $(T(\tau),a(\tau))$, the proper-time normalization gives
\begin{equation}
  -f(a)\dot T^2+\frac{\dot a^2}{f(a)}=-1,
\end{equation}
which yields 
\begin{equation}
  \dot T=\frac{\beta}{f(a)},\qquad \beta\equiv\sqrt{f(a)+\dot a^2}. \label{eq:hom-beta}
\end{equation}
It is also easy to show that
\begin{equation}
  \dot a=\bm{\omega}'.
  \label{eq:hom-clocks}
\end{equation}
We choose the future-directed trajectory and outward-pointing normal, and restrict to the exterior region $a>r_h$. The principal curvatures of the timelike boundary at $r=a(\tau)$ are
\begin{equation}
  K^\tau{}_\tau
  =\frac{\ddot a+\frac12 f'(a)}{\beta},
  \qquad
  K^\theta{}_\theta=K^\phi{}_\phi=\frac{\beta}{a}.
  \label{eq:hom-principal-curvatures}
\end{equation}
The angular components measure the instantaneous curvature of
the sphere, whereas the temporal component also depends on its
acceleration. The trace of the extrinsic curvature is given by,
\begin{equation}
  K=\frac{\ddot a+\frac12 f'(a)}{\beta}
  +\frac{2\beta}{a}.
  \label{eq:hom-CMC}
\end{equation}
Using \eqref{eq:hom-clocks} and \eqref{eq:hom-beta}, the boundary condition $K=K_R$ becomes an equation for the boundary Weyl factor,
\begin{equation}
  \begin{split}
    \bm{\omega}''
    ={}&
    a K_R
    \sqrt{
      1+\frac{a^2}{\ads^2}-\frac{\mu}{a}
      +(\bm{\omega}')^2
    }
    -2-\frac{3a^2}{\ads^2}
    +\frac{3\mu}{2a}
    -2(\bm{\omega}')^2,
    \qquad
    a=R\e^{\bm{\omega}}.
  \end{split}
  \label{eq:hom-exact-Weyl}
\end{equation}
This is a fully nonlinear equation for the boundary Weyl factor, with no small-amplitude expansion assumed. The nonlinearity arises both from the changing boundary radius and from the relation between its velocity and extrinsic curvature.

The evolution has a first integral,
\begin{equation}
  \mathcal C
  \equiv a^2\beta-\frac{K_R}{3}a^3.
  \label{eq:hom-charge}
\end{equation}
Its conservation follows directly from
\eqref{eq:hom-CMC},
\begin{equation}
  \frac{\dd\mathcal C}{\dd\tau}
  =
  a^2\dot a
  \left[
    \frac{\ddot a+\frac12 f'(a)}{\beta}
    +\frac{2\beta}{a}-K_R
  \right]
  =0.
  \label{eq:hom-charge-proof}
\end{equation}
Solving for $\beta$ gives a first-order radial equation,
\begin{equation}
  \dot a^2
  =
  \left(\frac{K_R^2}{9}-\frac{1}{\ads^2}\right)a^2
  -1+\frac{\mu+\frac23K_R\mathcal C}{a}
  +\frac{\mathcal C^2}{a^4}.
  \label{eq:hom-radial-first-integral}
\end{equation}
The relation
$\beta=K_Ra/3+\mathcal C/a^2>0$ selects the branch consistent
with the chosen orientation. Thus the radial dynamics can be
organized by the value of a single conserved quantity, fixed
by the initial position and velocity.

The geometric meaning of $\mathcal{C}$ follows from
\eqref{eq:hom-principal-curvatures} and \eqref{eq:hom-charge},
\begin{equation}
  K^m{}_n-\frac{K_R}{3}\delta^m{}_n
  =
  \frac{\mathcal C}{a^3}
  \operatorname{diag}(-2,1,1),
  \label{eq:hom-traceless-K}
\end{equation}
with entries ordered as $(\tau,\theta,\varphi)$. In this homogeneous sector, $\mathcal C$ therefore measures
the departure from total umbilicity.

We now compute the associated conformal energy using the conformal stress tensor \eqref{eq:CBY-definition}, whose time-time component is given by 
\begin{equation}
  T_{tt}
  =
  -\frac{f_R\mathcal C}{4\pi G_N R^3}.
  \label{eq:hom-stress-tt}
\end{equation}
The conserved
charge associated with $\partial_t$ is consequently
\begin{equation}
  E_{\rm conf}
  =
  \frac{R^2}{\sqrt{f_R}}
  \int_{S^2}\dd\Omega_2\,T_{tt}
  =
  -\frac{\Omega_R}{G_N}\mathcal C,
  \label{eq:hom-conformal-energy}
\end{equation}
providing another interpretation for the conserved quantity $\mathcal{C}$ as a rescaled conformal energy. Zero conformal energy is therefore equivalent, within this
sector, to a vanishing conformal stress tensor and to total
umbilicity of the moving conformal boundary. Since $\mathcal C$ is conserved,
the $\mathcal C=0$ condition defines an invariant sector of the evolution.\footnote{In general, conformal boundary conditions also admit initial data with nonzero $\mathcal C$.
They do not require the evolving boundary to remain umbilic.}

\subsection{Exact zero-energy evolution at the photon sphere}
\label{app:homogeneous-photon}
We now place the reference boundary at $R=\rph$. Using the particular value of $K_R$ in this case and the fact that $\mathcal{C}=0$, we find that \eqref{eq:hom-radial-first-integral}  simplifies to
\begin{equation}
  \dot a^2= \frac{(a-\rph)^2(a+2\rph)}{3\rph^2a}.
  \label{eq:hom-photon-radial}
\end{equation}
The static radius is a double zero of the radial equation.
Nontrivial trajectories can approach it asymptotically,
but cannot pass through it with zero velocity at a finite
proper time. The explicit dependence on
$\ads$ has cancelled. The AdS radius still determines the
relation between clocks and the location of the horizon.

Using \eqref{eq:hom-clocks}, we find that \eqref{eq:hom-photon-radial} yields the zero-energy constraint
\begin{align}
  (\bm{\omega}')^2
  &=
  \frac13
  \left(
    \e^{2\bm{\omega}}-3+2\e^{-\bm{\omega}}
  \right) 
   \label{eq:hom-zero-energy-Weyl}
\end{align}
Using this constraint, we find that the Weyl mode in this sector needs to satisfy
\begin{align}
  \bm{\omega}''
  &=
  \frac13
  \left(
    \e^{2\bm{\omega}}-\e^{-\bm{\omega}}
  \right).
  \label{eq:hom-zero-energy-acceleration}
\end{align}
The first-order equation \eqref{eq:hom-zero-energy-Weyl} can be integrated explicitly\footnote{For example, setting
$v=\sqrt{(1+2\e^{-\bm{\omega}})/3}$ reduces it to
$(v')^2=(1-v^2)^2/4$.}, yielding
\begin{equation}
  \e^{\bm{\omega}(s)}
  =
  \frac{(1+z(s))^2}{1-4z(s)+z(s)^2},
  \qquad
  z(s)=A\e^{\sigma s},
  \qquad
  \sigma=\pm1.
  \label{eq:hom-exact-solution}
\end{equation}
Here $A$ is real, and the solution is restricted to the
interval connected to $z=0$ on which the conformal factor
is positive and $a>r_h$. It is easy to check that \eqref{eq:hom-zero-energy-acceleration} is automatically satisfied.

The static solution corresponds
to $A=0$. For small $A$,
\begin{align}
   \bm{\omega}&=6 A e^{\pm \Omega_{\rm ph}t}+\mathcal{O}(A^2),
\end{align}
recovering the homogeneous growing and decaying perturbations. This also follows by linearizing
\eqref{eq:hom-zero-energy-acceleration}, which gives
$\bm{\omega}''=\bm{\omega}$ at leading order. We note that the same linearized result is obtained by formally setting $\mathfrak{L}=0$ in the boundary-branch dispersion relation \eqref{eq:umbilic-boundary-dispersion}, even though the latter was derived for non-homogeneous perturbations.

For $\sigma=+1$, a small positive $A$ gives an outward trajectory. As $s$ increases, $z=Ae^s$ grows and reaches $z=2-\sqrt{3}$, where the denominator of the exact solution vanishes. The conformal factor, and hence the boundary radius $a=\rph e^{\bm{\omega}}$, therefore diverges at a finite value of the conformal time, but infinite proper time. A small negative $A$ instead gives an inward trajectory that reaches the black-hole horizon, where the exterior description adopted here ceases to apply. The solutions with $\sigma=-1$ are the corresponding time-reversed trajectories. Let us emphasize though that the growing $l=0$ perturbation does not contradict the suppression of the large-angular-momentum instabilities in Section \ref{sec:large L analysis}.

The same evolution has an optical interpretation.
In the zero-energy sector, $\beta=\Omega_{\rm ph} a$.
Using \eqref{eq:hom-beta} and \eqref{eq:hom-photon-radial} to express
the trajectory in static bulk time gives
\begin{equation}
  \left(\frac{\dd a}{\dd T}\right)^2
  =
  f(a)^2
  \left[
    1-\frac{U(a)}{U(\rph)}
  \right],
  \label{eq:hom-optical-trajectory}
\end{equation}
where we remind that $U$ is the optical potential \eqref{eq:background-U}. This is precisely the radial equation obeyed by null geodesics with the critical impact parameter $1/\Omega_{\rm ph}$, namely the geodesics associated with the photon sphere. The conformal boundary itself, however, remains a timelike hypersurface. Its total umbilicity guarantees that any null geodesic initially tangent to the boundary continues to remain tangent as it evolves. In this sense, the moving conformal boundary defines a dynamical photon surface. Such surfaces in static spherical geometries, including
Schwarzschild-AdS, were characterized in
\cite{CederbaumGalloway2021}. Here they arise as the
zero-conformal-energy trajectories of the homogeneous sector of the conformal boundary problem. Near equilibrium, the Weyl growth rate $\Omega_{\rm ph}$ agrees with the photon-sphere Lyapunov exponent measured in
static time.

It is useful to compare this invariant zero-energy sector with
general homogeneous initial data near the static boundary. Keeping $K_R=K(\rph)$ fixed, the expansion of \eqref{eq:hom-charge} for $\bm{\omega},\bm{\omega}'=\Order(\varepsilon)$ gives
\begin{equation}
  \mathcal C
  =
  \frac{\rph^2}{2\sqrt{f_{\rm ph}}}
  \left[
    (\bm{\omega}')^2-\bm{\omega}^2
  \right]
  +\Order(\varepsilon^3),
  \label{eq:hom-quadratic-charge}
\end{equation}
and therefore no linear contribution. A generic homogeneous
perturbation therefore first departs from umbilicity at
second order. In particular, for the linear solution
$\bm{\omega}=A_+\e^s+A_-\e^{-s}+\Order(\varepsilon^2)$,
with $A_\pm=\Order(\varepsilon)$,
\begin{equation}
  \mathcal C
  =
  -\frac{2 \rph^2}{\sqrt{f_{\rm ph}}}A_+A_-
  +\Order(\varepsilon^3).
  \label{eq:hom-mode-energy}
\end{equation}
The pure growing and decaying perturbations each admit an exact nonlinear continuation within the zero-energy sector, given by \eqref{eq:hom-exact-solution}. A generic superposition of the two, however, acquires nonzero conformal energy at quadratic order.

For the homogeneous $l=0$ sector, the dynamics can thus be solved exactly beyond linear order within the zero-conformal-energy sector. The resulting solutions describe nonlinear motion of the umbilic boundary at fixed bulk mass, with the pure growing and decaying linear perturbations arising as their small-amplitude limits. Spherical symmetry is essential to this construction, since it removes propagating bulk gravitational degrees of freedom. The exact nonlinear evolution found here therefore does not imply a corresponding nonlinear extension of the bulk-boundary separation for general inhomogeneous perturbations.

\section{Linearized dynamics around the AdS$_4$ black brane}
\label{app:BB}
In this appendix, we list the scalar and vector master field equations and conformal boundary conditions for linearized perturbations around the AdS$_4$ black brane, following the Kodama-Ishibashi formalism
\cite{KodamaIshibashiSeto2000,KodamaIshibashi2003}. We then consider the asymptotic limit in both sectors, in which the conformal boundary conditions reproduce the boundary conditions appropriate to an undeformed boundary metric
\cite{MirandaMorganZanchin2008}. Finally, we consider the hydrodynamic regime, in which the sound and shear modes are obtained and shown to match the CFT$_3$ expressions reviewed in Section \ref{sec:review fluid gravity}.

\subsection{Master field equations and conformal boundary conditions}
\label{app:brane-CBC}
We consider the AdS$_4$ black brane metric \eqref{eq: black brane metric}, which we reproduce here for convenience,
\begin{align}
  \dd s^2
  &=-f_{\rm br}(r)\dd t^2+\frac{\dd r^2}{f_{\rm br}(r)}
  +\frac{r^2}{\ads^2}\left(\dd x^2+\dd y^2\right),
  \qquad
  f_{\rm br}(r)=\frac{r^2}{\ads^2}-\frac{r_h^3}{r\ads^2},
\end{align}
and place a timelike boundary at $r=\mathfrak r$. We introduce the radial coordinate
\begin{equation}
  z=\frac{r_h}{r},
  \label{eq:brane-compact-variables}
\end{equation}
where $r_h$ is the position of the black brane horizon. In this coordinate, the horizon is located at $z=1$, the asymptotic boundary at $z=0$, and the finite boundary at $\mathfrak z=r_h/\mathfrak r$.

We consider linearized perturbations around the black brane background using the Kodama-Ishibashi formalism. The scalar and vector perturbations are encoded in the master fields
\begin{align}
    \Phi^{(S,V)}(z,t)
    =\operatorname{Re}\; e^{-i\omega t}\phi^{(S,V)}(z).
\end{align}
The Kodama-Ishibashi decomposition involves eigenfunctions of the Laplacian on the two-dimensional flat space. The scalar harmonics are plane waves, $ \mathbb S(x,y)=e^{i(k_xx+k_yy)/\ads}$, and we define $k_S=\sqrt{k_x^2+k_y^2}>0$, which is the dimensionless momentum parameter in the scalar sector. The coordinates $x$ and $y$ have dimensions of length. In the vector sector, the vector harmonics satisfy $\nabla^i\mathbb V_i=0$ and $\nabla^2\mathbb V_i=-(k_V^2/\ads^2)\mathbb V_i$, where $k_V>0$ is the corresponding dimensionless momentum parameter.

It will be useful to introduce the rescaled frequency and momentum parameters
\begin{align}
    \varpi=\frac{\ads^2}{r_h}\omega,
    \qquad
    q=\frac{\ads}{r_h}k_{S,V},
\end{align}
where, by a slight abuse of notation, we use the same symbol $q$ in both the scalar and vector sectors. Since
\begin{align}\label{eq: for hydro regime def}
     \frac{\omega}{T}&=\frac{4\pi}{3}\varpi,\qquad \frac{k}{T}=\frac{4\pi}{3}q,
\end{align}
the hydrodynamic regime corresponds to
\begin{align}
  \varpi\ll 1,\qquad q\ll1.
\end{align}

\subsubsection{Scalar sector}
In the scalar sector, the master field equation is
\begin{equation}
  \phi^{(S)''}-\frac{3z^2}{g}\phi^{(S)'}
  +
  \left[
    \frac{\varpi^2}{g^2}
    -\frac{\mathcal V_S(z,q)}{g}
  \right]\phi^{(S)}=0,
  \label{eq:brane-master}
\end{equation}
where primes in this appendix denote derivatives with
respect to $z$, and we denote $g(z)=1-z^3$. The scalar potential is given by
\begin{equation}
  \mathcal V_S(z,q)=\frac{
    q^6+3q^4z+9q^2z^2+9z^3+18
  }{(q^2+3z)^2}.
  \label{eq:brane-master-potential}
\end{equation}
Reconstructing the metric perturbations in terms of the scalar master field and imposing the conformal boundary conditions lead to a Robin boundary condition for $\phi^{(S)}$,
\begin{equation}
  \mathcal C_1(\mathfrak{z};\varpi,q)\phi^{(S)}(\mathfrak{z})
  +
  \mathcal C_2(\mathfrak{z};\varpi,q)\phi^{(S)\prime}(\mathfrak{z})
  =0.
  \label{eq:brane-finite-CBC}
\end{equation}
The two coefficients are
\begin{align}
  \mathcal C_1={}&
  8q^8 \mathfrak{z} g^2
  +18\varpi^2\mathfrak{z}^3(9\mathfrak{z}^4+4\varpi^2)
  -q^6 \mathfrak{z} g(16\varpi^2-36\mathfrak{z}+45\mathfrak{z}^4)
  \notag\\
  &+2q^4
  \left[
    -3g(4-17\mathfrak{z}^3+22\mathfrak{z}^6)
    +3\varpi^2 \mathfrak{z}^2(-14+17 \mathfrak{z}^3)
    +4\varpi^4 \mathfrak{z}
  \right]
  \notag\\
  &+3q^2
  \left[
    9\mathfrak{z}^4g(2-5\mathfrak{z}^3)
    +4\varpi^2(2-13\mathfrak{z}^3+20\mathfrak{z}^6)+16\varpi^4\mathfrak{z}^2
  \right],
  \label{eq:brane-CBC-A}\\
  \mathcal C_2={}&
  -2q^2g^2(q^2+3\mathfrak{z})
  \left[
    -9\mathfrak{z}^4+q^2(4-7 \mathfrak{z}^3)-4\varpi^2
  \right],
  \label{eq:brane-CBC-B}
\end{align}
where $g=g(\mathfrak z)=1-\mathfrak z^3$ is evaluated at the boundary. We select the solution of the master-field equation that is ingoing at the horizon,
\begin{equation}
  \phi^{(S)}_{\rm in}(z)
  = c_S  (1-z)^{-i \varpi/3}
  \left[1+\Order(1-z)\right].
  \label{eq:brane-ingoing}
\end{equation}
 For each position of the conformal boundary, \eqref{eq:brane-finite-CBC} then determines the allowed frequencies.

\subsubsection{Vector sector}
\label{app:planar-vector}
In the vector sector, the master field equation is
\begin{equation}
  g^2 \phi^{(V)''}
  -3z^2g \phi^{(V)'}
  +\left[\varpi^2+g(3z-q^2)\right]\phi^{(V)}
  =0.
  \label{eq:app-planar-vector-master}
\end{equation}
Reconstructing the metric perturbations in terms of the vector master field and imposing the conformal boundary conditions lead to 
\begin{equation}
    \phi^{(V)}(\mathfrak{z})-\mathfrak{z} \phi^{(V)'}(\mathfrak{z})=0.
  \label{eq:app-planar-vector-CBC}
\end{equation}
The ingoing condition at the horizon selects
\begin{equation}
  \phi_{\rm in}^{(V)}(z)
  =c_V(1-z)^{-i\varpi/3}F(z),
  \qquad
  F(z)=1+\Order(1-z).
  \label{eq:app-planar-vector-ingoing}
\end{equation}
Factoring out this horizon behavior gives the equivalent boundary condition
\begin{equation}
  \mathcal R_V(\varpi;\mathfrak{z},q)
  =\left(1-\mathfrak{z}-\frac{i\varpi}{3}\mathfrak{z}\right)F(\mathfrak{z})
  -\mathfrak{z}(1-\mathfrak{z})F'(\mathfrak{z})=0.
  \label{eq:app-planar-vector-residual}
\end{equation}
This is the residual condition used in the numerical computations presented in the main text.

\subsection{Gravitational derivation of the CFT$_3$ sound mode}
\label{app:brane-limits}
We now consider the scalar sector in the regime $q\ll1$, $\varpi\ll1$, in which we recover the planar CFT$_3$ sound pole \eqref{eq:sound mode brane asy} \cite{Herzog2003Sound,MirandaMorganZanchin2008}. Starting from the scalar master-field equation, we introduce
\begin{align}
    \varpi=\epsilon\widetilde{\varpi},
    \qquad
    q=\epsilon\widetilde q,
\end{align}
to implement the hydrodynamic expansion. We also expand the master field, factoring out the ingoing behavior at the horizon,
\begin{align}
    \phi^{(S)}(z)
    =c_S(1-z)^{-i\epsilon\widetilde{\varpi}/3}
    \left[
        \phi_0(z)+\epsilon\phi_1(z)
        +\mathcal O(\epsilon^2)
    \right].
\end{align}
Solving the master field equation order by order in $\epsilon$, and at each order selecting the solution compatible with the ingoing horizon condition, gives
\begin{align}
    \phi_0(z)&=\frac{1}{z},\\
    \phi_1(z)&=
    \frac{\widetilde{\varpi}}{18z}
    \left[
        4\pi-i\left(
        \sqrt{3}\,\pi
        +6\log\left(z^2+z+1\right)
        \right)
    \right],
\end{align}
where the overall normalization has been absorbed into $c_S$. We expand the frequency as
\begin{align}
    \widetilde{\varpi}
    =\widetilde{\varpi}_0
    +\epsilon\widetilde{\varpi}_1
    +\mathcal{O}(\epsilon^2),
\end{align}
and substitute this together with the perturbative master field solution into the conformal boundary condition. Expanding the resulting equation in $\epsilon$ determines $\widetilde{\varpi}_0$ and $\widetilde{\varpi}_1$. Returning to the original variables, we find
\begin{align}
    \varpi_{\rm sound}
    &=
    \pm\frac{\sqrt{1-\mathfrak z^3}}{\sqrt{2}}\,q
    -\frac{i}{6}\left(1-\mathfrak z^3\right)q^2
    +\mathcal O(q^3).
    \label{eq: hydro sound mode from gravity}
\end{align}
This result is valid in the regime $\varpi\ll1$, $q\ll1$, for arbitrary $0<\mathfrak z<1$, and describes a sound mode throughout the cavity. Its frequency tends to zero as the conformal boundary approaches the horizon. Using the variable
\begin{align}
    \mathfrak u
    =1-\frac{r_h}{\mathfrak r}
    =1-\mathfrak z
\end{align}
introduced in the main text \eqref{eq:wall-compact-coordinate}, we find
\begin{equation}
    \varpi_{\rm sound}
    =
    \pm\sqrt{\frac{3\mathfrak u}{2}}\,q
    -\frac{i}{2}q^2\mathfrak u
    +\Order(\mathfrak u^{3/2}),
    \qquad
    \mathfrak u\rightarrow0.
    \label{eq:planar-scalar-stretched-sound}
\end{equation}
Writing $\mathfrak u=\delta/r_h +\mathcal O(\delta^2)$, where $\delta$ is the small parameter introduced in the stretched-horizon expansion \eqref{eq:near horizon expansion}, we see that \eqref{eq:planar-scalar-stretched-sound} reproduces the stretched-horizon sound mode \eqref{sound modes brane}.

On the other hand, in the asymptotic limit $\mathfrak z\to0$, using \eqref{eq: for hydro regime def}, \eqref{eq: hydro sound mode from gravity} reduces precisely to the CFT$_3$ sound mode \eqref{eq:sound mode brane asy}.

\subsection{Gravitational derivation of the CFT$_3$ shear mode}

We now consider the vector sector in the regime $q\ll1$, $\varpi\ll1$, in which we recover the planar CFT$_3$ shear pole \eqref{eq:vector shear derivation}. As in the scalar sector, we introduce
\begin{align}
    \varpi=\epsilon^2\widetilde{\varpi},
    \qquad
    q=\epsilon\widetilde q,
\end{align}
to implement the hydrodynamic expansion. We also expand the master field, factoring out the ingoing behavior at the horizon,
\begin{align}
    \phi^{(V)}(z)
    =c_V(1-z)^{-i\epsilon^2\widetilde{\varpi}/3}
    \left[
        \phi_0(z)+\epsilon^2\phi_1(z)
        +\mathcal O(\epsilon^4)
    \right].
\end{align}
Solving the master-field equation order by order in $\epsilon$, and at each order selecting the solution compatible with the ingoing horizon condition, gives
\begin{align}
    \phi_0(z)&=z,\\
    \phi_1(z)&=\frac{1}{6}
    \left[
        2\widetilde q^2-6i\widetilde{\varpi}
        -2i\sqrt{3}\,z\widetilde{\varpi}\,
        \arctan\!\left(\frac{1+2z}{\sqrt{3}}\right)
        +iz\widetilde{\varpi}\log\!\left(1+z+z^2\right)
    \right],
\end{align}
where the overall normalization has been absorbed into $c_V$. We substitute this solution into the conformal boundary condition \eqref{eq:app-planar-vector-CBC}. At the order considered, the boundary condition determines the leading term in the frequency. Determining the next correction would require the master field solution at order $\epsilon^4$. Returning to the original variables, we find
\begin{align}
    \varpi_{\rm shear}
    &=-\frac{i}{3}\left(1-\mathfrak z^3\right)q^2
    +\mathcal O(q^4).
    \label{eq: hydro shear mode from gravity}
\end{align}
This result is valid in the regime $\varpi\ll1$, $q\ll1$, for arbitrary $0<\mathfrak z<1$, and describes a diffusive mode throughout the cavity. Its frequency tends to zero as the conformal boundary approaches the horizon. In terms of the variable \eqref{eq:wall-compact-coordinate}, we find
\begin{equation}
    \varpi_{\rm shear}
    =
    -i q^2\mathfrak u
    +\Order(\mathfrak u^2),
    \qquad
    \mathfrak u\rightarrow0.
    \label{eq:planar-vector-stretched-shear}
\end{equation}
Using $\mathfrak u =\delta/r_h+\mathcal O(\delta^2)$, we see that \eqref{eq:planar-vector-stretched-shear} reproduces the leading-order term of the stretched-horizon shear mode \eqref{eq:shear mode near horizon brane}.

On the other hand, in the asymptotic limit $\mathfrak z\to0$, using \eqref{eq: for hydro regime def}, \eqref{eq: hydro shear mode from gravity} reduces precisely to the CFT$_3$ shear mode \eqref{eq:vector shear derivation}. At this asymptotic endpoint, the small-momentum dispersion relation is known beyond the leading diffusive term \cite{DilesEtAl2020},
\begin{equation}
    \varpi_{\rm shear}
    =
    -\frac{i}{3}q^2
    -\frac{i}{18}
    \left(
        1+\frac{\pi}{3\sqrt{3}}-\log 3
    \right)q^4
    +\Order(q^6).
    \label{eq:planar-vector-asymptotic-dispersion}
\end{equation}
Finally, we note that the dependence on the position of the boundary in \eqref{eq: hydro shear mode from gravity} has a simple interpretation in terms of local quantities. Converting to the proper frequency, momentum, and Tolman temperature,
\begin{align}
    \omega_{\rm loc}
    =\frac{\omega}{\sqrt{f_{\rm br}(\wallr)}},
    \qquad
    k_{\rm loc}
    =\frac{k\ads}{\wallr},
    \qquad
    T_{\rm loc}
    =\frac{T}{\sqrt{f_{\rm br}(\wallr)}},
    \label{eq: local quantities}
\end{align}
we obtain the standard diffusion law
\begin{equation}
    \omega_{\rm loc}
    =-iD_{\rm loc}k_{\rm loc}^2+\mathcal{O}(k_{\rm loc}^4),
    \qquad
    D_{\rm loc}T_{\rm loc}=\frac{1}{4\pi},
    \label{eq:planar-vector-local-diffusion}
\end{equation}
where $D_{\rm loc}$ is the diffusion constant measured in the local boundary frame. Thus, the leading diffusion coefficient in local thermal units is independent of the position of the boundary. At the asymptotic endpoint, returning to the $\omega,k$ variables,
\begin{align}
    \omega
    &=-iDk^2+\mathcal{O}(k^4),
    \qquad
    D=\frac{\ads^2}{3r_h}
    =\frac{1}{4\pi T}.
\end{align}
Using the hydrodynamic relation
\begin{align}
    D=\frac{\eta}{\epsilon+p}
    =\frac{\eta}{sT},
\end{align}
where the second equality holds at zero chemical potential, we recover the standard Einstein-gravity result for the shear-viscosity-to-entropy-density ratio \cite{KovtunSonStarinets2005},
\begin{equation}
    \frac{\eta}{s}
    =DT
    =\frac{1}{4\pi}.
    \label{eq:planar-vector-diffusion-viscosity}
\end{equation}

\section{Details of the large angular momentum WKB approximation for Schwarzschild-AdS$_4$}
\label{app:wkb}
In this appendix, we provide the details of the WKB analysis used in Section \ref{sec:large L analysis}. We focus on the leading large-angular-momentum master-field equation for the Schwarzschild-AdS$_4$ black hole, for which the scalar and vector Kodama--Ishibashi potentials coincide at leading order in $\mathfrak L$. Introducing the tortoise coordinate $\dd r^*/\dd r=f^{-1}$, the master-field equation takes the form \eqref{eq:eikonal-equation},
\begin{align}
    -\frac{1}{\mathfrak L^2}
    \partial_{r^*}^2\phi
    +\left(
        U(r)-\frac{\omega^2}{\mathfrak L^2}
    \right)\phi
    =\mathcal O(\mathfrak L^{-2})\phi,
    \qquad
    U(r)=\frac{f(r)}{r^2}.
    \label{eq:app-WKB-master}
\end{align}
The optical potential has a maximum at the photon sphere $r=\rph$. We consider frequencies below the maximum of the optical potential, for which an inner turning point exists. Depending on the frequency, a second turning point may appear on the asymptotic side of the photon sphere. When both turning points are present, we denote them by
\begin{align}
    U(r_1)=U(r_2)=\frac{\omega^2}{\mathfrak L^2},
    \qquad
    r_1<\rph<r_2.
\end{align}
Accordingly, here and in the following we consider frequencies for which $\omega^2/\mathfrak L^2$ remains finite in the large-$\mathfrak L$ limit. For Schwarzschild-AdS$_4$, the existence of the outer turning point additionally requires $\omega^2/\mathfrak L^2>\ads^{-2}$.

The radial solution can be constructed by matching WKB solutions to local turning-point approximations. When the two turning points remain parametrically separated, each is described by an Airy region of characteristic width $\Delta r^*\sim\mathfrak L^{-2/3}$. As the turning points approach the photon-sphere maximum, these regions eventually merge, and the quadratic variation of the potential must instead be retained. The corresponding near-maximum region has width $\Delta r^*\sim\mathfrak L^{-1/2}$ and is described by parabolic-cylinder functions.

We begin by discussing the WKB solution and its regimes of validity. We then consider isolated turning points and perform the Airy matching across the different WKB regions. Finally, we study the regime in which the turning points approach the photon-sphere maximum, derive the corresponding parabolic-cylinder solution, and match it to the WKB regions. In each case, the matching selects the solution that is ingoing at the horizon, on which the conformal boundary conditions of Section \ref{sec:large L analysis} are imposed.

\subsection{WKB solution and regimes of validity}
\label{app:WKB-validity}
Away from the turning points, we use the WKB ansatz
\begin{align}
    \phi(r^*)
    =
    c\exp\left[
        i\mathfrak L
        \int^{r^*}\dd\widetilde r^*
        \sum_{n=0}^{\infty}
        \frac{\sigma_n(\widetilde r^*)}{\mathfrak L^n}
    \right].
    \label{eq:app-WKB-ansatz}
\end{align}
Plugging it into the master field equation \eqref{eq:app-WKB-master}, we obtain at leading order,
\begin{align}
    \sigma_0
    =
    \pm\sqrt{
        \frac{\omega^2}{\mathfrak L^2}-U(r)
    }.
    \label{eq:app-WKB-sigma0}
\end{align}
In the classically allowed region connected to the horizon, we define
\begin{align}
    \kappa(r^*)
    \equiv
    \sqrt{
        \frac{\omega^2}{\mathfrak L^2}-U(r)
    }. \label{def kappa}
\end{align}
With the convention for time dependence $e^{-i\omega t}$ used in the main text, the solution ingoing at the future horizon behaves as $e^{-i\omega r^*}$. For $\mathrm{Re}\;\omega>0$, this selects the negative sign in \eqref{eq:app-WKB-sigma0}. At the next order, the master field equation fixes
\begin{align}
    \sigma_1
    =
    \frac{i\,\partial_{r^*}\sigma_0}{2\sigma_0},
\end{align}
so that, up to an overall normalization,
\begin{align}
    \phi_{\rm in}(r^*)
    \simeq
    \frac{1}{\sqrt{\kappa(r^*)}}
    \exp\left[
        -i\mathfrak L
        \int^{r^*}\dd\widetilde r^*\,
        \kappa(\widetilde r^*)
    \right].
    \label{eq:app-WKB-ingoing}
\end{align}
Here $\omega$ may itself admit a large-$\mathfrak L$ expansion, so the $\mathfrak L$ dependence contained in $\kappa$ is left implicit.

The validity of the WKB approximation requires
\begin{align}
    \kappa^2
    \gg
    \frac{1}{\mathfrak L}
    \left|\partial_{r^*}\kappa\right|.
    \label{eq:app-WKB-validity}
\end{align}
We consider first an isolated turning point $r_t^*$, namely
\begin{align}
    \partial_{r^*}U(r_t^*)=\Order(1).
\end{align}
Expanding the optical potential around the turning point,
\begin{align}
    U(r^*)
    =
    \frac{\omega^2}{\mathfrak L^2}
    +
    \partial_{r^*}U(r_t^*)
    (r^*-r_t^*)
    +\cdots,
\end{align}
the condition \eqref{eq:app-WKB-validity} gives
\begin{align}
    |r^*-r_t^*|
    \gg
    \frac{1}{
        \mathfrak L^{2/3}
        |\partial_{r^*}U(r_t^*)|^{1/3}
    }.
    \label{eq:app-WKB-Airy-width}
\end{align}
Thus, when the slope of the potential at the turning point remains finite, the WKB approximation breaks down in a layer of characteristic width
\begin{align}
    \Delta r^*_{\rm Airy}
    \sim
    \mathfrak L^{-2/3}.
\end{align}
The subscript anticipates that, within this region, the master-field equation reduces to the Airy equation.

The situation changes when the turning points approach the maximum of the optical potential at the photon sphere. Let
\begin{align}
    |r_t^*-\rph^*|
    \sim
    \mathfrak L^{-\beta},
    \qquad
    \beta>0.
    \label{eq:app-WKB-beta}
\end{align}
Since $\partial_{r^*}U(\rph^*)=0$,
\begin{align}
    \partial_{r^*}U(r_t^*)
    \simeq
    \partial_{r^*}^2U(\rph^*)
    (r_t^*-\rph^*)
    \sim
    \mathfrak L^{-\beta},
\end{align}
and therefore
\begin{align}
    \Delta r^*_{\rm Airy}
    \sim
    \mathfrak L^{-(2-\beta)/3}.
    \label{eq:app-WKB-Airy-width-near-max}
\end{align}
For $\beta<1/2$, this width remains parametrically smaller than the distance from the turning point to the maximum, and the two turning points can be treated independently by Airy matching. At $\beta=1/2$, the two scales become comparable. For $\beta\geq1/2$, the separate Airy regions merge and the quadratic expansion of the potential about the photon sphere must be retained.

The critical value $\beta=1/2$ can also be understood directly from the master-field equation. Introducing a coordinate that resolves the region near the maximum,
\begin{align}
    y^*=\mathfrak L^\beta(r^*-\rph^*),
\end{align}
the radial-derivative term in \eqref{eq:app-WKB-master} scales as $\mathfrak L^{2(\beta-1)}$, while the quadratic variation of the potential scales as
\begin{align}
    U(r^*)-U(\rph^*)
    \sim (r^*-\rph^*)^2
    \sim \mathfrak L^{-2\beta}.
\end{align}
The two contributions therefore balance precisely for $\beta=1/2$. The corresponding frequency correction is $\Order(1)$\footnote{If $\omega=\omega_0\mathfrak L+\omega_1\mathfrak L^\gamma+\cdots$, the turning-point condition gives $\omega_0=\Omega_{\rm ph}$ and $\gamma=1-2\beta$.}, so that
\begin{align}
    \omega=\Omega_{\rm ph}\mathfrak L+\Order(1),
\end{align}
which leads to the parabolic-cylinder problem considered below.

\subsection{Isolated turning points and Airy matching}
\label{app:WKB-Airy}
We first consider the regime $\beta<1/2$, in which the two turning points remain sufficiently separated that they can be treated independently. In a neighbourhood of each turning point, the potential can be expanded linearly and the master field equation reduces to the Airy equation. We determine the appropriate Airy solutions by matching them to the WKB solution that is ingoing at the horizon.\footnote{We perform the matching for real frequencies below the potential maximum and then analytically continue the resulting expressions to the complex frequencies relevant for the conformal boundary problem.}

\subsubsection{The inner turning point}
We start with the turning point on the horizon side of the photon sphere, which we denote by $r_1^*$. In the terminology of Section \ref{sec:large L analysis}, this is the turning point relevant to the interior problem. Since $\partial_{r^*}U(r_1^*)>0$, we introduce
\begin{align}
    \xi
    \equiv
    \left(\mathfrak L^2\partial_{r^*}U(r_1^*) \right)^{1/3} (r_1^*-r^*),
    \label{eq:app-inner-Airy-coordinate}
\end{align}
so that $\xi>0$ on the horizon side of the turning point. Expanding
\begin{align}
    U(r^*)
    =
    \frac{\omega^2}{\mathfrak L^2}
    +
    \partial_{r^*}U(r_1^*)
    (r^*-r_1^*)
    +\cdots,
\end{align}
the master field equation reduces to the Airy equation
\begin{align}
    \partial_\xi^2\phi+\xi\phi=0.
\end{align}
Hence, the solutions are given by 
\begin{align}
    \phi(\xi)
    =
    c_1\mathrm{Ai}(-\xi)+c_2\mathrm{Bi}(-\xi). \label{Airy solutions}
\end{align}
We select the solution that matches onto the ingoing WKB solution near the horizon. To do so, we expand \eqref{eq:app-WKB-ingoing} near the turning point. Since
\begin{align}
    \mathfrak L
    \int_{r^*}^{r_1^*}\dd\widetilde r^*\,\kappa
    \simeq
    \frac{2}{3}\xi^{3/2},
\end{align}
the corresponding Airy combination is
\begin{align}
    \phi_{\rm in}
    \propto
    \mathrm{Bi}(-\xi)+i\mathrm{Ai}(-\xi)
    \propto
    \mathrm{Ai}\left(-e^{2\pi i/3}\xi\right).
    \label{eq:app-inner-Airy-solution}
\end{align}
This is the rotated Airy solution used for the interior problem in Section \ref{sec:large L analysis}.

\subsubsection{The second WKB region}
We now assume the existence of both turning points. Between the two turning points, $r_1^*<r^*<r_2^*$, we have $U(r^*)>\omega^2/\mathfrak L^2$. We define
\begin{align}
    \lambda(r^*)
    \equiv
    \sqrt{
        U(r^*)-\frac{\omega^2}{\mathfrak L^2}
    }.
\end{align}
To continue the solution \eqref{eq:app-inner-Airy-solution} through the first turning point, we set $\chi=-\xi>0$. For $\chi\gg1$,
\begin{align}
    \mathrm{Ai}\left(e^{2\pi i/3}\chi\right)
    \propto
    \chi^{-1/4}
    \left[
        e^{\frac{2}{3}\chi^{3/2}}
        +\frac{i}{2}e^{-\frac{2}{3}\chi^{3/2}}
    \right].
\end{align}
The general WKB solution in this region is
\begin{align}
    \phi(r^*)
    =
    \frac{1}{\sqrt{\lambda(r^*)}}
    \left[
        c_+e^{-\mathfrak L S(r^*)}
        +c_-e^{\mathfrak L S(r^*)}
    \right],
    \qquad
    S(r^*)
    \equiv
    \int_{r_1^*}^{r^*}\dd\widetilde r^*\,\lambda(\widetilde r^*).
\end{align}
Near the first turning point,
\begin{align}
    \lambda(r^*)
    \simeq
    \sqrt{
        \partial_{r^*}U(r_1^*)
        (r^*-r_1^*)
    }.
\end{align}
Using $\chi=-\xi=\left(\mathfrak L^2\partial_{r^*}U(r_1^*)\right)^{1/3}(r^*-r_1^*)$, we then find
\begin{align}
    \mathfrak L S(r^*)
    \simeq
    \frac{2}{3}\chi^{3/2}.
\end{align}
Thus, the two WKB exponentials reproduce precisely the two exponential terms in the large-$\chi$ expansion of the Airy solution above. Matching their relative coefficients gives $c_+/c_-=i/2$, and hence, up to an overall normalization,
\begin{align}
    \phi(r^*)
    =
    \frac{c}{
        \left(
            U(r^*)-\omega^2/\mathfrak L^2
        \right)^{1/4}}
    \left[
        e^{-\mathfrak L S(r^*)}
        -2i\,e^{\mathfrak L S(r^*)}
    \right].
    \label{eq:app-second-WKB}
\end{align}
This gives the continuation of the solution that is ingoing at the horizon through the classically forbidden region between the two turning points.

\subsubsection{The outer turning point}
We now consider the turning point on the asymptotic side of the photon sphere, denoted by $r_2^*$, which is relevant for the exterior problem in Section \ref{sec:large L analysis}. Since $\partial_{r^*}U(r_2^*)<0$, we introduce
\begin{align}
    \zeta
    \equiv
    \left(
        \mathfrak L^2
        \left|\partial_{r^*}U(r_2^*)\right|
    \right)^{1/3}
    (r^*-r_2^*),
\end{align}
so that $\zeta<0$ inside the barrier. The master-field equation again reduces to the Airy equation, with general solution \eqref{Airy solutions}. To determine the appropriate combination that continues the ingoing solution, we match to the second WKB region. We write
\begin{align}
    S(r^*)
    &=
    \int_{r_1^*}^{r_2^*}\dd\widetilde r^*\,\lambda(\widetilde r^*)
    -
    \int_{r^*}^{r_2^*}\dd\widetilde r^*\,\lambda(\widetilde r^*) \equiv
    J-I(r^*).
\end{align}
Near the turning point, $\lambda(r^*)
    \simeq
    \sqrt{
        \left|\partial_{r^*}U(r_2^*)\right|
        (r_2^*-r^*)
    }$, and therefore
\begin{align}
    \mathfrak L I(r^*)
    \simeq
    \frac{2}{3}(-\zeta)^{3/2}.
\end{align}
Thus, $e^{\mathfrak L I}$ and $e^{-\mathfrak L I}$ match respectively onto the growing $\mathrm{Bi}(-\zeta)$ and decaying $\mathrm{Ai}(-\zeta)$ solutions, and \eqref{eq:app-second-WKB} becomes, up to overall normalization
\begin{align}
    \phi(\zeta)
    \propto
    \mathrm{Ai}(-\zeta)
    +\frac{i}{4}e^{-2\mathfrak L J}\mathrm{Bi}(-\zeta).
    \label{eq:app-outer-Airy}
\end{align}
For sufficiently separated turning points, $\mathfrak L J\gg1$, and the Bi component is exponentially suppressed. This remains true when the turning points approach the photon sphere as $\mathfrak L^{-\beta}$, provided $\beta<1/2$, since $\mathfrak L J\sim\mathfrak L^{1-2\beta}$. The leading solution near the outer turning point is therefore $\mathrm{Ai}(-\zeta)$, as used for the exterior problem in Section \ref{sec:large L analysis}.

\subsubsection{The outer WKB region}
Finally, continuing the leading Airy solution across the outer turning point and matching to the oscillatory WKB region $r^*>r_2^*$ selects the following combination, up to an overall normalization
\begin{align}
    \phi(r^*)
    \propto
    \frac{1}{\sqrt{\kappa(r^*)}}
    \cos\left[
        \mathfrak L
        \int_{r_2^*}^{r^*}\dd\widetilde r^*\,
        \kappa(\widetilde r^*)
        -\frac{\pi}{4}
    \right],
\end{align}
for $\kappa$ given in \eqref{def kappa}. 

\subsection{Coalescing turning points and the parabolic-cylinder solution}
\label{app:WKB-PC}
We now consider the regime $\beta\geq1/2$, in which the two Airy regions merge into a single region around the photon-sphere maximum. We introduce\footnote{To compare with Section \ref{sec:uniform}, note that the rescaled tortoise coordinate $y^*$ is related to the radial coordinate $y$ defined in \eqref{eq:uniform-scaling} by $y=f_{\mathrm{ph}}y^*+O(\mathfrak L^{-1/2})$. Using this relation, together with the definition of $a$ in \eqref{eq:a-q}, the solution \eqref{eq:app-PC-solution} reproduces \eqref{eq:parabolic-solution}.}
\begin{align}
    y^*=\sqrt{\mathfrak L}(r^*-\rph^*),
    \qquad
    \omega=\Omega_{\rm ph}\mathfrak L+W,
    \qquad
    W=\Order(1).
\end{align}
Expanding the optical potential about its maximum, the master-field equation becomes
\begin{align}
    \partial_{y^*}^2\phi
    +\left(
        \Omega_{\rm ph}^4 (y^*)^2
        +2\Omega_{\rm ph}W
    \right)\phi
    =0,
\end{align}
where we used $-\frac12\partial_{r^*}^2U(\rph)=\Omega_{\rm ph}^4$. Denoting $\nu_{\rm PC}=-\frac12+iW/\Omega_{\rm ph}$, the solution selected by the ingoing horizon condition is
\begin{align}
    \phi_{\rm in}(y^*)
    \propto
    D_{\nu_{\rm PC}}
    \left(
        -e^{-i\pi/4}\sqrt{2}\,\Omega_{\rm ph}y^*
    \right).
    \label{eq:app-PC-solution}
\end{align}
This is the parabolic-cylinder solution used in Section \ref{sec:large L analysis}.

\bibliographystyle{unsrtnat}
\bibliography{biblio}

@article{Anderson2008,
  author        = {Anderson, Michael T.},
  title         = {On boundary value problems for {Einstein} metrics},
  journal       = {Geom. Topol.},
  volume        = {12},
  year          = {2008},
  pages         = {2009--2045},
  doi           = {10.2140/gt.2008.12.2009},
  eprint        = {math/0612647},
  archivePrefix = {arXiv}
}

@inproceedings{Kodama:2003ck,
    author = "Kodama, Hideo and Ishibashi, Akihiro",
    title = "{Stability of generalized static black holes in higher dimensions}",
    booktitle = "{7th Hungarian Relativity Workshop (RW 2003)}",
    eprint = "gr-qc/0312012",
    archivePrefix = "arXiv",
    pages = "3--18",
    month = "12",
    year = "2003"
}

@article{Morgan:2009pn,
    author = "Morgan, Jaqueline and Cardoso, Vitor and Miranda, Alex S. and Molina, C. and Zanchin, Vilson T.",
    title = "{Gravitational quasinormal modes of AdS black branes in d spacetime dimensions}",
    eprint = "0907.5011",
    archivePrefix = "arXiv",
    primaryClass = "hep-th",
    doi = "10.1088/1126-6708/2009/09/117",
    journal = "JHEP",
    volume = "2009",
    number = "09",
    pages = "117",
    year = "2009"
}

@article{Banihashemi:2025qqi,
    author = "Banihashemi, Batoul and Shaghoulian, Edgar and Shashi, Sanjit",
    title = "{Thermal effective actions from conformal boundary conditions in gravity}",
    eprint = "2503.17471",
    archivePrefix = "arXiv",
    primaryClass = "hep-th",
    doi = "10.1088/1361-6382/adee72",
    journal = "Class. Quant. Grav.",
    volume = "42",
    number = "15",
    pages = "155004",
    year = "2025"
}

@article{Allameh:2025gsa,
    author = "Allameh, Kuroush and Shaghoulian, Edgar",
    title = "{Timelike Liouville theory and AdS$_{3}$ gravity at finite cutoff}",
    eprint = "2508.03236",
    archivePrefix = "arXiv",
    primaryClass = "hep-th",
    doi = "10.1007/JHEP09(2026)264",
    journal = "JHEP",
    volume = "09",
    pages = "264",
    year = "2026"
}

@article{Hashimoto:2023buz,
    author = "Hashimoto, Koji and Sugiura, Kakeru and Sugiyama, Katsuyuki and Yoda, Takuya",
    title = "{Photon sphere and quasinormal modes in AdS/CFT}",
    eprint = "2307.00237",
    archivePrefix = "arXiv",
    primaryClass = "hep-th",
    reportNumber = "KUNS-2971",
    doi = "10.1007/JHEP10(2023)149",
    journal = "JHEP",
    volume = "10",
    pages = "149",
    year = "2023"
}

@misc{GalanteGeorgescuSvesko,
    author = "Galante, Damián A. and Georgescu, Silvia and Svesko, Andrew",
    note = "To appear"
}

@article{Witten2021,
  author        = {Witten, Edward},
  title         = {A note on boundary conditions in {Euclidean} gravity},
  journal       = {Rev. Math. Phys.},
  volume        = {33},
  number        = {10},
  year          = {2021},
  pages         = {2140004},
  doi           = {10.1142/S0129055X21400043},
  eprint        = {1805.11559},
  archivePrefix = {arXiv},
  primaryClass  = {hep-th}
}

@article{AnAnderson2021,
  author        = {An, Zhongshan and Anderson, Michael T.},
  title         = {The initial boundary value problem and quasi-local Hamiltonians in general relativity},
  journal       = {Class. Quant. Grav.},
  volume        = {38},
  number        = {15},
  year          = {2021},
  pages         = {154001},
  doi           = {10.1088/1361-6382/ac0a86},
  eprint        = {2103.15673},
  archivePrefix = {arXiv},
  primaryClass  = {gr-qc}
}

@article{OdakSpeziale2021,
  author        = {Odak, Gloria  and Speziale, Simone},
  title         = {{Brown--York} charges with mixed boundary conditions},
  journal       = {JHEP},
  volume        = {11},
  year          = {2021},
  pages         = {224},
  doi           = {10.1007/JHEP11(2021)224},
  eprint        = {2109.02883},
  archivePrefix = {arXiv},
  primaryClass  = {hep-th}
}

@article{AnninosGalanteManeerat2023,
  author        = {Anninos, Dionysios and Galante, Dami\'an A. and Maneerat, Chawakorn},
  title         = {Gravitational observatories},
  journal       = {JHEP},
  volume        = {12},
  year          = {2023},
  pages         = {024},
  doi           = {10.1007/JHEP12(2023)024},
  eprint        = {2310.08648},
  archivePrefix = {arXiv},
  primaryClass  = {hep-th}
}

@article{AnninosAriasGalanteManeerat2025,
  author        = {Anninos, Dionysios and Arias, Ra\'ul and Galante, Dami\'an A. and Maneerat, Chawakorn},
  title         = {Gravitational observatories in {$\mathrm{AdS}_4$}},
  journal       = {JHEP},
  volume        = {07},
  year          = {2025},
  pages         = {234},
  doi           = {10.1007/JHEP07(2025)234},
  eprint        = {2412.16305},
  archivePrefix = {arXiv},
  primaryClass  = {hep-th}
}

@article{LiuReallSantosWiseman2025,
  author        = {Liu, Xiaoyi and Reall, Harvey S. and Santos, Jorge E. and Wiseman, Toby},
  title         = {Ill-posedness of the {Cauchy} problem for linearized gravity in a cavity with conformal boundary conditions},
  journal       = {Class. Quant. Grav.},
  volume        = {42},
  number        = {23},
  year          = {2025},
  pages         = {235003},
  doi           = {10.1088/1361-6382/ae1b60},
  eprint        = {2505.20410},
  archivePrefix = {arXiv},
  primaryClass  = {gr-qc}
}

@article{AnninosGalanteGeorgescuManeeratSvesko2026,
  author        = {Anninos, Dionysios and Galante, Dami\'an A. and Georgescu, Silvia and Maneerat, Chawakorn and Svesko, Andrew},
  title         = {The stretched horizon limit},
  journal       = {JHEP},
  volume        = {05},
  year          = {2026},
  pages         = {234},
  doi           = {10.1007/JHEP05(2026)234},
  eprint        = {2512.16738},
  archivePrefix = {arXiv},
  primaryClass  = {hep-th}
}

@article{KodamaIshibashiSeto2000,
  author        = {Kodama, Hideo and Ishibashi, Akihiro and Seto, Osamu},
  title         = {Brane world cosmology: gauge-invariant formalism for perturbation},
  journal       = {Phys. Rev. D},
  volume        = {62},
  year          = {2000},
  pages         = {064022},
  doi           = {10.1103/PhysRevD.62.064022},
  eprint        = {hep-th/0004160},
  archivePrefix = {arXiv}
}

@article{KodamaIshibashi2003,
  author        = {Kodama, Hideo and Ishibashi, Akihiro},
  title         = {A master equation for gravitational perturbations of maximally symmetric black holes in higher dimensions},
  journal       = {Prog. Theor. Phys.},
  volume        = {110},
  year          = {2003},
  pages         = {701--722},
  doi           = {10.1143/PTP.110.701},
  eprint        = {hep-th/0305147},
  archivePrefix = {arXiv}
}

@article{CardosoEtAl2009,
  author        = {Cardoso, Vitor and Miranda, Alex S. and Berti, Emanuele and Witek, Helvi and Zanchin, Vilson T.},
  title         = {Geodesic stability, {Lyapunov} exponents and quasinormal modes},
  journal       = {Phys. Rev. D},
  volume        = {79},
  year          = {2009},
  pages         = {064016},
  doi           = {10.1103/PhysRevD.79.064016},
  eprint        = {0812.1806},
  archivePrefix = {arXiv},
  primaryClass  = {hep-th}
}

@article{MichalogiorgakisPufu2007,
  author        = {Michalogiorgakis, Georgios and Pufu, Silviu S.},
  title         = {Low-lying gravitational modes in the scalar sector of the global {$\mathrm{AdS}_4$} black hole},
  journal       = {JHEP},
  volume        = {02},
  year          = {2007},
  pages         = {023},
  doi           = {10.1088/1126-6708/2007/02/023},
  eprint        = {hep-th/0612065},
  archivePrefix = {arXiv}
}

@article{Siopsis2007,
  author        = {Siopsis, George},
  title         = {Low frequency quasi-normal modes of {$\mathrm{AdS}$} black holes},
  journal       = {JHEP},
  volume        = {05},
  year          = {2007},
  pages         = {042},
  doi           = {10.1088/1126-6708/2007/05/042},
  eprint        = {hep-th/0702079},
  archivePrefix = {arXiv}
}

@article{KovtunSonStarinets2005,
  author        = {Kovtun, Pavel and Son, Dam T. and Starinets, Andrei O.},
  title         = {Viscosity in strongly interacting quantum field theories from black hole physics},
  journal       = {Phys. Rev. Lett.},
  volume        = {94},
  year          = {2005},
  pages         = {111601},
  doi           = {10.1103/PhysRevLett.94.111601},
  eprint        = {hep-th/0405231},
  archivePrefix = {arXiv}
}

@article{BanihashemiShaghoulianShashi2025,
  author        = {Banihashemi, Batoul and Shaghoulian, Edgar and Shashi, Sanjit},
  title         = {Flat space gravity at finite cutoff},
  journal       = {Class. Quant. Grav.},
  volume        = {42},
  number        = {3},
  year          = {2025},
  pages         = {035010},
  doi           = {10.1088/1361-6382/ada2d7},
  eprint        = {2409.07643},
  archivePrefix = {arXiv},
  primaryClass  = {hep-th}
}

@article{GalanteManeeratSvesko2025,
  author        = {Galante, Dami\'an A. and Maneerat, Chawakorn and Svesko, Andrew},
  title         = {Conformal boundaries near extremal black holes},
  journal       = {Class. Quant. Grav.},
  volume        = {42},
  number        = {19},
  year          = {2025},
  pages         = {195003},
  doi           = {10.1088/1361-6382/ae0408},
  eprint        = {2504.14003},
  archivePrefix = {arXiv},
  primaryClass  = {hep-th}
}

@article{GalanteMyersZikopoulos2026,
  author        = {Galante, Dami\'an A. and Myers, Robert C. and Zikopoulos, Themistocles},
  title         = {Conformal boundary conditions and higher curvature gravity},
  journal       = {JHEP},
  volume        = {06},
  year          = {2026},
  pages         = {268},
  doi           = {10.1007/JHEP06(2026)268},
  eprint        = {2512.10930},
  archivePrefix = {arXiv},
  primaryClass  = {hep-th}
}

@article{ClaudelVirbhadraEllis2001,
  author        = {Claudel, Clarissa-Marie and Virbhadra, K. S. and Ellis, George F. R.},
  title         = {The geometry of photon surfaces},
  journal       = {J. Math. Phys.},
  volume        = {42},
  number        = {2},
  year          = {2001},
  pages         = {818--838},
  doi           = {10.1063/1.1308507},
  eprint        = {gr-qc/0005050},
  archivePrefix = {arXiv}
}

@article{SchutzWill1985,
  author        = {Schutz, Bernard F. and Will, Clifford M.},
  title         = {Black hole normal modes: a semianalytic approach},
  journal       = {Astrophys. J. Lett.},
  volume        = {291},
  year          = {1985},
  pages         = {L33--L36},
  doi           = {10.1086/184453}
}

@article{IyerWill1987,
  author        = {Iyer, Sai and Will, Clifford M.},
  title         = {Black-hole normal modes: a {WKB} approach. {I}. Foundations and application of a higher-order {WKB} analysis of potential-barrier scattering},
  journal       = {Phys. Rev. D},
  volume        = {35},
  year          = {1987},
  pages         = {3621--3631},
  doi           = {10.1103/PhysRevD.35.3621}
}

@book{Olver1974,
  author    = {Olver, Frank W. J.},
  title     = {Asymptotics and Special Functions},
  publisher = {Academic Press},
  address   = {New York},
  year      = {1974},
  isbn      = {978-0-12-525850-0}
}

@article{IshibashiWald2004,
  author        = {Ishibashi, Akihiro and Wald, Robert M.},
  title         = {Dynamics in non-globally-hyperbolic static spacetimes. {III}. {Anti-de Sitter} spacetime},
  journal       = {Class. Quant. Grav.},
  volume        = {21},
  number        = {12},
  year          = {2004},
  pages         = {2981--3014},
  doi           = {10.1088/0264-9381/21/12/012},
  eprint        = {hep-th/0402184},
  archivePrefix = {arXiv}
}

@article{DiasSantos2013,
  author        = {Dias, {\'O}scar J. C. and Santos, Jorge E.},
  title         = {Boundary conditions for {Kerr--AdS} perturbations},
  journal       = {JHEP},
  volume        = {10},
  year          = {2013},
  pages         = {156},
  doi           = {10.1007/JHEP10(2013)156},
  eprint        = {1302.1580},
  archivePrefix = {arXiv},
  primaryClass  = {hep-th}
}

@article{AnninosGalanteManeerat2024,
  author        = {Anninos, Dionysios and Galante, Dami{\'a}n A.
                   and Maneerat, Chawakorn},
  title         = {Cosmological observatories},
  journal       = {Class. Quant. Grav.},
  volume        = {41},
  number        = {16},
  pages         = {165009},
  year          = {2024},
  doi           = {10.1088/1361-6382/ad5824},
  eprint        = {2402.04305},
  archivePrefix = {arXiv},
  primaryClass  = {hep-th}
}

@article{PerlickTsupko2022,
  author        = {Perlick, Volker and Tsupko, Oleg Yu.},
  title         = {Calculating black hole shadows:
                   Review of analytical studies},
  journal       = {Phys. Rept.},
  volume        = {947},
  pages         = {1--39},
  year          = {2022},
  doi           = {10.1016/j.physrep.2021.10.004},
  eprint        = {2105.07101},
  archivePrefix = {arXiv},
  primaryClass  = {gr-qc}
}

@article{CederbaumGalloway2021,
  author        = {Cederbaum, Carla and Galloway, Gregory J.},
  title         = {Photon surfaces with equipotential time-slices},
  journal       = {J. Math. Phys.},
  volume        = {62},
  number        = {3},
  pages         = {032504},
  year          = {2021},
  doi           = {10.1063/5.0031280},
  eprint        = {1910.04220},
  archivePrefix = {arXiv},
  primaryClass  = {math.DG}
}

@article{Herzog2003Sound,
  author        = {Herzog, Christopher P.},
  title         = {The Sound of {M}-Theory},
  journal       = {Phys. Rev. D},
  volume        = {68},
  pages         = {024013},
  year          = {2003},
  doi           = {10.1103/PhysRevD.68.024013},
  eprint        = {hep-th/0302086},
  archivePrefix = {arXiv},
  primaryClass  = {hep-th}
}

@article{KovtunStarinets2005,
  author        = {Kovtun, Pavel K. and Starinets, Andrei O.},
  title         = {Quasinormal modes and holography},
  journal       = {Phys. Rev. D},
  volume        = {72},
  pages         = {086009},
  year          = {2005},
  doi           = {10.1103/PhysRevD.72.086009},
  eprint        = {hep-th/0506184},
  archivePrefix = {arXiv},
  primaryClass  = {hep-th}
}

@article{MirandaMorganZanchin2008,
  author        = {Miranda, Alex S. and Morgan, Jaqueline
                   and Zanchin, Vilson T.},
  title         = {Quasinormal modes of plane-symmetric black holes
                   according to the {AdS/CFT} correspondence},
  journal       = {JHEP},
  volume        = {11},
  pages         = {030},
  year          = {2008},
  doi           = {10.1088/1126-6708/2008/11/030},
  eprint        = {0809.0297},
  archivePrefix = {arXiv},
  primaryClass  = {hep-th}
}

@article{DilesEtAl2020,
  author        = {Diles, Saulo M. and Mamani, Luis A. H.
                   and Miranda, Alex S. and Zanchin, Vilson T.},
  title         = {Third-order relativistic hydrodynamics:
                   dispersion relations and transport coefficients
                   of a dual plasma},
  journal       = {JHEP},
  volume        = {05},
  pages         = {019},
  year          = {2020},
  doi           = {10.1007/JHEP05(2020)019},
  eprint        = {1909.05199},
  archivePrefix = {arXiv},
  primaryClass  = {hep-th}
}

@article{HadarEtAl2022,
  author        = {Hadar, Shahar and Kapec, Daniel
                   and Lupsasca, Alexandru and Strominger, Andrew},
  title         = {Holography of the Photon Ring},
  journal       = {Class. Quant. Grav.},
  volume        = {39},
  number        = {21},
  pages         = {215001},
  year          = {2022},
  doi           = {10.1088/1361-6382/ac8d43},
  eprint        = {2205.05064},
  archivePrefix = {arXiv},
  primaryClass  = {gr-qc}
}

@misc{AnninosAriasToAppear,
  author = {Anninos, Dionysios and Arias, Ra{\'u}l
            and Galante, Dami{\'a}n A.},
  note   = {To appear}
}

@article{Kovtun:2003wp,
    author = "Kovtun, Pavel and Son, Dam T. and Starinets, Andrei O.",
    title = "{Holography and hydrodynamics: Diffusion on stretched horizons}",
    eprint = "hep-th/0309213",
    archivePrefix = "arXiv",
    reportNumber = "UW-PT-03-21, INT-PUB-03-17",
    doi = "10.1088/1126-6708/2003/10/064",
    journal = "JHEP",
    volume = "10",
    pages = "064",
    year = "2003"
}

@article{Donos:2015gia,
    author = "Donos, Aristomenis and Gauntlett, Jerome P.",
    title = "{Navier-Stokes Equations on Black Hole Horizons and DC Thermoelectric Conductivity}",
    eprint = "1506.01360",
    archivePrefix = "arXiv",
    primaryClass = "hep-th",
    reportNumber = "DCPT-15-31, IMPERIAL-TP-2015-JG-01",
    doi = "10.1103/PhysRevD.92.121901",
    journal = "Phys. Rev. D",
    volume = "92",
    number = "12",
    pages = "121901",
    year = "2015"
}

@article{Iqbal:2008by,
    author = "Iqbal, Nabil and Liu, Hong",
    title = "{Universality of the hydrodynamic limit in AdS/CFT and the membrane paradigm}",
    eprint = "0809.3808",
    archivePrefix = "arXiv",
    primaryClass = "hep-th",
    reportNumber = "MIT-CTP-3983",
    doi = "10.1103/PhysRevD.79.025023",
    journal = "Phys. Rev. D",
    volume = "79",
    pages = "025023",
    year = "2009"
}

@article{Fournodavlos:2021eye,
    author = "Fournodavlos, Grigorios and Smulevici, Jacques",
    title = "{The Initial Boundary Value Problem in General Relativity: The Umbilic Case}",
    eprint = "2104.08851",
    archivePrefix = "arXiv",
    primaryClass = "gr-qc",
    doi = "10.1093/imrn/rnab359",
    journal = "Int. Math. Res. Not.",
    volume = "2023",
    number = "5",
    pages = "3790--3807",
    year = "2023"
}

@article{Hamdan:2026olq,
    author = "Hamdan, Weam Abou and Maneerat, Chawakorn",
    title = "{Undulating Conformal Boundaries in 3D Gravity}",
    eprint = "2605.08058",
    archivePrefix = "arXiv",
    primaryClass = "hep-th",
    month = "5",
    year = "2026"
}

@article{Bredberg:2010ky,
    author = "Bredberg, Irene and Keeler, Cynthia and Lysov, Vyacheslav and Strominger, Andrew",
    title = "{Wilsonian Approach to Fluid/Gravity Duality}",
    eprint = "1006.1902",
    archivePrefix = "arXiv",
    primaryClass = "hep-th",
    doi = "10.1007/JHEP03(2011)141",
    journal = "JHEP",
    volume = "03",
    pages = "141",
    year = "2011"
}

@article{Anninos:2011zn,
	archiveprefix = {arXiv},
	author = {Anninos, Dionysios and Anous, Tarek and Bredberg, Irene and Ng, Gim Seng},
	doi = {10.1007/JHEP05(2012)107},
	eprint = {1110.3792},
	journal = {JHEP},
	pages = {107},
	primaryclass = {hep-th},
	title = {{Incompressible Fluids of the de Sitter Horizon and Beyond}},
	volume = {05},
	year = {2012}}

@article{York1972ConformalVfive,
 author = {York, Jr., James W.},
 title = {Role of Conformal Three-Geometry in the Dynamics of Gravitation},
 journal = {Phys. Rev. Lett.},
 volume = {28},
 number = {16},
 pages = {1082--1085},
 year = {1972},
 doi = {10.1103/PhysRevLett.28.1082}
}

@article{IshibashiKodama2003StabilityVfive,
 author = {Ishibashi, Akihiro and Kodama, Hideo},
 title = {Stability of Higher-Dimensional {Schwarzschild} Black Holes},
 journal = {Prog. Theor. Phys.},
 volume = {110},
 number = {5},
 pages = {901--919},
 year = {2003},
 doi = {10.1143/PTP.110.901},
 eprint = {hep-th/0305185},
 archivePrefix = {arXiv}
}

\end{document}